\documentclass[twoside,english,aps,prc,twocolumns,reprint,showpacs,preprintnumbers,superscriptaddress,nofootinbib,floatfix,10pt]{revtex4-1}
\usepackage[T1]{fontenc}
\usepackage[utf8]{inputenc}
\usepackage{color}
\usepackage{babel}
\usepackage{bm}
\usepackage{simpler-wick}
\usepackage{amsmath}
\usepackage{amsthm}
\usepackage{amssymb}
\usepackage{graphicx}
\usepackage{simplewick} 
\usepackage[unicode=true,
 bookmarks=true,bookmarksnumbered=false,bookmarksopen=false,
 breaklinks=false,pdfborder={0 0 0},pdfborderstyle={},backref=false,colorlinks=true]
 {hyperref}
\hypersetup{pdftitle={SU4Interaction},
 pdfauthor={BNLU},
 unicode=true, bookmarks=true,bookmarksnumbered=false,bookmarksopen=false, breaklinks=false,pdfborder={0 0 1},backref=false,colorlinks=true, allcolors=blue}

\makeatletter
\usepackage{amsfonts,amssymb,latexsym,amsmath}
\usepackage{graphicx}
\usepackage{times}
\usepackage{slashed}
\allowdisplaybreaks[4]

\makeatother

\begin{document}
\title{A Predictive Nuclear Effective Field Theory \\
from Wilsonian Renormalization Group Principles II \\
Hidden unitary transformations}

\author{Bing-Nan Lu}
\email{bnlv@gscaep.ac.cn}

\affiliation{Graduate School of China Academy of Engineering Physics, Beijing 100193,
China}
\author{Bao-Ge Deng}
\affiliation{Institute of Theoretical Physics, Chinese Academy of Sciences, Beijing
100190, China}
\affiliation{School of Physical Sciences, University of Chinese Academy of Sciences,
Beijing 100049, China}
\author{Jia-Ai Shi}
\affiliation{Graduate School of China Academy of Engineering Physics, Beijing 100193,
China}
\author{Chen-Can Wang}
\affiliation{Graduate School of China Academy of Engineering Physics, Beijing 100193,
China}

\begin{abstract}

A fundamental challenge in nuclear effective field theories (EFTs) is to demonstrate the equivalence of interactions built at different momentum cutoffs and calibrated to the same set of low-momentum observables. Focusing on contact terms, we explicitly construct the unitary transformation that connects these EFT Hamiltonians. By replacing the conventional \emph{relative}-momentum cutoff with a universal single-particle \emph{absolute}-momentum cutoff, we show that this transformation can be systematically implemented using the similarity renormalization group (SRG) technique. The key insight is that such a regularization scheme enables a manifest decoupling of the many-body Fock space, avoiding transitions of particles in intermediate states into high-momentum subspaces associated with conventional \emph{relative}-momentum regulators. The derivation closely parallels the momentum-shell integration formalism of the Wilsonian RG flow. Reformulating the path integral in terms of unitary transformations allows a more intuitive, energy-independent analysis of nonrelativistic many-body systems within the EFT framework. In particular, this framework ensures that effective interactions calibrated in few-body sectors remain unitarily equivalent when applied to many-body systems, a central requirement not yet fulfilled in nuclear many-body calculations. We also show that the breaking of Galilean invariance due to the \emph{absolute}-momentum regularization induces irrelevant operators controlled by the RG scaling behavior, without introducing further complications. We numerically demonstrate our theoretical analysis using a one-dimensional toy model of bosons. These results provide a useful tool for analyzing the RG evolution of nonrelativistic many-body systems.

\end{abstract}
\maketitle

\section{Introduction}
\label{sec:introduction}

Effective field theory (EFT) provides a systematic framework for describing physical systems with a clear hierarchy of scales~\cite{Weinberg1979PhysicaA96-327,Georgi1993ARNPS43-209}. Its applications span the Standard Model EFT in high-energy physics~\cite{Brivio2019PhysRept793-1}, chiral EFT in nuclear physics~\cite{Epelbaum2009RMP81-1773,Machleidt2011PhysRept503-1,Hammer2020RMP92-025004}, and the universal description of ultracold atomic gases~\cite{Braaten2006PhysRept428-259}. Practical EFT calculations generally require regularization, often implemented through an explicit momentum cutoff $\Lambda$ that separates resolved long-wavelength physics from unresolved short-distance dynamics. Wilsonian renormalization group (RG) theory formalizes this separation by systematically integrating out high-momentum degrees of freedom and encoding their effects in the running of low-energy constants (LECs)~\cite{Wilson1971PRB4-3174,Wilson1971PRB4-3184,Wilson1974PhysRept12-75,Polchinski1984NPB231-269}.
Renormalization of the EFT seeks regulator-independent predictions up to higher-order corrections.

Most studies of renormalization in nuclear EFT have focused on nucleon-nucleon (NN) scattering~\cite{Birse1999_PLB464-169,Valderrama2016_IJMPE25-1641007,Epelbaum2017_NPB925-161,Epelbaum2018_EPJA54-186,Epelbaum2020_EPJA56-152,Contessi2024_PLB856-138903} and the associated few-body problems~\cite{Bedaque2000_NPA676-357,Harada2013_PRD87-085006,Epelbaum2017_EPJA53-98}, for which analytical tools and accurate numerical solutions enable stringent consistency checks. The renormalization schemes developed in this context differ notably in their treatment of the ultraviolet cutoff. Some allow the cutoff to extend well beyond the EFT breakdown scale $\Lambda_b$~\cite{Nogga2005_PRC72-054006,Long2011_PRC84-057001,Long2012_PRC86-024001,Long2012_PRC85-034002,Kirscher2013_PLB721-335,Yang2025_PRC112-014004}, whereas others restrict it to scales at or below $\Lambda_b$~\cite{Epelbaum2009_EPJA41-341,Epelbaum2015_EPJA51-53,Reinert2018_EPJA54-86}. A conservative estimate for chiral nuclear EFT is $\Lambda_b\sim500$~MeV~\cite{Epelbaum2015_EPJA51-53}. The choice of cutoff range is closely connected to the treatment of subleading contact interactions and pion-exchange contributions, which may be included perturbatively~\cite{Kaplan1998_PLB424-390,Long2011_PRC84-057001,Long2012_PRC86-024001,Long2012_PRC85-034002,Song2017_PRC96-024002,Valderrama2017_PRC95-054001} or treated nonperturbatively through iteration~\cite{Kaplan1996_NPB478-629,Kirscher2010_EPJA44-239,Entem2003_PRC68-041001,Epelbaum2005_NPA747-362,Entem2017_PRC96-024004,Reinert2018_EPJA54-86}. Both strategies have been applied to selected few-body observables~\cite{Bedaque1999_NPA646-444,Bedaque1999_PRL82-463,Platter2004_PRA70-052101,Platter2005_PLB607-254,Song2017_PRC96-024002,Bazak2019_PRL122-143001,Avraham2026_PLB872-140100}. Extensions to larger systems remain less extensive, although recent studies have explored nuclear matter and medium-mass nuclei, including $^{16}$O, through next-to-leading order (NLO)~\cite{Schiavilla2021_PRC103-054003,Yang2021_PRC103-054304,Sanchez2023_PRC107-064307,Yang2023_EPJA59-233,Contessi2025_arXiv2505-09299}. Nevertheless, achieving systematic control of theoretical uncertainties in many-body predictions at this order remains challenging.

For practical \textit{ab initio} nuclear applications, an equally important question is whether predictions remain stable over a finite range of cutoffs, as expected from renormalization group (RG) invariance. Modern calculations commonly combine nonperturbative many-body solvers with high-precision chiral NN interactions at N$^3$LO or higher and employ relatively low regulator cutoffs of approximately $300$ to $500$~MeV~\cite{Entem2003_PRC68-041001,Epelbaum2005_NPA747-362,Epelbaum2015_EPJA51-53,Entem2017_PRC96-024004,Reinert2018_EPJA54-86}, partly for computational feasibility, since the required model space grows rapidly for hard interactions at a relatively large cutoff~\cite{Roth2014_PRC90-024325}. Constructed at a fixed chiral order and constrained mainly by few-body data, these interactions have enabled increasingly accurate calculations across the nuclear chart~\cite{Stroberg2021_PRL126-022501,Soma2020_PRC101-014318,Maris2022_PRC106-064002}. The sensitivity to the cutoff, however, remains a major source of theoretical uncertainty~\cite{Dyhdalo2016_PRC94-034001,Epelbaum2019_PRC99-024313,Maris2021_PRC103-054001,Maris2022_PRC106-064002}.

Our focus is on the conditions under which EFT Hamiltonians defined at different regulator cutoffs yield equivalent low-energy predictions in many-body systems, up to EFT truncation uncertainties. Such equivalence is expected from consistent Wilsonian RG evolution when the required operators are retained, but its realization in practical \textit{ab initio} calculations requires explicit verification. Although chiral nuclear interactions can be constructed over an appropriate range of cutoffs, many calculations employ only a limited set of cutoff values, which restricts systematic tests of their RG evolution.

Varying the EFT regulator must be distinguished from applying transformations designed to solve a specified Hamiltonian. 
For example, the in-medium similarity renormalization group (IMSRG) decouples a reference state or a selected valence space from the remaining Hilbert space~\cite{Tsukiyama2011_PRL106-222502,Hergert2016_PhysRept621-165}, while wavefunction matching (WFM) transforms an interaction so that its short-range wavefunctions match those of a simpler reference Hamiltonian~\cite{Elhatisari2024_Nature630-59}. When implemented consistently, these methods preserve the predictions of the input Hamiltonian. Their use does not, by itself, establish the equivalence of EFT Hamiltonians built and calibrated at different regulator cutoffs. Truncating the induced interactions introduces approximation errors in the many-body algorithms rather than in the original EFT Hamiltonian itself. A controlled test of EFT regulator independence therefore requires comparing predictions from consistently calibrated Hamiltonians at different cutoffs, with the numerical uncertainties of the many-body methods fully quantified. In the present theoretical study, we assume that the many-body problem can be solved exactly for a given input Hamiltonian and introduce a one-dimensional toy model that meets this requirement.

A further complication arises when the Hamiltonian is readjusted to compensate for discrepancies with experiment, for example, by changing the cutoff and recalibrating three-nucleon couplings using selected observables in medium-mass nuclei and nuclear matter~\cite{Ekstrom2015_PRC91-051301,Drischler2019_PRL122-042501,Huther2020_PLB808-135651,Jiang2024_PRC109-L061302,Elhatisari2024_Nature630-59,Hildenbrand2026_PRL136-062501}. Several studies have reported improved agreement with experiment for particular cutoff choices~\cite{Drischler2019_PRL122-042501,Hoppe2019_PRC100-024318,Huther2020_PLB808-135651}. Such recalibration can absorb both EFT truncation uncertainties and errors associated with the many-body approximation into the same set of couplings. Unless both contributions are independently quantified, the cutoff independence of EFT predictions cannot be unambiguously established. A stringent test in realistic many-body systems should therefore employ interactions calibrated to few-body data alone, without introducing observables from medium-mass nuclei or nuclear matter as additional constraints.

We first examine the conventional implementation of nuclear EFT. In Weinberg's framework, the two-nucleon potential is constructed order by order from two-nucleon-irreducible diagrams and then iterated in the Lippmann-Schwinger equation to obtain NN scattering amplitudes or used in many-body calculations to predict nuclear observables~\cite{Weinberg1990_PLB251-288,Weinberg1991_NPB363-3}. Chiral nuclear interactions constructed in this way have been extended to high orders~\cite{Entem2017_PRC96-024004,Reinert2018_EPJA54-86,Saha2023_PRC107-034002,Somasundaram2024_PRC109-034005,Wu2025_PRC112-014009,Hu2026_PRL137-142502}. Calibrated to NN phase shifts and few-body observables, these interactions have increasingly replaced phenomenological potentials as the primary input to nuclear \textit{ab initio} calculations. In practice, their construction and many-body applications have largely been treated as separate stages, following a workflow similar to that used for phenomenological forces. The two-nucleon interaction is typically calibrated in the center-of-mass frame and subsequently embedded in many-body systems as a Galilean-invariant potential. As shown below, this procedure is not fully consistent with the field-theoretic formulation of EFT renormalization.

Starting from an EFT Lagrangian containing all operators allowed by the underlying symmetries, renormalization can be organized within two distinct but complementary frameworks. The first expresses bare parameters in terms of renormalized couplings order by order in perturbation theory, or equivalently absorbs the ultraviolet divergences into counterterms~\cite{tHooft1972_NPB44-189,Weinberg1979PhysicaA96-327,Gasser1984_AnnPhys158-142,Kaplan1998_PLB424-390}. Leading interactions may need to be resummed to capture nonperturbative phenomena such as bound states. Once renormalization is carried out consistently, the regulator can be removed from the resulting amplitudes or observables at each order. In nuclear EFT, this framework has been extensively applied to NN scattering and few-body systems~\cite{Kaplan1998_PLB424-390,Kaplan1998_NPB534-329,Fleming2000_NPA677-313,Valderrama2011_PRC83-024003,Long2011_PRC84-057001,Long2012_PRC85-034002,Long2012_PRC86-024001,Vanasse2013_PRC88-044001,Song2017_PRC96-024002} and more recently extended to many-body calculations~\cite{Dyhdalo2016_PRC94-034001,Yang2023_EPJA59-233,Schiavilla2021_PRC103-054003,Yang2021_PRC103-054304,Sanchez2023_PRC107-064307,Contessi2025_arXiv2505-09299}. However, power counting prescribes the number of subleading-operator insertions at each order. The resulting higher-order perturbative calculations are technically demanding for many \textit{ab initio} methods, particularly beyond next-to-leading order. Implementations using dimensional regularization or very large cutoffs present additional challenges for numerical many-body solvers.

The second framework employs an explicit cutoff and solves the theory nonperturbatively, with cutoff-dependent bare couplings fixed by matching to the underlying theory or by fitting to physical observables. This approach is standard in lattice field theories, including lattice QCD~\cite{Wilson1974_PRD10-2445,Kogut1975_PRD11-395,Kogut1983_RMP55-775} and lattice nonrelativistic QCD (NRQCD)~\cite{Thacker1991_PRD43-196,Lepage1992_PRD46-4052,Davies1994_PRD50-6963}, the latter being an EFT for heavy quarks. Whereas lattice QCD admits a continuum limit, an EFT such as lattice NRQCD does not require complete removal of the regulator when its truncated action is treated nonperturbatively. Nevertheless, Wilsonian RG implies that low-momentum observables can remain regulator independent up to truncation errors within an appropriate cutoff window, provided that the required operators are retained and their couplings are consistently matched.
This formulation is particularly suited to analyzing how predictions respond to finite or infinitesimal variations of the cutoff.
In particular, the variations arising from different operators must cancel out under the same change in cutoff, which imposes a stringent constraint on the regularization of individual components.
Although simple momentum truncation with operator-specific cutoffs can yield finite predictions comparable to experiment, it generally does not permit such cancellation.
When both perturbative formulations with order-by-order renormalization and lattice formulations with a finite cutoff are available and consistently implemented, the two should yield predictions that agree with each other within their estimated uncertainties.

It is interesting to consider lattice renormalization schemes in nuclear EFT. Although nuclear lattice EFT has been developed as a broad \textit{ab initio} tool~\cite{Lee2004_PRC70-014007,Borasoy2007_EPJA31-105,Lee2009_PPNP63-117,Epelbaum2011_PRL106-192501,Epelbaum2014_PRL112-102501,Lahde2014_PLB732-110,Elhatisari2015_Nature528-111,Elhatisari2016_PRL117-132501,Elhatisari2017_PRL119-222505,Lu2019_PLB797-134863,Lu2020_PRL125-192502,Lu2022_PRL128-242501,Elhatisari2024_Nature630-59,Wu2025_PRC112-014009,Niu2025_PRL135-222504}, calculations are typically performed at a single lattice spacing, and renormalization has been considered only for NN and few-body systems~\cite{Klein2015_PLB747-511,Alarcon2017_EPJA53-83,Klein2018_EPJA54-121,Li2018_PRC98-044002,Harada2016_PRC94-024004,Shi2026_PLB874-140303}. Nevertheless, lattice regularization has implications beyond any specific many-body method. A defining feature of lattice regularization is that it regulates the underlying field modes consistently across all particle-number sectors. This can also be implemented in the continuum by consistently truncating all single-particle momenta. In Ref.~\cite{Lu2026_in_preparation}, we systematically examined the use of a consistent single-particle momentum cutoff in many-body calculations with NLEFT at a small lattice spacing, $a\sim1$~fm, which minimizes discretization errors and permits comparison with continuum calculations. The predicted binding energies were nearly cutoff independent for $300 \leq \Lambda \leq 450$~MeV, with small residual variations associated with omitted operators, including Galilean-invariance-restoration terms. We also show that such a regulator suppresses overbinding and leads to qualitatively different predictions for nuclear matter. The underlying argument is based on the path-integral formulation of Wilsonian RG. Its central principle is that, when a complete operator basis is retained, coarse graining preserves the form of the Lagrangian, with all changes absorbed into the running couplings or LECs.

For nonrelativistic EFTs, including nuclear EFT, however, integrating out high-momentum modes can generate interactions involving time derivatives. Such terms complicate the connection to conventional many-body methods, which are usually formulated in terms of an energy-independent Hamiltonian~\cite{Shankar1994_RMP66-129,Birse1999_PLB464-169,Harada2006_PLB636-305}. Although these interaction terms can be eliminated order by order via field redefinitions~\cite{Kamefuchi1961_NP28-529,Arzt1995_PLB342-189,Furnstahl2001_NPA689-846}, a formulation based directly on unitary transformations of the Hamiltonian would provide a more transparent physical picture of nonrelativistic many-body systems.
In this regard, conventional approaches such as the Okubo~\cite{Okubo1954_PTP12-603} and Lee-Suzuki transformations~\cite{Lee1980_PLB91-173,Suzuki1980_PTP64-2091,Suzuki1982_PTP68-246} and the similarity renormalization group (SRG)~\cite{Glazek1993_PRD48-5863,Wegner1994_AnnPhys506-77,Bogner2007_PRC75-061001,Jurgenson2009_PRL103-082501} are typically formulated in terms of matrix elements of the Hamiltonian and other observables. They do not generally preserve the original local-operator structure explicitly, making it difficult to track the closure and evolution of the EFT operator basis. Although these methods are powerful numerical tools, a formulation directly in terms of local field operators is better suited to the theoretical analysis pursued here and in related works.


In this work, we use unitary transformations to derive the Wilsonian RG flow of a nonrelativistic EFT Hamiltonian with zero-range contact interactions. Our construction parallels momentum-shell integration in the path integral~\cite{Wegner1973_PRA8-401,Wilson1974PhysRept12-75,Polchinski1984NPB231-269} while keeping the Hamiltonian energy independent throughout the flow. We separate the single-particle modes into low- and high-momentum sectors and decompose each contact operator in terms of the corresponding fields. A unitary SRG flow~\cite{Anderson2008_PRC77-037001,RuizArriola2014_PLB728-596} then decouples the low-momentum model space from states containing at least one high-momentum particle. By retaining the operator decomposition at every step, we show that, within the adopted truncation, the low-momentum Hamiltonian preserves its original operator form. The resulting coupled flow equations specify how each successive momentum shell contributes to the running LECs and provide a basis for analyzing operator mixing and RG scaling.

Exactly solvable models offer a stringent test because they eliminate uncertainties associated with approximate many-body methods. Discrepancies with the underlying theory and residual violations of RG invariance can then be traced directly to the EFT formulation and its operator truncation.
For example, one-dimensional models have been applied to illustrate the essential points of EFT renormalizations~\cite{Szpigel1999_arXivnucl-th9906031,Felline2003_PRC68-034003,Valiente2015_FBS56-845,Yong2018_PRA97-043630,Klein2018_EPJA54-233,Jurgenson2009_NPA818-152}.
In this work, we construct a one-dimensional EFT with a microscopic toy model serving as the underlying theory. Observables computed in the toy model provide pseudodata for calibrating the EFT and independent benchmarks for testing its predictions. This controlled setting allows us to systematically assess the predictive power and RG consistency of the EFT under the regularization scheme introduced here.
Here we focus on contact interactions among particles, while long-range forces such as pion-exchange potentials can be studied in an analogous way by applying the same regularization to the pion-nucleon and pion-pion vertices.

This article is organized as follows. Section~\ref{sec:rg-flow} introduces the Hamiltonian and the regularization scheme, describes the decomposition of Fock space and the Hamiltonian, and establishes the closure of the local operator basis and the convergence of the RG flow. Section~\ref{sec:symmetries} discusses the implications of symmetries. Section~\ref{sec:regulator-effects} examines regulator effects. Section~\ref{sec:toy-model} presents numerical results for a one-dimensional toy model. Section~\ref{sec:summary} summarizes the results and discusses their implications.

\section{Wilsonian renormalization group flow}
\label{sec:rg-flow}

In this section, we derive the Wilsonian RG flow using unitary transformations. As a simple representative example, we consider an EFT of identical bosons in one spatial dimension with zero-range contact interactions. The extension to three dimensions and to fermions is straightforward. In the momentum-shell construction adopted here, the cutoff is applied to the single-particle momentum modes entering all interaction operators. This choice differs from commonly used relative-momentum regulators that preserve Galilean invariance. We discuss the consequences of this regulator choice in the following sections and illustrate them through numerical calculations for the bosonic model.

The derivation presented here is formal, in the spirit of the Wilsonian momentum-shell arguments commonly used in field theory textbooks. It is intended to provide physical insight into the RG evolution of nonrelativistic many-body systems and does not constitute a mathematically rigorous proof. Its conclusions should therefore be interpreted at a practical level, with their applicability assessed through numerical verification in the systems of interest.

\subsection{Hamiltonian}

Without loss of generality, we consider a continuum EFT for identical spinless bosons in one spatial dimension, defined by the Hamiltonian
\begin{equation}
H=\int dx:\!-\frac{\Phi^{\dagger}\partial_x^{2}\Phi}{2m}+\frac{C_{2}}{2}(\Phi_{\Lambda}^{\dagger}\Phi_{\Lambda})^{2}+\frac{C_{3}}{6}(\Phi_{\Lambda}^{\dagger}\Phi_{\Lambda})^{3}+\cdots\!:,
\label{eq:Hamiltonian}
\end{equation}
where $\Phi$ and $\Phi^{\dagger}$ are the annihilation and creation operators, respectively, and $m$ and $C_{2,3}$ are the particle mass and coupling constants. The colons denote normal ordering. The ellipsis stands for higher-body forces and interactions containing spatial derivatives. The subscript $\Lambda$ indicates that the interactions act only on particles with momentum below the cutoff $\Lambda$,
\begin{equation}
\Phi_{\Lambda}^{\dagger}(x)=\int\frac{dp}{2\pi}\theta(\Lambda-|p|)\exp(ipx)a_{p}^{\dagger},
\label{eq:op_cutoff}
\end{equation}
where $a_{p}^{\dagger}$ is the creation operator in momentum representation and $\theta$ is the step function serving as a momentum regulator. $\Phi_{\Lambda}^{\dagger}$ creates a particle with a finite extension $\sim\Lambda^{-1}$. Note that the cutoff scheme in Eq.~\eqref{eq:op_cutoff} is imposed on single-particle momenta, in contrast to the conventional EFT regularization that acts on relative momenta. The use of soft regulators such as Gaussian functions is equally valid.

\begin{figure}
\begin{centering}
\includegraphics[width=1\linewidth]{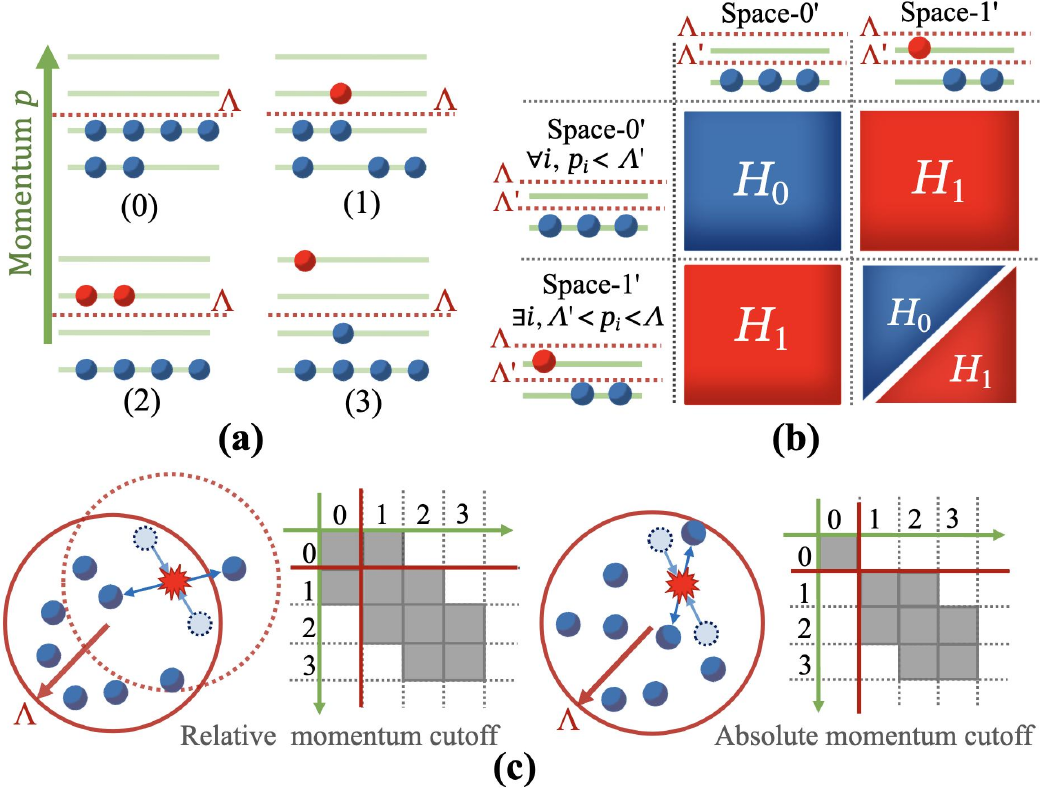}
\par\end{centering}
\caption{(a) Schematic representation of typical many-body configurations. Blue (red) circles denote particles with momenta below (above) the single-particle cutoff $\Lambda$.
(b) Block structure of the Hamiltonian in a momentum-space basis. 
$H_0$ is block diagonal, and $H_1$ vanishes in the low-momentum diagonal block.
(c) Comparison of two regularization schemes. A conventional \emph{relative}-momentum cutoff couples low- and high-momentum states, whereas an \emph{absolute}-momentum cutoff strictly forbids such mixing.} \label{fig:partition}
\end{figure}

Consistent with the EFT framework, we assume that the Hamiltonian in Eq.~\eqref{eq:Hamiltonian} includes all possible local Hermitian interactions compatible with the fundamental spacetime symmetries of one-dimensional translation. Particle-number conservation is enforced by requiring an equal number of creation and annihilation operators in each term. We further impose discrete parity and time-reversal symmetries. Crucially, because the momentum-space regulator explicitly breaks Galilean invariance, terms such as $(\Phi_\Lambda^\dagger \partial_x \Phi_\Lambda)^2$ are permitted. Our regularization scheme, which truncates the \emph{absolute} single-particle momentum rather than the widely adopted \emph{relative}-momentum cutoff used in continuum EFT calculations, plays an essential role in our analysis. We note that this scheme is closely related to lattice field theories~\cite{Wilson1974_PRD10-2445,Lee2009_PPNP63-117,Lahde:2019npb,Dean:2025}. The use of a soft cutoff provides a direct bridge between lattice and continuum formulations, allows for a continuous variation of the cutoff, and thereby enables a systematic analysis of the RG flow~\cite{Klein2015_PLB747-511}.

\subsection{Fock space decomposition}

Fock space is spanned by momentum eigenstates of the form $a_{p_{1}}^{\dagger}a_{p_{2}}^{\dagger}\cdots a_{p_{N}}^{\dagger}|0\rangle$, where $|0\rangle$ denotes the vacuum state and $N\in[0,\infty)$ is the particle number. We decompose this space into a low-momentum subspace $\mathcal{S}_0$ and a complementary high-momentum subspace $\mathcal{S}_1$. The subspace $\mathcal{S}_0$ is spanned by all many-body states $|p_1,\dots,p_N\rangle$ in which every single-particle momentum satisfies $|p_i|<\Lambda$,
\begin{equation*}
\mathcal{S}_0 = \mathrm{span}\big\{ |p_1,\dots,p_N\rangle \,\big|\, |p_i|<\Lambda\ \forall i \big\},
\end{equation*}
while the complementary subspace $\mathcal{S}_1$ contains at least one momentum with $|p_i|\geq\Lambda$,
\begin{equation*}
\mathcal{S}_1 = \mathrm{span}\big\{ |p_1,\dots,p_N\rangle \,\big|\, \exists i,\ |p_i|\geq\Lambda \big\}.
\end{equation*}
The full Fock space is the direct sum $\mathcal{F}=\mathcal{S}_0\oplus\mathcal{S}_1$. Since the analysis is carried out in the full second-quantized formalism, this decomposition is independent of the total particle number and applies separately to each sector of fixed $N$.

For any state $|\phi\rangle\in\mathcal{S}_0$, the action of the Hamiltonian gives $H|\phi\rangle\in\mathcal{S}_0$. This follows because $H$ is built from $\Phi_\Lambda$ and $\Phi_\Lambda^\dagger$, which contain only modes with $|p|<\Lambda$, so no particle is excited above the cutoff. Consequently, $H$ is block diagonal with respect to the decomposition $\mathcal{F}=\mathcal{S}_0\oplus\mathcal{S}_1$, and $\mathcal{S}_1$ can be discarded without affecting the low-momentum physics confined to $\mathcal{S}_0$.

To see this block-diagonal structure explicitly, consider a typical matrix element between two many-body momentum eigenstates,
\begin{align*}
    \mathcal{M} =& \langle 0 | a_{p^\prime_N} a_{p^\prime_{N-1}} \cdots a_{p^\prime_1} 
     \int dx \, \Phi_\Lambda^\dagger (x) \Phi_\Lambda^\dagger(x)  \Phi_\Lambda(x) \Phi_\Lambda(x) \\
     &\times  a^\dagger_{p_1} a^\dagger_{p_2} \cdots a^\dagger_{p_N} |0 \rangle,
\end{align*}
where $p_1,\dots,p_N$ and $p^\prime_1,\dots,p^\prime_N$ denote the initial and final momenta. Suppose that all these momenta lie below $\Lambda$ except for $p^\prime_1$ with $|p^\prime_1|\geq\Lambda$. This matrix element then connects $\mathcal{S}_0$ with $\mathcal{S}_1$. The local operators $\Phi_\Lambda(x)$ and $\Phi_\Lambda^\dagger(x)$ contain only annihilation and creation operators with $|p|<\Lambda$. Therefore, the annihilation operator $a_{p^\prime_1}$ cannot contract with any creation operator to its right, since the latter involve only momenta below $\Lambda$. Hence, the matrix element vanishes identically. By the same reasoning, every matrix element of $H$ between $\mathcal{S}_0$ and $\mathcal{S}_1$ is zero, confirming that the Hamiltonian is block diagonal. The kinetic energy is already diagonal in the momentum basis and does not affect this conclusion.

The block-diagonal structure induced by an \emph{absolute}-momentum cutoff is a central prerequisite for our subsequent analysis. In contrast, the widely used \emph{relative}-momentum cutoff scheme does not lead to a clean decomposition of the Hamiltonian.

In Fig.~\ref{fig:partition}(a), we illustrate several representative many-body configurations. State (0), in which all particle momenta lie below $\Lambda$, resides entirely within subspace 0. The other three states contain at least one excited particle above $\Lambda$ and therefore belong to subspace 1. Fig.~\ref{fig:partition}(c) schematically contrasts the matrix elements of EFT Hamiltonians under different regularization schemes for these states. In conventional cutoff schemes, only the \emph{relative} momentum of an interacting pair is constrained. As a result, particles with nearly identical momenta near $\Lambda$ can scatter to states above $\Lambda$, producing off-diagonal matrix elements between subspaces 0 and 1. Such mixing is \emph{strictly forbidden} in our \emph{absolute}-momentum cutoff scheme, which enforces a block-diagonal Hamiltonian. This momentum leakage of conventional regulators becomes increasingly severe in larger systems, where successive scatterings can couple low-momentum states to configurations containing particles with extraordinarily high momenta.

\subsection{Block structure of the Hamiltonian}

We now introduce a new cutoff $\Lambda^{\prime}$ slightly below $\Lambda$. The original low-momentum subspace $0$ can be further decomposed into a direct sum of subspace $0^{\prime}$, spanned by states with all $|p_i|\leq\Lambda^{\prime}$, and subspace $1^{\prime}$, spanned by states containing at least one momentum $p_i$ satisfying $\Lambda^{\prime}<|p_i|\leq\Lambda$. This decomposition induces a corresponding block-matrix partition of the Hamiltonian, as illustrated in Fig.~\ref{fig:partition}(b).

Following Wilson's approach, we decompose the field operator into low- and high-momentum components,
\begin{equation*}
\Phi_{\Lambda}^{\dagger}=\Phi_{\Lambda^{\prime}}^{\dagger}+\delta\Phi^{\dagger},
\end{equation*}
where $\Phi_{\Lambda^{\prime}}^{\dagger}$ is defined by Eq.~\eqref{eq:op_cutoff} with $\Lambda$ replaced by $\Lambda^{\prime}$. Here, $\Phi_{\Lambda^{\prime}}^{\dagger}$ creates soft modes below $\Lambda^{\prime}$, while $\delta\Phi^{\dagger}$ creates hard modes between $\Lambda^{\prime}$ and $\Lambda$. Substituting this decomposition into $H$ yields an expansion in $\Phi_{\Lambda^{\prime}}^{\dagger}$, $\delta\Phi^{\dagger}$, and their conjugates. We then split the Hamiltonian into two parts, $H=H_{0}+H_{1}$, whose leading terms are
\begin{eqnarray}
H_{0} & = & \int dx:\!-\frac{\Phi^{\dagger}\partial_x^{2}\Phi}{2m}+\frac{C_{2}}{2}(\Phi_{\Lambda^{\prime}}^{\dagger}\Phi_{\Lambda^{\prime}})^{2}+\frac{C_{3}}{6}(\Phi_{\Lambda^{\prime}}^{\dagger}\Phi_{\Lambda^{\prime}})^{3}\nonumber \\
 &  & +2C_{2}(\Phi_{\Lambda^{\prime}}^{\dagger}\Phi_{\Lambda^{\prime}})(\delta\Phi^{\dagger}\delta\Phi)+C_{2}(\Phi_{\Lambda^{\prime}}^{\dagger}\delta\Phi)(\delta\Phi^{\dagger}\delta\Phi)\nonumber \\
 &  & +\frac{3C_{3}}{2}(\Phi_{\Lambda^{\prime}}^{\dagger}\Phi_{\Lambda^{\prime}})^{2}\delta\Phi^{\dagger}\delta\Phi+\frac{C_{3}}{2}(\Phi_{\Lambda^{\prime}}^{\dagger}\delta\Phi)^{2}\delta\Phi^{\dagger}\delta\Phi+\cdots\!:,\nonumber \\
H_{1} & = & \int dx:\!C_{2}(\Phi_{\Lambda^{\prime}}^{\dagger}\Phi_{\Lambda^{\prime}})(\Phi_{\Lambda^{\prime}}^{\dagger}\delta\Phi)+\frac{C_{2}}{2}(\delta\Phi^{\dagger}\Phi_{\Lambda^{\prime}})^{2}\nonumber \\
 &  & +\frac{C_{3}}{2}(\Phi_{\Lambda^{\prime}}^{\dagger}\Phi_{\Lambda^{\prime}})^{2}\Phi_{\Lambda^{\prime}}^{\dagger}\delta\Phi+\frac{C_{3}}{6}(\delta\Phi^{\dagger}\Phi_{\Lambda^{\prime}})^{3}+\cdots\!:.\label{eq:H_decomposition}
\end{eqnarray}
The separation is defined as follows. For each term in $H$, let $m$ and $n$ count the number of $\delta \Phi^\dagger$ and $\delta \Phi$, respectively. We collect all terms with ($m = 0$, $n = 0$) or ($m > 0$, $n > 0$) into $H_0$. The remaining terms with either ($m = 0$, $n > 0$) or ($m > 0$, $n = 0$) constitute $H_1$.

Local operators that do not contain $\delta\Phi$ and $\delta\Phi^\dagger$ remain block diagonal with respect to the decomposition into subspaces $0^{\prime}$ and $1^{\prime}$ by the same reasoning as before. A local operator containing both $\delta\Phi$ and $\delta\Phi^\dagger$ can have nonzero matrix elements only between two states in subspace $1^{\prime}$. This is because $\delta\Phi$ and $\delta\Phi^\dagger$ must contract with high-momentum particles satisfying $\Lambda^{\prime}<|p|<\Lambda$ in the initial and final states, otherwise the matrix element vanishes exactly. Direct contractions between the initial and final states simply produce momentum $\delta$-functions. All such terms are therefore collected in the block-diagonal part $H_0$. A typical nonzero matrix element between two subspace $1^{\prime}$ states reads
\begin{align*}
    \langle 0 | 
    \wick{ 
        \c1 a_{\overline{p^\prime_N}} 
        \c2 a_{p^\prime_{N-1}} 
        \dots 
        \c3 a_{p^\prime_1} 
        \,
        \c1 \delta\Phi^\dagger 
        \c3 \Phi^\dagger  
        \c4 \Phi 
        \c5 \delta\Phi 
        \,
        \c4 a^\dagger_{p_1} 
        \c2 a^\dagger_{p_2} 
        \dots 
        \c5 a^\dagger_{\overline{p_N}} 
    } 
    |0 \rangle,
\end{align*}
where the overlined momenta satisfy $\Lambda^{\prime}<|p|<\Lambda$ and the remaining momenta satisfy $|p|<\Lambda^{\prime}$. The operators $\delta\Phi$ and $\delta\Phi^\dagger$ contract exclusively with the particles carrying overlined momenta.

Similarly, operators that contain $\delta\Phi$ but no $\delta\Phi^\dagger$, or vice versa by Hermitian conjugation, can yield nonzero matrix elements only when the initial or the final state supplies a high-momentum particle to contract with these local operators. Note, however, that this does not require the other state to lie entirely in subspace $0^{\prime}$, because a high-momentum particle in the initial state may also contract directly with one in the final state. As an illustration, consider the following matrix element between two subspace $1^{\prime}$ states,
\begin{align*}
    \langle 0 | 
    \wick{ 
        \c1 a_{\overline{p^\prime_N}} 
        \c2 a_{p^\prime_{N-1}} 
        \dots 
        \c3 a_{p^\prime_1} 
        \;
        \c2 \Phi^\dagger 
        \c3 \Phi^\dagger  
        \c4 \Phi 
        \c5 \delta\Phi 
        \;
        \c5 a^\dagger_{\overline{p_1}} 
        \c4 a^\dagger_{p_2} 
        \dots 
        \c1 a^\dagger_{\overline{p_N}} 
    } 
    |0 \rangle,
\end{align*}
which can be nonzero due to complete contraction. Here, the high-momentum particle with momentum $\overline{p_N}$ remains as a spectator throughout the contraction process. Consequently, operators in $H_1$ vanish when both states lie in subspace $0^{\prime}$, but a diagonal block within subspace $1^{\prime}$ persists. Because the interaction cannot fully eliminate spectator high-momentum particles, a clean separation into purely diagonal and off-diagonal blocks is generally not achievable in this operator basis. Nevertheless, we shall show that this residual block-diagonal part of $H_1$ does not impede the suppression of these operators under SRG flow.

$H_0$ and $H_1$ have distinct block structures with respect to the decomposition of space-$0$ into $0'$ and $1'$. For any state $|\phi\rangle$ in space-$0'$, $H_0|\phi\rangle$ contains only particles with momenta $|p|\leq\Lambda^{\prime}$ and thus remains in $0'$. In contrast, $H_1|\phi\rangle$ excites one or more particles into the momentum shell $\Lambda^{\prime}<|p|<\Lambda$, placing it in space-$1'$. Consequently, $H_0$ is block diagonal, connecting $0'$ to itself and $1'$ to itself, while $H_1$ vanishes in the low-momentum diagonal block, as illustrated in Fig.~\ref{fig:partition}(b). In block-matrix form,
\begin{align}
H_0 = \begin{pmatrix} 
A_0 &  \\ 
 & A_1 
\end{pmatrix}
,
\quad 
H_1 = 
\begin{pmatrix} 
 & b \\ 
b^\dagger & \epsilon 
\end{pmatrix}
,
\label{eq:H0H1matrices}
\end{align}
where the dominant low-energy physics resides in $A_0$, and the remaining blocks originate from the thin momentum shell between $\Lambda^{\prime}$ and $\Lambda$. Both the decomposition of the Fock space and that of the Hamiltonian operators are essential for our construction. Importantly, every operator in $H_1$ contributes to the off-diagonal blocks $b$ or $b^\dagger$. Therefore, applying the standard SRG technique to eliminate these off-diagonal matrix elements also guarantees the removal of the whole $H_1$.

\subsection{Closure of the operator basis}

The construction of a Wilsonian RG flow requires the set of local operators to be closed under evolution. The closure of the operator basis is an essential ingredient in generating the SRG flow. To establish this property, we first prove that the commutator of any two terms in the operator decomposition of Eq.~\eqref{eq:H_decomposition} can be expressed as a linear combination of the same set of operators, that is, they span an infinite-dimensional Lie algebra,
\begin{align}
[O_i, O_j] = \sum_{k \in H} \Gamma_{ij}^k O_k,
\end{align}
where $O_{i,j,k}$ denote the operators in Eq.~\eqref{eq:H_decomposition} without their coefficients and $\Gamma_{ij}^k$ are constants. The proof proceeds by first rewriting the commutator in normal-ordered form using Wick's theorem. Terms without operator contractions commute trivially under normal ordering and thus cancel exactly. The surviving terms each contain at least one short-range, $\delta$-like function arising from operator contractions. For example, the contraction of $\Phi_{\Lambda^\prime}$ and $\Phi_{\Lambda^\prime}^\dagger$ yields a short-range function with range $\sim\Lambda^{\prime-1}$. These short-range functions force the participating local operators to act at the same spatial point, allowing the resulting contributions to be expanded in terms of the local operators already present in Eq.~\eqref{eq:H_decomposition}.

We now demonstrate how the commutator of two local operators leads to an expansion in terms of similar local operators. We apply a Gaussian regulator instead of the hard cutoff introduced earlier,
\begin{align}
  \Phi^\dagger (x) = \int \frac{dp}{2\pi} a^\dagger_p \exp\left(-\frac{p^2}{2\Lambda^2}\right) \exp(ipx),
\label{eq:gaussian-field}
\end{align}
where $a^\dagger_p$ creates a particle with momentum $p$. The use of the smooth cutoff eliminates the long-range oscillatory tails arising from the hard momentum cutoff, making the interactions more local. We note that the specific form of the regulator does not alter our general discussion.

As an example, we choose two operators $O_1$ and $O_2$ appearing in the expansion of the Hamiltonian $H$,
\begin{align}
O_1 &= \int dx \; \phi^\dagger(x) \delta^\dagger(x) \phi(x) \delta(x), \nonumber \\
O_2 &= \int dy \; \phi^\dagger(y) \phi^\dagger(y) \phi(y) \delta(y),
\end{align}
where we set $\phi(x) = \Phi_{\Lambda'}(x)$ and $\delta(x) = \delta\Phi(x)$. The operators $\Phi_{\Lambda'}(x)$ are the truncated local operators defined in Eq.~\eqref{eq:gaussian-field}. Consequently, the only nonzero contractions are
\begin{align*}
\wick{\c1 \phi(x) \c1 \phi^\dagger(y)} = F(x - y), \quad 
\wick{\c1 \delta(x) \c1 \delta^\dagger(y)} = f(x - y).
\end{align*}
The correlation functions in this case become
\begin{align}
F(y) &= \int \frac{dp}{2\pi} \exp\left(-\frac{p^2}{2\Lambda'^2}\right) e^{ipy} 
      = \frac{\Lambda'}{\sqrt{2\pi}} \exp\left( -\frac{\Lambda'^2 y^2}{2} \right), \nonumber \\
f(y) &= \int \frac{dp}{2\pi} \left[ \exp\left(-\frac{p^2}{2\Lambda^2}\right) 
      - \exp\left(-\frac{p^2}{2\Lambda'^2}\right) \right] e^{ipy} \nonumber \\
      &= \frac{\Lambda \exp(-\Lambda^2 y^2 / 2) - \Lambda' \exp(-\Lambda'^2 y^2/2)}{\sqrt{2\pi}}.
\end{align}

We now calculate the commutator of $O_1$ and $O_2$ by expressing their products in normal-ordered form using Wick's theorem,
\begin{align}
    O_1O_2 &= \int dx dy \Big\{ \phi^\dagger(x) \delta^\dagger(x) \phi(x) \delta(x) \phi^\dagger(y) \phi^\dagger(y) \phi(y) \delta(y) \Big\} \nonumber \\
    &= \int dx dy :\Big\{ \phi^\dagger(x) \delta^\dagger(x) \phi(x) \delta(x) \phi^\dagger(y) \phi^\dagger(y) \phi(y) \delta(y)  \nonumber \\
    &\quad + \wick{ \phi^\dagger(x) \delta^\dagger(x) \c1 \phi(x) \delta(x) \c1 \phi^\dagger(y) \phi^\dagger(y) \phi(y) \delta(y) } \nonumber \\
    &\quad + \wick{ \phi^\dagger(x) \delta^\dagger(x) \c1 \phi(x) \delta(x) \phi^\dagger(y) \c1 \phi^\dagger(y) \phi(y) \delta(y) } \Big\}:, \nonumber \\[10pt]
    O_2O_1 &= \int dx dy \Big\{ \phi^\dagger(y) \phi^\dagger(y) \phi(y) \delta(y) \phi^\dagger(x) \delta^\dagger(x) \phi(x) \delta(x) \Big\} \nonumber \\
    &= \int dx dy :\Big\{ \phi^\dagger(y) \phi^\dagger(y) \phi(y) \delta(y) \phi^\dagger(x) \delta^\dagger(x) \phi(x) \delta(x) \nonumber \\
    &\quad + \wick{ \phi^\dagger(y) \phi^\dagger(y) \c1 \phi(y) \delta(y) \c1 \phi^\dagger(x) \delta^\dagger(x) \phi(x) \delta(x) } \nonumber \\
    &\quad + \wick{ \phi^\dagger(y) \phi^\dagger(y) \phi(y) \c1 \delta(y) \phi^\dagger(x) \c1 \delta^\dagger(x) \phi(x) \delta(x) } \nonumber \\
    &\quad + \wick{ \phi^\dagger(y) \phi^\dagger(y) \c1 \phi(y) \c2 \delta(y) \c1 \phi^\dagger(x) \c2 \delta^\dagger(x) \phi(x) \delta(x) } \Big\}:. \nonumber 
\end{align}
The terms without contractions cancel exactly in the commutator. This procedure is commonly applied in many-body methods such as the coupled-cluster method, where the uncontracted terms similarly vanish. Substituting the contractions with the correlation functions yields the commutator
\begin{align}
    [O_1, O_2] &= \int dx dy \Big\{ 2F(x-y) \phi^\dagger(x) \phi^\dagger(y) \delta^\dagger(x) \phi(y) \delta(x) \delta(y) \nonumber \\
    &\quad - F(x-y) \phi^\dagger(y)^2 \delta^\dagger(x) \phi(x) \delta(x) \delta(y) \nonumber \\
    &\quad - f(x-y) \phi^\dagger(x) \phi^\dagger(y)^2 \phi(x) \phi(y) \delta(x) \nonumber \\
    &\quad - F(x-y)f(x-y) \phi^\dagger(y)^2 \phi(x) \delta(x) \Big\}, \nonumber
\end{align}
where the double contraction contributes a term involving the product of local functions.

To express this operator as a series of local operators, we perform a Taylor expansion of the variable $y$ around the point $x$. Taking the second term in the commutator as an example, we find that this expansion generates a series of local derivative terms,
\begin{align*}
& \int dx dy \, F(x-y) \, \phi^\dagger(y)^2 \delta^\dagger(x) \phi(x) \delta(x) \delta(y) \\
= & \int dx dy \, F(x-y) : \delta^\dagger(x) \phi(x) \delta(x) \left\{ \phi^\dagger(y)^2 \delta(y) \right\}: \\
= & \sum_{n=0}^\infty \frac{1}{n!} \int dx dy \, F(x-y) (y-x)^n \\
& \qquad \qquad \times : \delta^\dagger(x) \phi(x) \delta(x) \, \partial_x^n \left\{ \phi^\dagger(x)^2 \delta(x) \right\} : \\
= & \sum_{n=0}^\infty \frac{1}{n!} \left\{ \int dy \, F(y) y^n \right\} \\
 & \qquad \qquad  \times \int dx : \delta^\dagger(x) \phi(x) \delta(x) \, \partial_x^n \left\{ \phi^\dagger(x)^2 \delta(x) \right\} :,
\end{align*}
where we have shifted the integration variable to absorb the local functions into the coupling constants. The remaining terms in the commutator can be treated in a similar manner.

\subsection{Convergence of the SRG flow equation}

The unitary transformation that decouples the low- and high-momentum sectors is constructed using the similarity renormalization group (SRG) flow equation~\cite{Wegner1994_AnnPhys506-77,Glazek1993_PRD48-5863,Bogner2007_PRC75-061001,Tsukiyama2011_PRL106-222502}. This implements a continuous unitary transformation $U(t)$ parameterized by $t \geq 0$. The evolved Hamiltonian $H(t) = U(t)^{-1} H U(t)$ satisfies a flow equation analogous to the Heisenberg equation,
\begin{equation}
i\partial_{t} H(t) = [\eta(t), H(t)], \label{eq:SRGflowequation}
\end{equation}
where we adopt a Hermitian generator of Wegner type, $\eta(t) = i[H_{0}(t), H_{1}(t)]$. We stress that the SRG flow in the parameter $t$ is distinct from the Wilsonian RG flow induced by lowering the cutoff $\Lambda$.

Due to the closure property established above, expanding the commutators in $\eta(t)$ and in Eq.~\eqref{eq:SRGflowequation} within the operator basis of Eq.~\eqref{eq:H_decomposition} shows that the SRG flow induces a continuous evolution of the corresponding coefficients. This evolution is governed by a set of coupled nonlinear differential equations,
\begin{align}
\partial_t c_i = \sum_{j,k \in H} \sum_{m \in H_0} \sum_{n \in H_1} \Gamma^i_{jk} \Gamma^j_{mn} c_k c_m c_n,
\label{eq:SRGflowEq}
\end{align}
where $c_{k, m, n}$ denote the individual coefficients of the operators in Eq.~\eqref{eq:H_decomposition}, and the sums over $m$ and $n$ are restricted to operators belonging to $H_0$ and $H_1$, respectively. The initial conditions at $t = 0$ are set by the LECs $C_{2}, C_{3}, \dots$ defined at the original cutoff $\Lambda$.

We now transition from the operator representation to the matrix representation. Substituting the block forms in Eq.~\eqref{eq:H0H1matrices} into the flow equation Eq.~\eqref{eq:SRGflowequation} shows that the off-diagonal blocks $b$ and $b^\dagger$ in $H_1$ are exponentially suppressed during the SRG evolution,
\begin{align}
   \partial_t \mathrm{Tr}(b^\dagger b) \approx -2 \mathrm{Tr}[(bA_0)^\dagger bA_0] < 0,
\end{align}
where higher-order terms have been neglected. Since every term in $H_1(t)$ possesses nonzero off-diagonal matrix elements, this suppression implies the eventual elimination of $H_1(t)$ itself. In the limit $t\rightarrow\infty$, all terms in $H_{1}(t)$ vanish, the flow reaches a fixed point characterized by $H_{1}=\eta=0$, and the unitary transformation $U(t)$ converges to a well-defined limit $U(\infty)$. Consequently, the resulting Hamiltonian $H(\infty)=U(\infty) H U(\infty)^{-1}$ contains only the block-diagonal part $H_0$, whose coefficients have been renormalized.

We now establish that the SRG flow converges to the fixed point $H_{1} = \eta = 0$ as $t \rightarrow \infty$. The proof follows a logic similar to the derivation of the $V_{\text{low-}k}$ potential~\cite{Bogner2010_PPNP65-94}, with minor adjustments. A key distinction here is that our formulation operates within an abstract Fock space, where an explicit representation of matrix elements is generally unattainable. Nevertheless, the convergence of the flow can be established through a general argument based on the inherent structure of the flow equation.

We decompose the Hamiltonian as $H = H_0 + H_1$, where $H_0$ is block diagonal. Formally, we write
\[
H_{0}=\begin{pmatrix}
A_{0} &  \\
 & A_{1}
\end{pmatrix},\qquad H_{1}=\begin{pmatrix}
 & b\\
b^{\dagger} & \epsilon
\end{pmatrix},
\]
where $A_{0}$, $A_{1}$, and $\epsilon$ are Hermitian matrices. The generator $i\eta$ is then defined by
\[
i\eta = [H_{0},H_{1}] = \begin{pmatrix}
 & A_{0}b - bA_{1} \\
A_{1}b^{\dagger} - b^{\dagger}A_{0} & A_{1}\epsilon - \epsilon A_{1}
\end{pmatrix}.
\]
Substituting $i\eta$ into the Wegner flow equation, $\partial_t H = [i\eta, H]$, we obtain
\begin{align*}
\frac{dH}{dt} = \left[ \begin{pmatrix}
 & A_{0}b - bA_{1} \\
A_{1}b^{\dagger} - b^{\dagger}A_{0} & A_{1}\epsilon - \epsilon A_{1}
\end{pmatrix}, \begin{pmatrix}
A_{0} & b \\
b^{\dagger} & A_{1} + \epsilon
\end{pmatrix} \right].
\end{align*}
Extracting the $(2,1)$ matrix element yields the evolution of the off-diagonal block $b^\dagger$,
\begin{align*}
\frac{db^{\dagger}}{dt}=(A_{1}\epsilon-\epsilon A_{1})b^{\dagger}+(A_{1}b^{\dagger}-b^{\dagger}A_{0})A_{0} \\
-(A_{1}+\epsilon)(A_{1}b^{\dagger}-b^{\dagger}A_{0}).
\end{align*}
The flow of the squared magnitudes of the off-diagonal elements, characterized by $\mathrm{Tr}(b^\dagger b)$, is given by
\begin{align}
\frac{d}{dt} \mathrm{Tr}(b^{\dagger}b) &=  -2 \mathrm{Tr}\left[ (A_{1}b^{\dagger} - b^{\dagger}A_{0} + \frac{1}{2} \epsilon b^\dagger)^{\dagger} \right. \nonumber \\ 
& \left. \times (A_{1}b^{\dagger} - b^{\dagger}A_{0} + \frac{1}{2} \epsilon b^\dagger) \right] + \frac{1}{2}\mathrm{Tr}(\epsilon^2 b^\dagger b).
\label{eq:SRGmagnitude}
\end{align}
Because the block-diagonal part is not perfectly isolated in this decomposition, the resulting expression is not strictly negative definite. However, in the context of a thin momentum shell where $\epsilon$ is small compared to the matrix $H_0$, the terms involving $\epsilon$ can be safely neglected. Alternatively, one can further reduce the difference $\Lambda - \Lambda^\prime$ to suppress the $\epsilon^2$ term. Consequently, the right-hand side is dominated by negative-definite terms quadratic in $A_0$, leading to an exponential suppression of the left-hand side.

The preceding arguments rely on the assumption that the density of many-body states within the momentum shell is sufficiently small. However, this assumption may fail in subspaces with large particle numbers. For instance, consider a one-dimensional system with a uniform momentum-level distribution. The number of many-body states for $N=100$ particles distributed below a cutoff $\Lambda$ scales as $\Lambda^{100}$, neglecting exchange symmetry. Consequently, the ratio of states between $\Lambda^\prime$ and $\Lambda$ scales as $1 - (\Lambda^\prime / \Lambda)^{100}$, which implies that for large $N$ the contribution from a thin momentum shell can become dominant. One could further constrain the shell thickness $\Lambda - \Lambda^\prime$, but requiring an infinitesimal interval is both finely tuned and impractical.

Fortunately, this issue is mitigated for the EFT interactions of primary interest, such as chiral EFT in nuclear physics. In these frameworks, the Hamiltonian is governed by a clear hierarchy in which two- and three-body forces dominate while higher-order terms act as perturbative corrections. For simplicity, consider a case where only two-body forces are nonperturbative. By first decoupling the interactions within the two-body subspace, where three-body and higher-order forces naturally vanish, the preceding argument remains rigorous. Crucially, the elimination of two-body mixing terms in $H_1$, while demonstrated in the two-body subspace, persists for any particle number. Since all remaining higher-body terms in $H_1$ are perturbative, the off-diagonal block $\epsilon$ is significantly suppressed relative to $A_1$, which is dominated by the kinetic energy. Thus, the final term in Eq.~\eqref{eq:SRGmagnitude} remains negligible, ensuring that the SRG flow drives the system toward the fixed point $b = 0$ and $H_1 = 0$. Under these conditions, the requirement for an extremely narrow momentum shell is relaxed, and convergence is instead guaranteed by the few-body dominance of the EFT Hamiltonian. This reasoning extends naturally to the application of soft cutoff functions.

It is important to note that the SRG flow does not eliminate the high-momentum components in $H_0$ that involve both $\delta\Phi$ and $\delta\Phi^{\dagger}$. Since these components are decoupled from the low-momentum sector, they can be removed by applying the projector $P_{0'}$ onto space-$0'$. The projected Hamiltonian $H' = P_{0'} H(\infty) P_{0'}$ retains the same operator structure as the original Hamiltonian in Eq.~\eqref{eq:Hamiltonian}, but with the cutoff $\Lambda$ lowered to $\Lambda'$ and with all coefficients ($m$, $C_2$, $C_3$, etc.) replaced by their renormalized counterparts $m'$, $C_2'$, $C_3'$, etc. Owing to the unitarity of $U(\infty)$, $H'$ and $H$ share identical low-lying spectra. This entire procedure can be iterated with successively lower cutoffs ($\Lambda'' < \Lambda'$, etc.), thereby fully establishing the Wilsonian RG flow. Our derivation is nonperturbative and applies to general interactions.

\section{Consequences of symmetry constraints}
\label{sec:symmetries}

In this section, we discuss the constraints imposed by various symmetries, including momentum conservation, particle-number conservation, Galilean invariance, and Lorentz invariance. The RG flow constructed above explicitly gives the unitary transformations between successive cutoffs. The operator basis remains invariant under this evolution, with only the LECs changed. It is therefore important to examine which terms can be generated in this process. In principle, all terms compatible with the symmetries can emerge, with their LECs constrained by the same symmetries. A particularly important issue is that our regularization scheme enforces a clear separation of the Fock space but violates Galilean invariance, which is essential for nonrelativistic systems. We discuss the generation and evolution of such Galilean-invariance-breaking terms. In particular, Galilean invariance requires the single-particle dispersion to be exactly quadratic, whereas RG evolution can in principle generate terms proportional to $p^4$ or even higher powers of momentum. We will show that particle-number conservation rigorously forbids terms that modify the single-particle dispersion. We also discuss a future extension to Lorentz invariance. Finally, we show that the locality of the unitary transformation also preserves the invariance of the scattering $S$ matrix, thereby completing our proof that low-energy physics remains invariant under the RG flow.

\subsection{Dispersion relation and mass renormalization}
\label{subsec:dispersion}

As in the general RG analysis, we show that symmetries can impose important constraints on the RG flow. In particular, we demonstrate that momentum conservation and particle-number conservation together uniquely fix the single-particle dispersion relation and forbid particle mass renormalization.

We first note that the SRG generator $\eta(t)$ is constructed as a spatial integral of local operators containing equal numbers of creation and annihilation operators. Consequently, the generator commutes with the particle number operator $N$ and the total momentum operator $P$,
\begin{equation}
[N, \eta(t)] = 0, \quad [P, \eta(t)] = 0.
\end{equation}
This implies that the induced SRG unitary transformation
\[
U(t) = \mathcal{T} \exp \left( i \int_0^t \eta(\tau) d\tau \right)
\]
preserves the Hilbert space sectors defined by these quantum numbers, where $\mathcal{T}$ denotes time ordering along the SRG evolution time. In the single-particle subspace, a state with momentum $p$ below the cutoff, $|p\rangle = a_{p}^{\dagger}|0\rangle$, is the \textit{unique} state characterized by these eigenvalues. Since the unitary transformation $U(t)$ changes neither these quantum numbers nor the energy, $|p\rangle$ must be a common eigenstate of the initial Hamiltonian $H$ and the evolved Hamiltonian $H_{\Lambda^\prime} = U(\infty)H_{\Lambda}U^\dagger(\infty)$. Its energy eigenvalue remains unchanged because the transformation is unitary,
\begin{equation}
H_{\Lambda^\prime} |p\rangle = E_p |p\rangle, \quad E_p = \frac{p^2}{2m}.
\end{equation}
This leads to two critical observations. First, the effective mass remains unrenormalized ($m' = m$). Second, any one-body operators that would modify the dispersion relation, such as higher-order derivative terms $\Phi^{\dagger}\partial_x^{4}\Phi$, are strictly forbidden.
Thus, the single-particle spectrum and its dispersion relation are protected by the symmetries and remain intact throughout the flow. Only the interactions are renormalized, consistent with our expectation for a nonrelativistic RG transformation.

\subsection{Galilean invariance breaking interactions}

\begin{figure} 
\includegraphics[width=1\columnwidth]{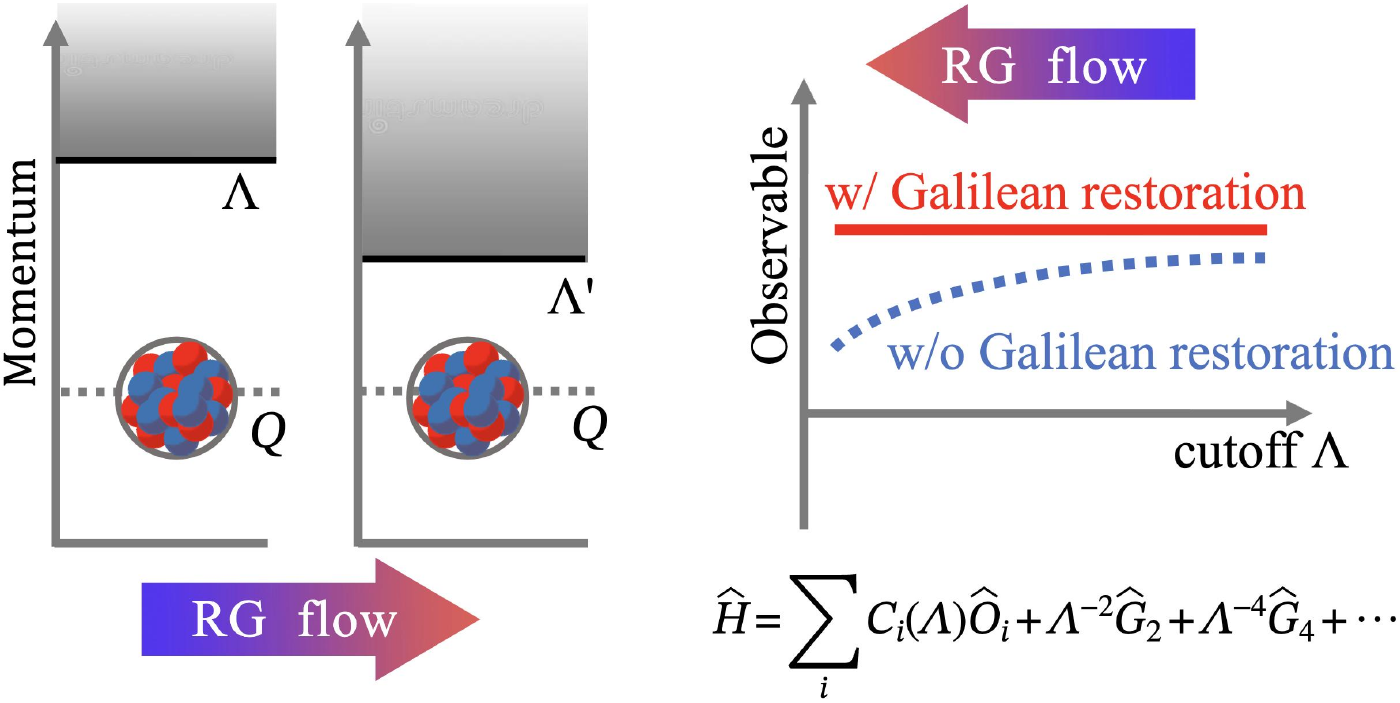}
\caption{Lowering the momentum cutoff necessitates the inclusion of symmetry-restoring terms that counteract the explicit breaking of Galilean invariance. 
Neglecting these terms introduces residual cutoff dependence. However, this dependence is systematically suppressed and eventually vanishes in the large-cutoff (UV) limit.}
\label{fig:GIR}
\end{figure}

A remaining challenge is that the RG flow can still generate contact terms
that explicitly break Galilean invariance, as illustrated in
Fig.~\ref{fig:GIR}. This situation is analogous to lattice field theories,
such as lattice QCD~\cite{Wilson1974_PRD10-2445} and nuclear lattice
EFT~\cite{Lee2009_PPNP63-117,Lahde:2019npb,Dean:2025}, where
discretization inherently violates certain continuous
symmetries~\cite{Nielsen1981NPB185-20,Lu2014PRD90-034507,Lu2015PRD92-014506}.
In those contexts, the Symanzik improvement program restores such symmetries
by augmenting the action with irrelevant operators that cancel discretization
artifacts~\cite{symanzik1983NPB226-187}. Similarly, in our regularization
scheme, the regulator-induced breaking of Galilean invariance scales as
powers of $1/\Lambda$. In a typical EFT, these symmetry-breaking
contributions can also be eliminated systematically, either by formally
extrapolating results from a finite $\Lambda$ window to the
$\Lambda\to\infty$ limit or by incorporating appropriate symmetry-restoring
counterterms into the
Hamiltonian~\cite{Klein2018_EPJA54-233,Li2019_PRC99-064001,Elhatisari2024_Nature630-59}.
These restoration methods have been demonstrated numerically in recent
nuclear chiral EFT calculations using our regularization
scheme~\cite{Shi2026_PLB874-140303}. Neglecting these corrections generally
leads to percent-level cutoff dependence in nuclear binding energies, which
diminishes rapidly with increasing $\Lambda$. Here, we use both
symmetry-restoration terms and symmetry-breaking terms to refer to the same
set of operators generated during the RG flow that explicitly break the
symmetry, whose breaking effects cancel exactly against those induced by the
regulators.

The leading-order Galilean-invariance-restoration (GIR) contact term involves the center-of-mass momentum $P_{\rm cm}$. For a Gaussian regulator, we introduce
\begin{equation}
    V_{\rm GIR} = C_{\rm GIR} P_{\rm cm}^2 \exp\left( -\frac{p_1^2 + p_2^2 + p_1^{\prime 2} + p_2^{\prime 2}}{2\Lambda^2} \right),
\end{equation}
where the same \textit{absolute}-momentum regulator is applied to maintain internal consistency. In practical calculations, $P_{\rm cm}$ can be expanded using single-particle momenta and transformed into spatial derivatives implemented using finite differences or fast Fourier transforms~\cite{Elhatisari2024_Nature630-59,Li2019_PRC99-064001,Shi2026_PLB874-140303}. Notably, this term vanishes by construction in the two-body center-of-mass frame. The corresponding LECs, denoted here by $C_{\rm GIR}$, can be determined by comparing observables across different inertial frames. Typically, these LECs are functions of the cutoff $\Lambda$ and scale as $\mathcal{O}(\Lambda^{-2})$. Note that the system remains strictly confined to the low-momentum Hilbert space after adding the GIR terms due to the associated \emph{absolute}-momentum regulator. Consequently, the Hamiltonian remains block diagonal, which distinguishes this approach from those using Galilean-invariant regulators that inevitably induce mixing between low- and high-momentum states.

We note that challenges regarding Galilean invariance are well documented in nuclear lattice effective field theory (NLEFT)~\cite{Elhatisari2024_Nature630-59,Li2019_PRC99-064001,Shi2026_PLB874-140303}. A discrete lattice naturally defines a preferred stationary frame, thereby breaking the equivalence of motion across different inertial frames. Consequently, nonrelativistic calculations in NLEFT frequently incorporate GIR terms into the Hamiltonian to maintain physical consistency. For instance, Ref.~\cite{Klein2018_EPJA54-233} used a one-dimensional lattice toy model to demonstrate that including lattice GIR terms suppresses the dependence of results on the lattice spacing. More complex GIR terms have also been incorporated into NLEFT calculations to improve the agreement with experimental data~\cite{Elhatisari2024_Nature630-59}. Notably, while conventional NLEFT calculations treat GIR terms primarily as a technique for removing lattice artifacts, this work views them as a more general tool for addressing RG challenges in continuum EFT calculations. See Sec.~\ref{subsec:NLEFT} for a discussion of the role of lattice spacing. A recent NLEFT calculation using a smooth cutoff shows that the $^4$He binding energy can be reproduced to an accuracy of $100$~keV and remains independent of the momentum cutoff when leading-order GIR terms are included. Without these terms, the $^4$He binding energy deviates by approximately $1$~MeV, and this deviation vanishes as $\mathcal{O}(\Lambda^{-2})$, consistent with theoretical expectations~\cite{Shi2026_PLB874-140303}. Since these calculations employ a small lattice spacing, they can be viewed as based on a continuum EFT.

In a companion paper, we extend this investigation to medium-mass nuclei and find that Galilean invariance breaking effects are highly suppressed~\cite{Lu2026_in_preparation}. The binding energies of nuclei up to $^{40}$Ca can be reproduced independently of the cutoff using a single three-body force, which implies a significant simplification for nuclear chiral EFT calculations. In these calculations, the GIR terms vanish in two-body scattering and interplay with the three-body force within three-body systems. If the GIR terms are omitted, their effects can largely be absorbed by refitting the three-body force to the triton binding energy, thereby maintaining the predictive power of the EFT. This phenomenon is consistent with the expectations of the Wilsonian RG, where the effects of irrelevant operators are absorbed by relevant ones, with residual effects vanishing for increasing cutoffs~\cite{schwartz2014book}. 

In summary, we do not require Galilean invariance for each operator from
the outset. A single-particle cutoff scheme that relaxes Galilean invariance can
significantly improve RG invariance, as demonstrated theoretically in this
work and verified numerically in our companion
paper~\cite{Lu2026_in_preparation} and related
literature~\cite{Klein2018_EPJA54-233,Shi2026_PLB874-140303}. Currently, no
regularization scheme for Hamiltonian-based EFT simultaneously preserves
Galilean invariance, consistently suppresses high-momentum modes, and induces
no time-derivative interactions. Given the fundamental role of RG invariance
in EFT construction and the fact that Galilean invariance is only an
approximation to exact Lorentz symmetry, adopting this trade-off may be
advantageous in some cases.

\subsection{Relativistic corrections}

In this work, we consider only the nonrelativistic Hamiltonian. In principle, the path-integral formalism is more convenient for relativistic systems. Nevertheless, for low-energy systems such as nuclei, we can employ a semirelativistic approximation by including relativistic corrections proportional to powers of $M^{-1}$, with $M$ the particle mass. In these calculations the relativistic effects only modify the form of the Hamiltonian, and particle-number conservation is retained, i.e., particle generation and annihilation in the vacuum are neglected. This principle has been applied to construct covariant chiral EFT and has shown certain advantages in nuclear structure calculations~\cite{Ren2018cpc,li2016prd,SHEN2019PPNP}.

For the kinetic energy term, relativistic corrections can be derived by expanding the relativistic dispersion relation,
\begin{align*}
  E(p) = \sqrt{p^2 + M^2} = M + \frac{p^2}{2M} - \frac{p^4}{8M^3} + \frac{p^6}{16M^5} + \mathcal{O}(p^8),
\end{align*}
where for $p \ll M$ the first two terms dominate, corresponding to the nonrelativistic limit. Higher-order terms are suppressed by powers of $p/M$ and modify the single-particle dispersion relation. According to the reasoning presented in Sec.~\ref{subsec:dispersion}, the unitary RG transformation is constrained by particle number and momentum conservation, so it does not alter the dispersion relation and leaves this expression intact. Note that some quantum field theory textbooks (e.g., Schwartz's book~\cite{schwartz2014book}) present this as an example of a nonrenormalizable field theory. In this sense, we may treat the relativistic correction terms similarly to ordinary interaction vertices.

For Lorentz-invariant interactions, we can also perform a low-momentum expansion to express the covariant vertices as a power series in the normal momenta. As an example, consider the Lorentz-invariant dot product of two incoming covariant momenta,
\begin{align}
       p_1^\mu p_{2\mu} =& \sqrt{p_1^2 + M^2}\,\sqrt{p_2^2 + M^2} - p_1p_2 \nonumber \\
       =& M^2 + \frac{1}{2}(p_1-p_2)^2 - \frac{(p_1^2 - p_2^2)^2}{8M^2} + \mathcal{O}(p^6),
       \label{eq:CovariantDot}
\end{align}
where the first two terms correspond to the nonrelativistic limit and respect Galilean invariance. The third term depends on both the relative momentum and the center-of-mass momentum and therefore violates Galilean invariance. Nevertheless, since our regularization scheme does not impose Galilean invariance from the outset, these corrections can be treated in exactly the same manner as the other interaction terms.

When expanded in the local operator basis involving normal momenta, both Galilean invariance and Lorentz invariance impose certain relations between the coefficients. In the low-momentum expansion, these symmetries yield a series of linear relations among the LECs. If we impose a regulator that breaks these symmetries, the RG evolution automatically generates restoration terms in the Hamiltonian to compensate for this breaking. Since this symmetry-breaking effect originates entirely from the regulator, it is reasonable to infer that the corresponding LECs scale as negative powers of $\Lambda$ and vanish for large cutoff values. We therefore write down all terms resulting from the expansion of an exactly Lorentz-invariant action. The resulting LECs are not free but are connected by several linear relations. We then impose an \emph{absolute} single-particle momentum regulator on the normal momenta. In this case many Galilean-invariance-breaking terms are already present and need not be added to the Hamiltonian. Nevertheless, a complete operator basis is still required to fulfill exact RG invariance. The remaining LECs not fixed by Lorentz invariance can be determined by fitting to experimental data. The predictions should be fully RG invariant provided the EFT interaction is taken to sufficiently high orders. If RG invariance is found to be broken, we may either supplement the Hamiltonian with additional higher-order local operators constrained by the underlying Lorentz invariance or simply extrapolate to large cutoffs to suppress the contributions of omitted irrelevant operators.

The procedure for restoring Lorentz invariance described above is closely analogous to the framework of nonrelativistic QCD (NRQCD)~\cite{Bodwin1995_PRD51-1125,Lepage1992_PRD46-4052,Luke1992_PLB286-348,Manohar1997_PRD56-230}. In NRQCD, heavy quarks are initially treated as nonrelativistic degrees of freedom interacting with the gauge field at leading order. As one moves to higher orders in the nonrelativistic expansion, Lorentz covariance is systematically recovered by incorporating higher-dimensional operators suppressed by powers of $1/M$. The Wilson coefficients of these higher-order terms are not independent but are uniquely constrained by the underlying Lorentz symmetry of the full theory, a feature often manifested as reparameterization invariance. Provided that a Lorentz-invariant regularization scheme is employed, these symmetry-enforced relations among coefficients remain robust against quantum corrections.

However, a distinction arises when comparing NRQCD with nuclear chiral EFT. In the former, the dominant relativistic corrections typically manifest in the kinetic energy sector, while contact interactions remain relatively weak. In contrast, contact terms in nuclear chiral EFT play a foundational role in describing the short-range nuclear force. Consequently, the restoration of Lorentz invariance in the chiral EFT framework necessitates the inclusion of more intricate operator structures, leading to a significantly more complex organizational scheme for the higher-order contact Lagrangian.

\subsection{Invariance of the scattering $S$ matrix}

An important feature of the SRG generator $\eta(t)$ is its locality. It is composed of operators that act over a characteristic range of $\Lambda^{-1}$. Consequently, the cumulative transformation $U(\infty)$ is quasilocal. Consider a many-body state consisting of two wave packets $|\phi_{1}\rangle$ and $|\phi_{2}\rangle$ separated by a distance $R \gg \Lambda^{-1}$. Due to quasilocality, the transformation $U(\infty)$ acts independently on each cluster,
\begin{equation}
U(\infty) \bigl( |\phi_{1}\rangle \otimes |\phi_{2}\rangle \bigr) \approx U(\infty)|\phi_{1}\rangle \otimes U(\infty)|\phi_{2}\rangle.
\end{equation}
This factorization property is essential for maintaining the extensivity of the energy and the consistency of the cluster-decomposition principle in the evolved theory.

In a typical scattering experiment, the asymptotic \textit{in} and \textit{out} states at real time $t \to \pm \infty$ consist of well-separated noninteracting clusters. Formally, if $H_{\Lambda^\prime}$ and $H_\Lambda$ are related by $H_{\Lambda^\prime} = U H_{\Lambda} U^\dagger$, then the evolved $S$ matrix $S_{\Lambda^\prime}$ is related to the original $S_\Lambda$ by
\begin{equation}
S_{\Lambda^\prime} = \mathbb{I} + 2\pi i \delta(E_i - E_f) T_{\Lambda^\prime} = U S_\Lambda U^\dagger,
\end{equation}
where $T_{\Lambda^\prime}$ is the scattering $T$ matrix. In the asymptotic limit where the particles are far apart, the operators $U$ transform the eigenstates of $H_\Lambda$ into those of $H_{\Lambda^\prime}$ while leaving the eigenenergies invariant. In computing the $S$-matrix elements, the transformations on the states and the Hamiltonian cancel precisely. As a result, all $S$-matrix elements and phase shifts remain invariant,
\begin{align*}
\langle \psi_{\rm out}' | H_{\Lambda^\prime} | \psi_{\rm in}' \rangle 
=& \langle \psi_{\rm out} | U^\dagger (U H_\Lambda U^\dagger) U | \psi_{\rm in} \rangle \\ =& \langle \psi_{\rm out} | H_\Lambda | \psi_{\rm in} \rangle,
\end{align*}
where $|\psi_{\rm in/out}\rangle$ denote the asymptotic incoming and outgoing states, respectively.

\begin{figure}[htbp]
\centering
\includegraphics[width=1.0\columnwidth]{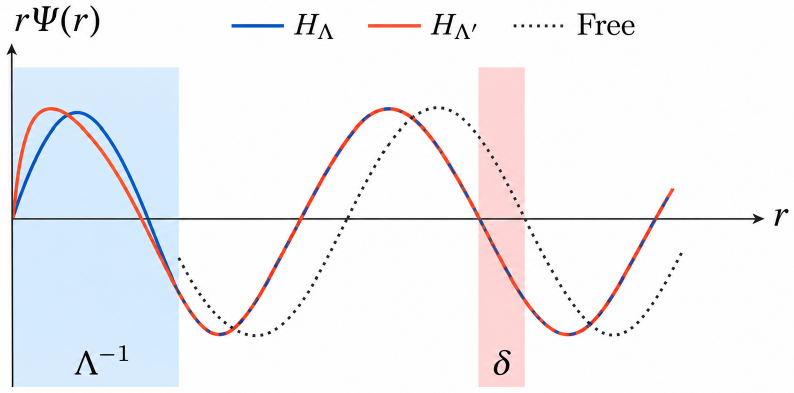}
\caption{Schematic plot of radial wave functions for two-particle scattering. The blue and red lines denote the results calculated with $H_\Lambda$ and $H_{\Lambda^\prime}$, respectively. In the asymptotic region $r \gg \Lambda^{-1}$, the two wave functions coincide, and their common shift relative to the free wave function (dotted line) defines the invariant phase shift $\delta$.}
\label{fig:radial}
\end{figure}

This invariance is illustrated schematically in Fig.~\ref{fig:radial}, where we compare the radial wave functions $u(r)$ for a two-particle scattering process. The effect of the quasilocal transformation $U(\infty)$ is confined to the short-range region $r \lesssim \Lambda^{-1}$, where the wave function is strongly modified as the basis is unitarily rotated to decouple high-energy modes. In contrast, in the long-range region $r \gg \Lambda^{-1}$, the evolved wave function $u'(r)$ asymptotically coincides with the original $u(r)$. Since the phase shift $\delta$ is determined solely by the asymptotic behavior at $r \to \infty$, the SRG flow leaves the scattering observables unchanged. This logic extends naturally to more complex multichannel reactions and inelastic processes, confirming that the local unitary transformation only decouples high-momentum details and does not modify low-momentum scattering and reaction processes.

The unitary transformation introduced here shares similarities with the wave-function matching (WFM) method~\cite{Elhatisari2024_Nature630-59}, which employs local transformations to facilitate perturbative calculations. In contrast, while WFM is designed to bridge the gap between realistic and easily solvable Hamiltonians and primarily focuses on two-body interactions, our work constructs local unitary transformations to bridge the gap between EFTs defined with different cutoffs and operates in a general many-body Fock space.

\section{Regulator effects}
\label{sec:regulator-effects}

\subsection{Momentum leakage of relative-momentum regulators}
\label{subsec:momentum-leakage}

We examine the standard Galilean-invariant regulator defined in Eq.~\eqref{eq:reg_rel}. For a general two-body contact interaction, this regulator permits a pair of particles with momenta $(P_{\mathrm{cm}}/2+p, P_{\mathrm{cm}}/2-p)$ to scatter into $(P_{\mathrm{cm}}/2+p', P_{\mathrm{cm}}/2-p')$, where the relative momenta satisfy $|p|,|p'|<\Lambda$. As illustrated by the configurations in Fig.~\ref{fig:partition}, assuming a uniform momentum distribution such as in a one-dimensional periodic box, a cutoff $\Lambda$ greater than the level spacing allows transitions between successive energy levels. Consequently, excitations from configurations (0) through (3) are allowed, leading to matrix elements structured as shown in the left panel of Fig.~\ref{fig:partition}(c). Although total momentum conservation eventually enforces decoupling above a certain maximum scale $\Lambda_{\mathrm{max}}\gg\Lambda$, the impact of the regulator within many-body subspaces is inherently diffuse. For large systems such as nuclei with numerous nucleons, low-lying states can couple to extremely high-momentum states by concentrating the available momentum onto a single particle. This coupling becomes particularly problematic in the nonperturbative regime, where high-momentum states indirectly influence low-momentum physics through higher-order corrections. This effect is an intrinsic property of the regulator itself and is independent of the chosen basis or the specific many-body method.

\subsection{Comparison of absolute- and relative-momentum regulators}

Both \emph{absolute}- and \emph{relative}-momentum regulators can be formulated using an identical cutoff scale $\Lambda$. Employing a soft Gaussian regulator, these expressions are given by
\begin{align}
    f_{\mathrm{rel}}  &= \exp\left(-\frac{p^2 + p^{\prime2}}{\Lambda^2}\right), \\
    f_{\mathrm{abs}}  &= \exp\left(-\frac{p_1^2 + p_2^2 + p_1^{\prime2} + p_2^{\prime2}}{2\Lambda^2 } \right) \nonumber \\
    &= \exp\left(-\frac{P_{\mathrm{cm}}^2}{2\Lambda^2}\right) \exp\left(-\frac{p^2 + p^{\prime2}}{\Lambda^2}\right),
\end{align}
where we have applied total momentum conservation, and $p=(p_1-p_2)/2$ and $p'=(p_1'-p_2')/2$ denote the incoming and outgoing relative momenta, respectively. In the specific case of two-body scattering within the center-of-mass frame, where $P_{\mathrm{cm}}=p_1+p_2=p_1'+p_2'=0$, the two regulators become identical and yield exactly the same LECs.

However, in many-body calculations, the additional exponential factor in $f_{\mathrm{abs}}$ suppresses states with large single-particle momentum, which effectively weakens the strength of the interaction. For systems with strongly attractive interactions, such as those characterized by nuclear forces, the \emph{absolute}-momentum regulator typically results in lower binding energies than the conventional \emph{relative}-momentum regulator. This difference highlights the nontrivial impact of the regulator choice on many-body observables, even when the two regulators are constrained to yield identical results in the two-body center-of-mass frame.

\subsection{Implications for EFT consistency}

The momentum leakage in many-body calculations has received limited explicit discussion in the
literature for several reasons. First, EFT principles primarily guide the
construction of interactions, whereas their subsequent use in many-body
calculations often follows established procedures developed for
phenomenological potentials, so the distinct role of the EFT cutoff can
remain less explicit at this stage. Second, the effect becomes more
pronounced in larger systems, while in few-body systems much of its
contribution can be absorbed by readjusting three-body couplings. This may
help explain why many early nuclear chiral EFT calculations obtained good
agreement for light systems with $A\le4$~\cite{nogga2000prl,epelbaum2001prl,Kamada2001PRC,Binder2016_PRC93-044002},
whereas descriptions of medium-mass and heavier nuclei often required
further
adjustments~\cite{Hu2026_PRL137-142502,Jiang2024_PRC109-L061302,wang2024prc,Soma2020_PRC101-014318,Huther2020_PLB808-135651,Drischler2019_PRL122-042501,Hoppe2019_PRC100-024318,Shirokov2016_PLB761-87,Logoteta2016_PLB758-449,Ekstrom2015_PRC91-051301,Ekstrom2013_PRL110-192502}.
In particular, some interactions slightly underbind $^{3}\mathrm{H}$ while
substantially overbinding symmetric nuclear matter. Such discrepancies have
often motivated the inclusion of higher-order contributions and more general
three-body interactions.

For phenomenological interactions without an explicit momentum cutoff,
imposing Galilean invariance is a natural part of the potential formulation.
Similarly, a regulated EFT interaction can be employed as an effective force
model in many-body calculations without explicitly separating all
single-particle modes at $\Lambda$. An additional consistency requirement
arises, however, when these calculations are interpreted within a Wilsonian
RG framework. The separation of resolved and unresolved modes must then be
implemented consistently across particle-number sectors. Although the LECs
are calibrated in the two-body sector to encode unresolved physics above
$\Lambda$, a relative-momentum regulator can still couple many-body states to
configurations containing particles with momenta well above this scale. The
few-body calibration therefore does not, by itself, ensure a common
decoupling of high-momentum modes in larger systems. This mismatch must be
addressed when establishing a consistent Wilsonian RG evolution across
sectors.

In our companion paper~\cite{Lu2026_in_preparation}, we directly compare the
two regularization schemes and find that the \emph{absolute}-momentum cutoff
removes the severe nuclear-matter overbinding at the two-body level and
qualitatively changes the equation of state. These results quantify the
momentum-leakage effect and indicate that the regulator choice can account
for part of the overbinding observed in medium-mass nuclei without
introducing additional many-body operator structures or extending the
calibration beyond the lightest
nuclei~\cite{Binder2014_PLB736-119,Lahde2014_PLB732-110}. These findings
complement the progress in reproducing experimental data and motivate
further tests of the underlying regulator consistency. Many nuclear chiral
EFT calculations employ only one or two cutoff values, leaving the broader
regulator dependence less constrained. Systematic comparisons of closely
related EFT implementations are therefore particularly informative when
small differences in their construction lead to qualitatively different
many-body predictions.

\begin{figure} 
\includegraphics[width=1\columnwidth]{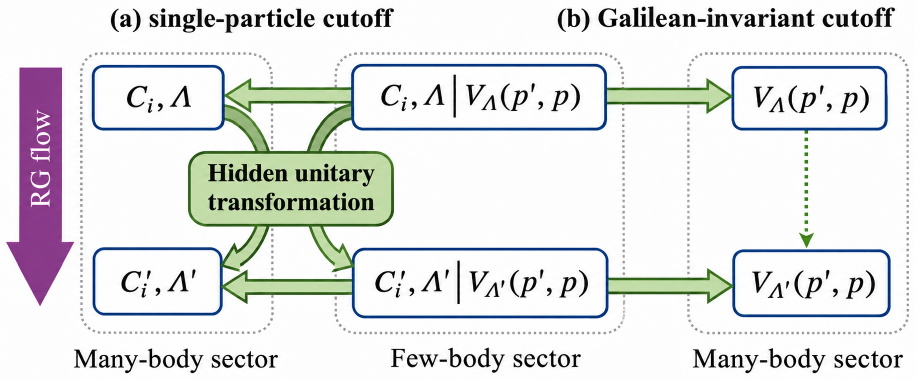}
\caption{Two paradigms for applying many-body EFTs. (a) Global regularization. A consistent single-particle cutoff $\Lambda$ is imposed across both few- and many-body sectors. This global consistency implies the existence of a hidden unitary transformation linking different cutoffs, ensuring that few-body-calibrated interactions remain unitarily equivalent in many-body applications. (b) Potential Model. The EFT potential is constructed within the few-body sector and then promoted to many-body systems by assuming Galilean invariance. While both schemes coincide in the few-body regime, they yield diverging many-body predictions.}
\label{fig:RGflow}
\end{figure}

A common two-stage workflow first parameterizes few-body potentials in the
center-of-mass frame and then embeds them in many-body systems by invoking
Galilean invariance. The relative-momentum regularization used in such
constructions~\cite{Entem2003_PRC68-041001,Epelbaum2005_NPA747-362} can be
illustrated by the sharp-cutoff form
\begin{equation}
    f_{\mathrm{rel}}\bigl(\{p_i,p_i'\}\bigr)=
    \theta\bigl(\Lambda-|\bm p_1-\bm p_2|/2\bigr)\,
    \theta\bigl(\Lambda-|\bm p_1'-\bm p_2'|/2\bigr),
    \label{eq:reg_rel}
\end{equation}
where $\bm p_{1,2}$ and $\bm p'_{1,2}$ denote incoming and outgoing momenta,
respectively. Although interactions constructed at different cutoffs can be
unitarily equivalent in the two-body sector, the transformations relating
them need not commute with the conventional embedding of the regulated
potentials into many-body spaces, as illustrated conceptually in
Fig.~\ref{fig:RGflow}. Few-body equivalence alone therefore does not
establish equivalent many-body predictions. Indeed, interactions yielding
nearly identical few-body fits can produce substantially different many-body
results, a persistent challenge in nuclear \textit{ab initio}
calculations~\cite{Machleidt2023FewBodySyst64-77}. In addition to omitted
higher-order contributions and truncated many-body forces, our analysis
identifies a consistency condition on the embedding itself. Hamiltonians
constructed at different cutoffs must be connected by a unitary
transformation that acts consistently across particle-number sectors. The
existence of such a common transformation cannot be assumed for an arbitrary
regularization prescription and requires explicit construction or proof. We
establish such a connection using a globally consistent \emph{absolute}-
momentum cutoff, which permits the construction of the corresponding hidden
unitary transformation. From a field-theoretic perspective, the common
cutoff supports systematic cancellations of cutoff-dependent variations
among contact interactions. If different sectors implement different
separations of momentum modes, these cancellations need not carry over from
few-body to many-body systems, leaving residual cutoff dependence
unconstrained by the few-body calibration. Similar consistency requirements
arise in relativistic and statistical field theories.

Generally, embedding a regulated two-body interaction as a Galilean-invariant
potential does not imply an unambiguous separation of many-body Hilbert
spaces. Such an interaction can still serve as an effective many-body
Hamiltonian, but its predictions depend on the embedding scheme, and its
cutoff independence must be demonstrated separately.

\subsection{Connection to lattice effective field theory}
\label{subsec:NLEFT}

Our proposed regularization scheme is closely connected to lattice-based frameworks. In particular, RG invariance under the \emph{absolute}-momentum cutoff has been demonstrated numerically within the \textit{ab initio} framework of nuclear lattice effective field theory (NLEFT)~\cite{Shi2026_PLB874-140303,Lu2026_in_preparation}. In these implementations, we employ smooth momentum regulators with a cutoff scale $\Lambda$ significantly smaller than the intrinsic lattice UV cutoff, i.e., $\Lambda \ll \pi/a$. This separation of scales ensures that lattice artifacts are systematically suppressed. In this regime, the lattice does not serve as the primary EFT regulator but rather as a robust discretization tool for numerical computation. Consequently, the continuum limit $a \to 0$ can in principle be taken, rendering the resulting interactions applicable to a wide range of many-body methods beyond the lattice, such as the Brueckner-Hartree-Fock (BHF) approach.
A key advantage of using a smooth regulator is that it enables a continuous variation of the cutoff scale, thereby facilitating a rigorous Wilsonian RG analysis. This methodology parallels the connection between Wilson's continuous RG transformations and Kadanoff's discrete block-spin renormalization.
These results carry implications that extend well beyond lattice simulations. We anticipate that this study will provide a general framework for constructing RG-invariant EFTs across diverse branches of many-body physics.



\subsection{Connection to SRG calculation}

SRG resolution-scale invariance should be distinguished from the regulator independence of EFT predictions. Standard SRG calculations apply \textit{explicit} unitary transformations to a chosen initial Hamiltonian, represented, for example, in a momentum or harmonic-oscillator basis. When all induced many-body interactions are retained, the resulting Hamiltonians are unitarily equivalent to the initial one by construction. The evolved interactions are determined by the flow itself, rather than specified independently at each scale.
In contrast, our work examines \textit{implicit}, or hidden, unitary transformations between EFT Hamiltonians constructed independently at different cutoffs and calibrated to the same low-energy observables. Establishing such a mapping is nontrivial because its existence depends on the operator content, regulator prescription, and matching conditions. In particular, exact unitary equivalence requires identical spectra in the space being compared. In many-body systems, the relation must also hold consistently across particle-number sectors. Our work addresses this problem by identifying a regularization prescription for which the corresponding unitary mapping can be constructed.



\subsection{Numerical verifications}


Our results suggest a systematically improvable framework for \textit{ab initio} EFT calculations. By including local operators relevant at the momentum scale of interest, the large-$\Lambda$ limit of many-body predictions converges to a value fixed by the few-body inputs used for calibration. The further inclusion of irrelevant operators to restore symmetries, such as Galilean invariance, accelerates this convergence but does not alter this limit. A key advantage is that the predictions are uniquely determined, with no subsequent readjustment permitted. It is therefore of interest to investigate whether these predictions agree with experiments in realistic nuclear systems when combined with long-range pion-exchange potentials. To date, calculations under this paradigm have been performed up to N$^{2}$LO and have yielded promising results, while the extension to N$^{3}$LO is currently being pursued. Our results also motivate a systematic comparison of regulator effects under identical conditions, particularly using the same many-body algorithms.

We suggest that these results be further validated using complementary many-body methods. For instance, while SRG evolution is traditionally employed to soften interactions to facilitate convergence in harmonic-oscillator bases~\cite{Hebeler2011_PRC83-031301,Simonis2017_PRC96-014303}, the potentials constructed within our framework are inherently soft due to the prescribed low-momentum cutoff. Consequently, these interactions are already well suited for basis-expansion methods and can be applied directly without further numerical transformation.

\section{Numerical verification with a one-dimensional boson EFT}
\label{sec:toy-model}

Although calculations for more realistic systems, such as finite nuclei and
nuclear matter, have been performed in Refs.~\cite{Shi2026_PLB874-140303,Lu2026_in_preparation},
the impact of various operators, particularly the higher-order ones, may be
obscured by numerical uncertainties. Therefore, to examine the underlying
principles when higher-order interactions are included, it is desirable to
test the framework using an exactly solvable model. In this section, we
examine the cutoff dependence of the predictions from a one-dimensional
boson EFT.

\subsection{Theoretical framework}

We consider a system of identical bosons in a one-dimensional periodic box of length $L$, interacting through zero-range contact interactions. Before regularization, the Hamiltonian reads
\begin{align*}
H={}&-\frac{1}{2M}\Psi^\dagger \partial_x^2\Psi
+\frac{1}{2}C_{1_{2N}}(\Psi^\dagger\Psi)^2
\\
&+\frac{1}{2}C_{q^2_{2N}}(\Psi^\dagger\Psi)(-\partial_x^2)(\Psi^\dagger\Psi)
\\
&+\frac{1}{2}C_{Q_{2N}^2}
\bigl(\Psi^\dagger\Psi^\dagger\bigr)(-\partial_x^2)\bigl(\Psi\Psi\bigr)
+\frac{1}{3!}C_{1_{3N}}(\Psi^\dagger\Psi)^3
\\
&+\frac{1}{4!}C_{1_{4N}}(\Psi^\dagger\Psi)^4
\\
&+\frac{1}{3!}C_{q_{3N}^2}
(\Psi^\dagger\Psi)^2(-\partial_x^2)(\Psi^\dagger\Psi)
\\
&+\frac{1}{3!}C_{Q^2_{3N}}
\bigl(\Psi^\dagger\Psi^\dagger\Psi^\dagger\bigr)
(-\partial_x^2)\bigl(\Psi\Psi\Psi\bigr)
\\
&+\frac{1}{5!}C_{1_{5N}}(\Psi^\dagger\Psi)^5+\cdots,
\end{align*}
where $M=938.92$~MeV is the boson mass, and $\Psi$ and $\Psi^\dagger$ are the bosonic annihilation and creation field operators, respectively. For brevity, the spatial integral and normal-ordering symbols are left implicit.

Denoting the incoming and outgoing momenta by $p_i$ and $p_i^\prime$, respectively, in momentum space the operators are
\begin{align}
   O_{1_{2N}} &= \delta\left(\sum_{i=1}^2p_i-\sum_{i=1}^2p_i^\prime\right), \nonumber\\
   O_{q^2_{2N}} &= (p_1^\prime-p_1)^2
   \delta\left(\sum_{i=1}^2p_i-\sum_{i=1}^2p_i^\prime\right), \nonumber\\
   O_{Q^2_{2N}} &= (p_1+p_2)^2
   \delta\left(\sum_{i=1}^2p_i-\sum_{i=1}^2p_i^\prime\right), \nonumber\\
   O_{1_{3N}} &= \delta\left(\sum_{i=1}^3p_i-\sum_{i=1}^3p_i^\prime\right), \nonumber\\
   O_{1_{4N}} &= \delta\left(\sum_{i=1}^4p_i-\sum_{i=1}^4p_i^\prime\right), \nonumber\\
   O_{q^2_{3N}} &= \sum_{i=1}^3(p_i^\prime-p_i)^2
   \delta\left(\sum_{i=1}^3p_i-\sum_{i=1}^3p_i^\prime\right), \nonumber\\
   O_{Q^2_{3N}} &= (p_1+p_2+p_3)^2
   \delta\left(\sum_{i=1}^3p_i-\sum_{i=1}^3p_i^\prime\right), \nonumber\\
   O_{1_{5N}} &= \delta\left(\sum_{i=1}^5p_i-\sum_{i=1}^5p_i^\prime\right).   \label{eq:operator_set}
\end{align}
Here, the delta functions enforcing total momentum conservation are retained explicitly to distinguish operators with similar momentum dependence but different numbers of interacting particles.

Here, we consider contact operators whose LECs have mass dimensions $d_O\geq -3$. We impose invariance under spatial translations, parity, and time reversal. Translation invariance gives rise to the overall delta functions enforcing total momentum conservation, while parity invariance allows only even powers of momenta. For a Hermitian Hamiltonian, parity and time-reversal invariance further imply symmetry under interchange of the incoming and outgoing momentum sets. Bose statistics additionally require invariance under independent permutations within each set. The operators displayed explicitly in Eq.~\eqref{eq:operator_set} form a complete basis through order $\mathcal{O}(Q^3)$. Specifically, $O_{1_{2N}}$ appears at order $\mathcal{O}(Q^0)$, and $O_{1_{3N}}$ at order $\mathcal{O}(Q^1)$. At order $\mathcal{O}(Q^2)$, the allowed operators are $O_{q^2_{2N}}$, $O_{Q^2_{2N}}$, and $O_{1_{4N}}$. At order $\mathcal{O}(Q^3)$, they are $O_{q^2_{3N}}$, $O_{Q^2_{3N}}$, and $O_{1_{5N}}$. In one spatial dimension, the density operator $\Psi^\dagger\Psi$ has mass dimension one, compared with three in three spatial dimensions. Consequently, many-body contact interactions are less suppressed in this dimensional power counting and enter at lower orders than in three dimensions.

We choose this operator basis to make its transformation properties under Galilean boosts explicit. For example, the three-body sector at order $\mathcal{O}(Q^3)$ contains $O_{q^2_{3N}}$, which preserves Galilean invariance, and $O_{Q^2_{3N}}$, which breaks it. No additional independent quadratic momentum structures are allowed by the permutation symmetries. To demonstrate this, consider all quadratic monomials formed from the momenta $\{p_1,p_2,p_3,p_1^\prime,p_2^\prime,p_3^\prime\}$. For terms involving only incoming momenta, permutation symmetry allows only $\sum_i p_i^2$ and $\sum_{i<j}p_ip_j$, with analogous expressions for the outgoing momenta. Symmetry under interchange of the incoming and outgoing states then restricts these terms to linear combinations of
\begin{align*}
A &= \sum_{i=1}^3\left(p_i^2+p_i^{\prime\,2}\right),\\
B &= \sum_{i<j}\left(p_ip_j+p_i^\prime p_j^\prime\right).
\end{align*}
Momentum conservation gives
\[
A+2B=2Q_{3N}^2,
\qquad
Q_{3N}\equiv\sum_{i=1}^3p_i=\sum_{i=1}^3p_i^\prime.
\]
Meanwhile, the only mixed quadratic structure invariant under independent permutations of the incoming and outgoing momenta is
\[
\left(\sum_{i=1}^3p_i\right)
\left(\sum_{j=1}^3p_j^\prime\right)=Q_{3N}^2.
\]
Thus, only two independent quadratic three-body structures remain. A convenient basis is
\[
\sum_{i,j=1}^3(p_i^\prime-p_j)^2
=3A-2Q_{3N}^2,
\qquad
Q_{3N}^2.
\]
After bosonic symmetrization, these correspond, up to overall normalization factors, to $O_{q^2_{3N}}$ and $O_{Q^2_{3N}}$, respectively.



As described above, we apply a smooth cutoff uniformly to all single-particle momentum modes by smearing the field operator,
\begin{equation}
    \Psi(x)\rightarrow\Psi_\Lambda(x)
    \equiv\int_{-\infty}^{\infty}dx^\prime\,
    f_\Lambda(x-x^\prime)\Psi(x^\prime),
\end{equation}
where the smearing kernel is defined by the Fourier transform of the momentum-space regulator,
\begin{equation}
    f_\Lambda(x)=\int_{-\infty}^{\infty}\frac{dp}{2\pi}
    \exp\left(-\frac{p^6}{2\Lambda^6}\right)\exp(ipx).
\end{equation}

We discretize space on a uniform lattice with spacing $a=0.395$~fm, corresponding to a lattice momentum cutoff $\Lambda_a=\pi\hbar c/a\simeq 1570$~MeV. The resulting lattice expressions are evaluated using fast Fourier transforms (FFTs). We have numerically verified that discretization errors are negligible for $\Lambda\ll\Lambda_a$. In this regime, the framework can therefore be regarded as a continuum EFT, with the lattice serving solely as a numerical tool. We obtain the ground-state energies of systems containing two to six bosons by diagonalizing the lattice Hamiltonian using the Lanczos algorithm. 

We have also examined finite-volume effects (FVEs) arising from periodic boundary conditions. 
These effects induce systematic energy shifts for many-boson systems.
For convenience we use $L$ to denote the box size in unit of lattice spacing in what follows.
Nevertheless, as we are only interested in the dependence of the energies on the soft cutoff $\Lambda$, these FVEs have a cutoff dependence similar to that of the exact energy in infinite volume and do not affect the scaling behavior with $\Lambda$, provided that the box is sufficiently large.
We tested the results for different box sizes $L$, finding no essential differences for $N = 4$, $L = 16$~fm, $N = 5$, $L = 12$~fm, $N = 6$, $L = 8$~fm. 
We adopt these settings to reduce computational cost. 
Note that the binding energy per particle increases significantly with more particles, implying that the wave functions decay much faster and a smaller box is sufficient.
Tests with larger boxes did not alter these conclusions.

\subsection{Pseudodata and EFT calibration}
\label{subsec:toy-calibration}

We now consider a system of bosons interacting through a $\delta$-function potential.
On the lattice, the two-body interaction is represented by
\begin{equation}
V_{n,n^\prime}=V_0\delta_{n,n^\prime},  \label{eq:delta_potential}
\end{equation}
where $n$ and $n^\prime$ denote lattice sites and $V_0=-50$~MeV
is the interaction strength.
We refer to this interaction as the \textit{bare} potential, which defines
the underlying theory.
Direct calculations with this potential generate a set of pseudodata.
We then assume that only these pseudodata are available, while the
underlying interaction is unknown.
To construct a low-energy EFT, we introduce a momentum cutoff $\Lambda$
above the relevant binding momenta and below the lattice momentum
cutoff $\Lambda_a$.
The EFT parameters are calibrated to selected pseudodata, and the
remaining pseudodata are predicted.
This setup allows us to assess whether the EFT yields reliable
predictions without explicit
knowledge of the underlying interaction at high momentum scales.

Direct diagonalization of the Hamiltonian with the bare potential
yields the ground-state energies $E_{2N,0}=-2.26$~MeV, $E_{3N}=-9.04$~MeV, $E_{4N,0}=-22.61$~MeV,
$E_{5N,0}=-45.21$~MeV, and $E_{6N,0}=-79.24$~MeV.
The corresponding two-body scattering length and effective range are $a_0=4.28$~fm and $r_0=-4.6\times10^{-5}$~fm.
We also use the calculated two-body phase shifts to constrain the
EFT parameters.
To further assess the predictive power of the EFT, we compute the
first excited-state energies of the three- through six-boson systems, 
$E_{3N,1}=-2.26$~MeV, $E_{4N,1}=-9.13$~MeV, $E_{5N,1}=-22.84$~MeV, and $E_{6N,1}=-46.28$~MeV.
These excited-state energies are close to the separation thresholds and serve as additional benchmarks.

For each fixed cutoff $\Lambda$, we determine the LECs order by order
by fitting to a selected set of pseudodata.
The order assigned to an operator specifies when it first enters
the EFT expansion.
The LECs of operators already present at lower
orders are refitted as higher-order terms are included.

At order $\mathcal{O}(Q^0)$, the only interaction is the two-body contact operator
$O_{1_{2N}}$.
Its LEC, $C_{1_{2N}}$, is fixed by the two-body scattering length
in the center-of-mass frame.
At order $\mathcal{O}(Q^1)$, the three-body contact operator $O_{1_{3N}}$
is also included.
We first fix $C_{1_{2N}}$ using the two-body scattering length
and then determine $C_{1_{3N}}$ from the three-body ground-state
energy $E_{3N,0}$.

At order $\mathcal{O}(Q^2)$, the additional operators are
$O_{q^2_{2N}}$, $O_{Q^2_{2N}}$, and $O_{1_{4N}}$.
We jointly fit $C_{1_{2N}}$ and $C_{q^2_{2N}}$ to the two-body scattering length $a_0$ and effective range $r_0$ in the center-of-mass frame.
Using these LECs, we then calculate $a_0$
in moving frames with nonzero total momentum.
Because the regulator breaks Galilean invariance, the extracted
scattering length generally depends on the total momentum.
We determine $C_{Q^2_{2N}}$ by requiring the scattering length
to be independent of the total momentum through order $\mathcal{O}(Q^2)$.
With the two-body LECs fixed, we fit $C_{1_{3N}}$ to $E_{3N}$
and subsequently fit $C_{1_{4N}}$ to the four-body ground-state
energy $E_{4N}$.

At order $\mathcal{O}(Q^3)$, two additional three-body operators,
$O_{q^2_{3N}}$ and $O_{Q^2_{3N}}$, enter together with
the five-body contact operator $O_{1_{5N}}$.
We first determine the two-body LECs using the same procedure
as at order $\mathcal{O}(Q^2)$.
We then jointly fit $C_{1_{3N}}$ and the coefficient of the
Galilean-invariant operator $O_{q^2_{3N}}$ to the three-body
ground-state energy $E_{3N, 0}$ and first excited-state energy
$E_{3N,1}$ in the center-of-mass frame.
The coefficient $C_{Q^2_{3N}}$ is subsequently fixed by requiring
the intrinsic three-body ground-state energy,
$E_{3N}(P_\mathrm{cm})-P_\mathrm{cm}^2/(6M)$, to be independent of the total momentum $P_\mathrm{cm}$
to the order considered.
Finally, we fit $C_{1_{4N}}$ to $E_{4N, 0}$ and then determine
$C_{1_{5N}}$ from the five-body ground-state energy $E_{5N, 0}$.

\begin{figure}
\centering
\includegraphics[width=1.0\columnwidth]{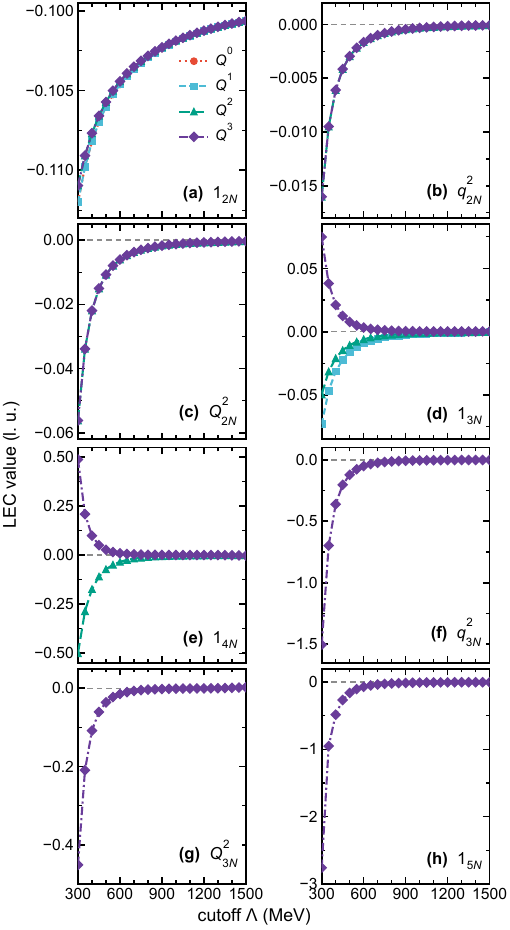}
\caption{LECs for the one-dimensional boson EFT fitted to few-body pseudodata generated by a delta potential. Horizontal dashed lines at zero are shown to guide the eye. All results are given in lattice units (l.u.).}
\label{fig:LECs_delta}
\end{figure}

The LECs determined from pseudodata generated by the $\delta$-function
potential are shown in Fig.~\ref{fig:LECs_delta}.
All LECs are expressed in dimensionless lattice units using appropriate
powers of the lattice spacing $a$.
We repeat the calculations for soft cutoffs $\Lambda$ ranging from
$300$ to $1500$~MeV.
At low cutoffs, below approximately $600$ to $700$~MeV, all LECs are
appreciably nonzero.
As $\Lambda$ approaches the lattice momentum cutoff, most LECs tend
toward zero, while the leading two-body coefficient $C_{1_{2N}}$
remains finite.
This trend is consistent with the expected recovery of the bare
$\delta$-function interaction as the soft regulator is removed.
Comparisons across EFT orders show substantial changes only in
$C_{1_{3N}}$ and $C_{1_{4N}}$.
The nonzero coefficients of the additional operators reflect
interactions induced by integrating out high-momentum modes,
as expected in the Wilsonian RG framework.
We next examine whether these operators improve the predictive
power of the EFT.

\begin{figure}
\centering
\includegraphics[width=1.0\columnwidth]{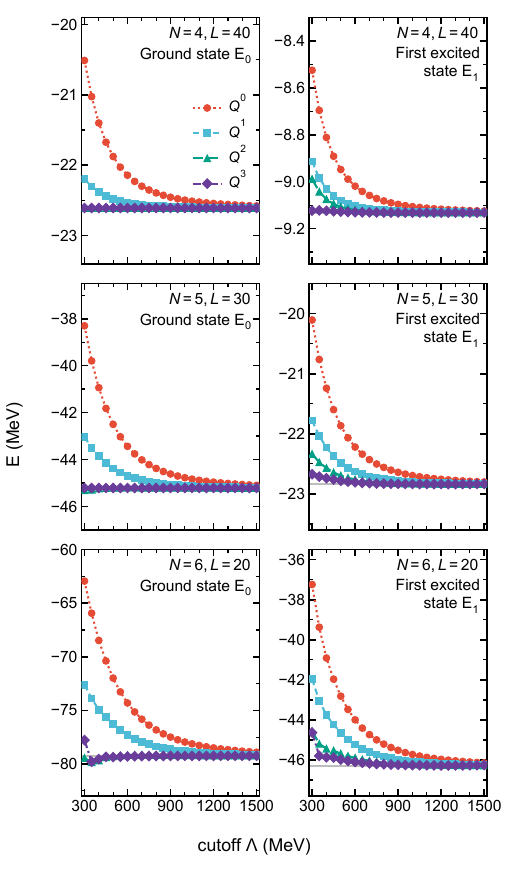}
\caption{Ground-state and first excited-state energies of four, five, and six bosons in periodic boxes as functions of the momentum cutoff $\Lambda$, calculated with EFT interactions calibrated to few-body pseudodata generated by a delta potential. Corresponding LECs are shown in Fig.~\ref{fig:LECs_delta}. Horizontal gray lines denote the corresponding many-body binding energies from the pseudodata.}
\label{fig:prediction_delta}
\end{figure} 

Fig.~\ref{fig:prediction_delta} shows the ground-state and first excited-state energies
of four-, five-, and six-boson systems calculated at successive
EFT orders.
The four-body ground-state energy fixes $C_{1_{4N}}$ at orders
$\mathcal{O}(Q^2)$ and $\mathcal{O}(Q^3)$, while the five-body ground-state energy fixes
$C_{1_{5N}}$ at order $\mathcal{O}(Q^3)$.
These energies are therefore calibration observables at the
corresponding orders. All other energies shown are predictions.
Horizontal gray lines indicate the pseudodata generated from
the bare $\delta$-function potential.
At each fixed cutoff $\Lambda$, the predicted energies exhibit
clear order-by-order convergence toward the pseudodata.
This convergence persists even at cutoffs below $600$~MeV,
where higher-order corrections are substantial.
Because no six-body operator is included, all six-body energies
are predictions based entirely on parameters fitted in systems
with fewer particles.
These results demonstrate that, for this benchmark, an EFT with
a regulator acting on the absolute single-particle momenta can
systematically reproduce the pseudodata without explicit knowledge
of the underlying interaction.
The regime of low cutoffs is particularly informative because the EFT is
calibrated solely to low-energy phase shifts and few-body energies,
while the high-momentum details of the underlying interaction
remain unresolved.

\subsection{Absorption of GIR terms into three-body forces}
\label{subsec:gir-absorption}

In Sec.~\ref{sec:symmetries}, we argue that operators that break Galilean
invariance are irrelevant and that their effects on the low-energy
observables considered here can be effectively absorbed into
a redefinition of the LECs of Galilean-invariant operators.
We examine this claim within our one-dimensional EFT through
order $\mathcal{O}(Q^2)$, where the only operator of this type is
$O_{Q^2_{2N}}$.
Its coefficient compensates for the symmetry breaking induced
by the regulator, and we therefore refer to it as the Galilean
invariance restoration (GIR) term.
We compare three prescriptions for treating this term.
In the first, we retain it explicitly and determine its LEC
by requiring the two-body scattering length to be independent
of the total momentum, following the procedure described above.
In the second, we first determine all LECs with the GIR term
included, but subsequently omit this term without refitting
the remaining coefficients.
The difference from the first prescription quantifies the effect
of omitting the GIR term at fixed values of all other LECs.
In the third, we omit the GIR term and refit the three-body
coefficient $C_{1_{3N}}$ to the three-body ground-state energy.
In all three prescriptions, $C_{1_{2N}}$ and $C_{q^2_{2N}}$
remain unchanged, and the two-body observables in the
center-of-mass frame are unaffected.
Within a three-body system, however, an interacting pair generally
carries nonzero total momentum even when the system as a whole
is at rest.
The GIR term therefore contributes to the three-body energy.
In the third prescription, its effect on this energy is absorbed
by readjusting $C_{1_{3N}}$.
These prescriptions can yield different predictions for larger
systems, which we examine numerically below.
We expect their differences to diminish as $\Lambda$ increases
and the regulator-induced symmetry breaking weakens, while
substantial differences may remain at low cutoffs.

\begin{figure} 
\centering
\includegraphics[width=1.0\columnwidth]{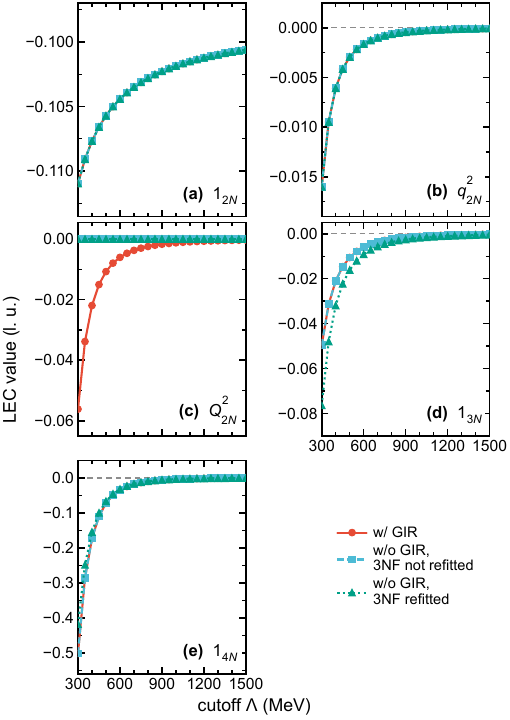}
\caption{LECs for the one-dimensional boson EFT fitted to few-body pseudodata generated by a delta potential. Horizontal dashed lines at zero are shown to guide the eye. The results at order $\mathcal{O}(Q^2)$ are obtained using three different strategies for handling the GIR terms. All results are given in lattice units (l.u.).}
\label{fig:LECs_from_3_strategies_delta}
\end{figure} 

\begin{figure}
\centering
\includegraphics[width=1.0\columnwidth]{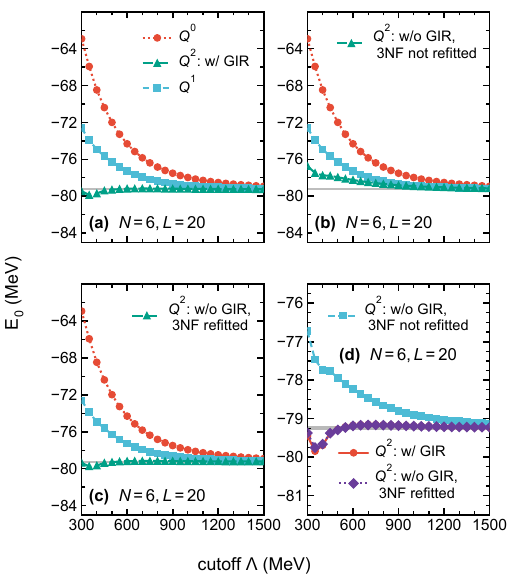}
\caption{Ground-state binding energy of six bosons in a periodic box, calculated with EFT interactions calibrated to data generated by a delta potential. The results at order $\mathcal{O}(Q^2)$ are obtained using three different strategies for handling the GIR terms. (a) Full GIR terms are included in both the fitting and the calculations. (b) GIR terms are included in the fitting but neglected in the calculations. (c) GIR terms are neglected in both the fitting and the calculations, with the three-body force refitted to the three-body binding energy to absorb their effects. (d) Direct comparison of the results at order $\mathcal{O}(Q^2)$ from the three strategies. Corresponding LECs are shown in Fig.~\ref{fig:LECs_from_3_strategies_delta}.}
\label{fig:energies_from_3_strategies_delta}
\end{figure} 

Fig.~\ref{fig:LECs_from_3_strategies_delta} shows the LECs used in the three prescriptions,
and Fig.~\ref{fig:energies_from_3_strategies_delta} presents the corresponding six-body ground-state
energies.
Because all three prescriptions coincide at orders $\mathcal{O}(Q^0)$ and
$\mathcal{O}(Q^1)$, we focus on the results at order $\mathcal{O}(Q^2)$.
With the GIR term included, the results in Fig.~\ref{fig:energies_from_3_strategies_delta}(a) exhibit
very weak cutoff dependence, consistent with our earlier findings.
Omitting this term without refitting the remaining LECs,
as shown in Fig.~\ref{fig:energies_from_3_strategies_delta}(b), produces systematic deviations of a few
MeV from both the pseudodata and the full results at low cutoffs.
After refitting $C_{1_{3N}}$ to the three-body ground-state energy, the six-body predictions in Fig.~\ref{fig:energies_from_3_strategies_delta}(c) recover close
agreement with the pseudodata and exhibit very weak cutoff
dependence even at low cutoffs.
Fig.~\ref{fig:energies_from_3_strategies_delta}(d) directly compares the three sets of results
at order $\mathcal{O}(Q^2)$.
Although omitting the GIR term shifts the six-body energy
by a few MeV at low cutoffs, recalibrating the three-body
interaction largely removes both the deviation from the
pseudodata and the additional cutoff dependence.
Thus, for the observable considered here, the effect of the
omitted GIR term can be largely absorbed into the three-body LEC.
The small residual deviations near $\Lambda\simeq300$~MeV
can be further reduced by higher-order corrections, as illustrated
by the order $\mathcal{O}(Q^3)$ calculations in the preceding section.

\subsection{EFT from a Gaussian potential}
\label{subsec:gaussian-model}

Next, we replace the underlying two-body $\delta$-function potential
with an interaction of finite width.
This provides a closer analogue of three-dimensional applications
with positive $S$-wave effective ranges, where finite-range physics
can constrain the cutoff range accessible to a contact EFT truncated
at a fixed order and treated nonperturbatively.
We adopt a Gaussian two-body interaction as the underlying theory,
\begin{equation}
    V_{n,n^\prime}=V_0\exp\left[-\frac{(x_n - x_{n^\prime})^2}{2R^2}\right],    \label{eq:Gaussian_potential}
\end{equation}
where $n$ and $n^\prime$ label lattice sites, $x_n$ and $x_{n^\prime}$ are the corresponding Cartesian coordinates in fm, and $V_0=-16.0$~MeV and $R=0.593$~fm specify the strength and range of the force.
The coordinate difference in Eq.~\eqref{eq:Gaussian_potential} is evaluated with periodic boundary conditions.
The resulting ground-state energies are
$E_{2N}=-2.50$~MeV, $E_{3N}=-9.64$~MeV, $E_{4N}=-23.07$~MeV and $E_{5N}=-44.01$~MeV.
We also use the two-body scattering parameters $a_0=4.14$~fm and $r_0=-0.134$~fm to constrain the two-body LECs.
The power counting and calibration procedure follow those adopted
in the preceding sections.
For simplicity, we restrict the calculations to order $\mathcal{O}(Q^2)$,
which includes contact interactions involving up to four particles,
and consider cutoffs $\Lambda\leq700$~MeV.
The five- and six-body energies are therefore predictions.

\begin{figure}
\centering
\includegraphics[width=1.0\columnwidth]{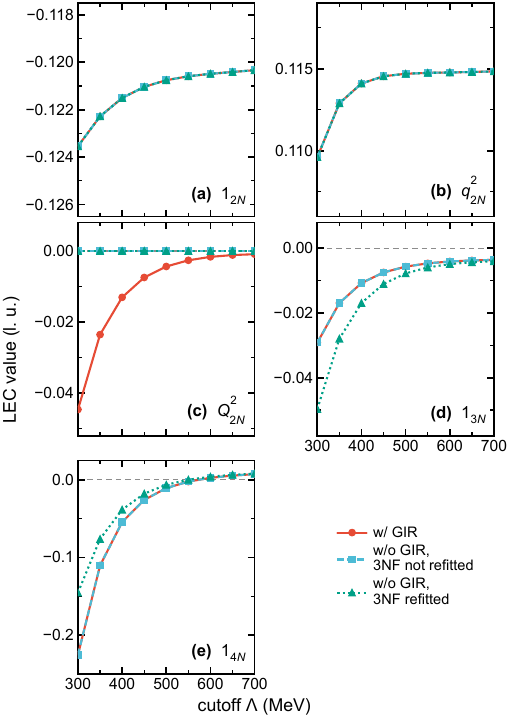}
\caption{LECs for the one-dimensional boson EFT fitted to pseudodata generated by a Gaussian potential. Horizontal dashed lines at zero are shown to guide the eye. All results are given in lattice units (l.u.).}
\label{fig:LECs_from_3_strategies_Gaussian}
\end{figure}

\begin{figure} 
\centering
\includegraphics[width=1.0\columnwidth]{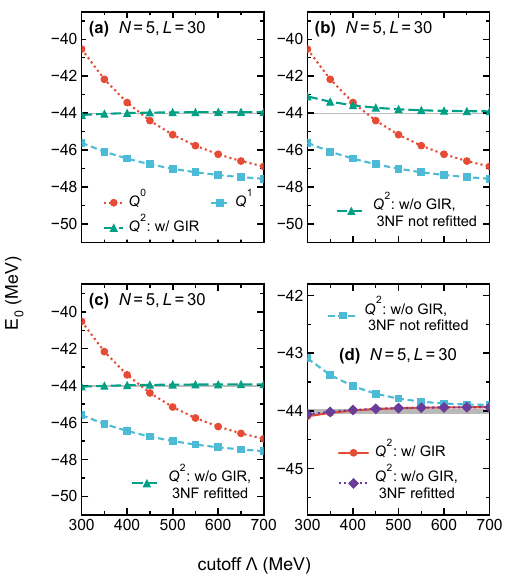}
\caption{Ground-state binding energy of five bosons in a periodic box, calculated with EFT interactions calibrated to pseudodata generated by a Gaussian potential. The results at order $\mathcal{O}(Q^2)$ are obtained using three different strategies for handling the GIR terms. (a) Full GIR terms are included in both the fitting and the calculations. (b) GIR terms are included in the fitting but neglected in the calculations. (c) GIR terms are neglected in both the fitting and the calculations, with the three-body force refitted to the three-body binding energy to absorb their effects. (d) Direct comparison of the results at order $\mathcal{O}(Q^2)$ from the three strategies. Corresponding LECs are shown in Fig.~\ref{fig:LECs_from_3_strategies_Gaussian}.}
\label{fig:energies_from_3_strategies_Gaussian}
\end{figure} 

Fig.~\ref{fig:LECs_from_3_strategies_Gaussian} shows the LECs fitted to the pseudodata generated from
this underlying interaction.
Several coefficients, including $C_{1_{2N}}$ and $C_{q^2_{2N}}$,
remain appreciably nonzero at the upper end of the cutoff range studied.
In particular, the persistence of the derivative coupling
$C_{q^2_{2N}}$ distinguishes this case from the $\delta$-potential
benchmark.
Panels (a), (b), and (c) of Fig.~\ref{fig:energies_from_3_strategies_Gaussian} show the five-body ground-state energies
obtained using the three prescriptions for treating
$O_{Q^2_{2N}}$ described in the preceding section.
Horizontal gray lines indicate the pseudodata.
The results at orders $\mathcal{O}(Q^0)$ and $\mathcal{O}(Q^1)$ are identical in all three panels.
All three prescriptions yield reasonable agreement with the pseudodata
at order $\mathcal{O}(Q^2)$.
Omitting the GIR term without refitting the remaining LECs produces
a deviation of approximately $1$~MeV, as shown in Fig.~\ref{fig:energies_from_3_strategies_Gaussian}(b).
Refitting the three-body coupling to the three-body ground-state
energy largely removes this deviation in the five-body system,
yielding close agreement with the pseudodata and very weak
cutoff dependence.
Fig.~\ref{fig:energies_from_3_strategies_Gaussian}(d) directly compares the three sets of results
at order $\mathcal{O}(Q^2)$.
The additional cutoff dependence induced by omitting the GIR term
is largely removed by recalibrating the three-body interaction,
showing that its effects on the five-body energy can be effectively
absorbed into the three-body LEC.
The corresponding numerical results are listed in Table~\ref{tab:lecs-energies-cutoff}.

\begin{table*}[t]
\caption{Low-energy constants (LECs) at order $\mathcal{O}(Q^2)$ and ground-state energies
for $N=5$ and $L=30$, as functions of the cutoff $\Lambda$.
Scheme A includes Galilean-invariance restoration (GIR).
Scheme B omits GIR without refitting the three-body force (3NF).
Scheme C omits GIR with the 3NF refitted to the three-body binding energy.
\textit{All} indicates values common to A, B, and C within the displayed precision.
Rows that vanish at all cutoffs are omitted.
LECs are in lattice units, and energies are in MeV.}
\label{tab:lecs-energies-cutoff}
\centering
\begingroup
\fontsize{9}{10.5}\selectfont
\setlength{\tabcolsep}{2.5pt}
\renewcommand{\arraystretch}{1.12}
\begin{tabular*}{\textwidth}{@{\extracolsep{\fill}}ll*{9}{r}@{}}
\hline\hline
\multicolumn{2}{l}{$\Lambda$ (MeV)} & \multicolumn{1}{c}{$300$} & \multicolumn{1}{c}{$350$} & \multicolumn{1}{c}{$400$} & \multicolumn{1}{c}{$450$} & \multicolumn{1}{c}{$500$} & \multicolumn{1}{c}{$550$} & \multicolumn{1}{c}{$600$} & \multicolumn{1}{c}{$650$} & \multicolumn{1}{c}{$700$} \\
\hline
\multicolumn{11}{l}{\textit{LECs at order $\mathcal{O}(Q^2)$ (lattice units)}} \\[1pt]
$1_{2N}$ & All & $-0.124$ & $-0.122$ & $-0.122$ & $-0.121$ & $-0.121$ & $-0.121$ & $-0.120$ & $-0.120$ & $-0.120$ \\
$q^2_{2N}$ & All & $0.110$ & $0.113$ & $0.114$ & $0.115$ & $0.115$ & $0.115$ & $0.115$ & $0.115$ & $0.115$ \\
$Q^2_{2N}$ & A & $-0.0446$ & $-0.0235$ & $-0.0130$ & $-0.00747$ & $-0.00438$ & $-0.00264$ & $-0.00168$ & $-0.00117$ & $-0.000929$ \\
$1_{3N}$ & A, B & $-0.0292$ & $-0.0169$ & $-0.0107$ & $-0.00746$ & $-0.00566$ & $-0.00465$ & $-0.00408$ & $-0.00375$ & $-0.00357$ \\
$1_{3N}$ & C & $-0.0497$ & $-0.0279$ & $-0.0169$ & $-0.0110$ & $-0.00775$ & $-0.00591$ & $-0.00488$ & $-0.00431$ & $-0.00402$ \\
$1_{4N}$ & A, B & $-0.225$ & $-0.110$ & $-0.0546$ & $-0.0260$ & $-0.0108$ & $-0.00235$ & $0.00247$ & $0.00537$ & $0.00721$ \\
$1_{4N}$ & C & $-0.145$ & $-0.0763$ & $-0.0385$ & $-0.0179$ & $-0.00646$ & $0.0000751$ & $0.00395$ & $0.00637$ & $0.00800$ \\
\hline
\multicolumn{11}{l}{\textit{Ground-state energies (MeV)}} \\[1pt]
$\mathcal{O}(Q^0)$ & All & $-40.528$ & $-42.163$ & $-43.421$ & $-44.401$ & $-45.163$ & $-45.757$ & $-46.223$ & $-46.590$ & $-46.882$ \\
$\mathcal{O}(Q^1)$ & All & $-45.600$ & $-46.084$ & $-46.459$ & $-46.765$ & $-47.007$ & $-47.197$ & $-47.345$ & $-47.461$ & $-47.551$ \\
$\mathcal{O}(Q^2)$ & A & $-44.092$ & $-44.033$ & $-43.991$ & $-43.967$ & $-43.953$ & $-43.945$ & $-43.940$ & $-43.937$ & $-43.935$ \\
$\mathcal{O}(Q^2)$ & B & $-43.088$ & $-43.378$ & $-43.571$ & $-43.701$ & $-43.786$ & $-43.840$ & $-43.871$ & $-43.888$ & $-43.896$ \\
$\mathcal{O}(Q^2)$ & C & $-44.059$ & $-44.017$ & $-43.983$ & $-43.963$ & $-43.951$ & $-43.943$ & $-43.939$ & $-43.936$ & $-43.934$ \\
$E_{\mathrm{bare}}$ & All & $-44.014$ & $-44.014$ & $-44.014$ & $-44.014$ & $-44.014$ & $-44.014$ & $-44.014$ & $-44.014$ & $-44.014$ \\
\hline\hline
\end{tabular*}
\endgroup
\end{table*}

\subsection{Characteristic momentum}

A central question in applying EFT is how to identify the
characteristic momentum that governs the low-energy expansion.
For free-space scattering, the relevant scales are set by the
incoming and outgoing relative momenta.
These momenta should lie well below the EFT breakdown scale,
with the regulator cutoff chosen sufficiently large to resolve them.
For a bound system of $N$ particles with ground-state energy $E<0$,
one possible estimate is $\kappa_{\mathrm{tot}}=\sqrt{2M|E|}$,
where $M$ is the particle mass.
If this estimate determined the applicability of the EFT,
the growth of the total binding energy with particle number
would eventually make $\kappa_{\mathrm{tot}}$ exceed a fixed
cutoff $\Lambda$, even when the binding energy per particle
remains approximately constant.
Such a criterion would impose increasingly restrictive conditions
on applications to heavier nuclei, which can contain hundreds
of nucleons.
Identifying the appropriate momentum scale is therefore essential
for assessing the applicability of nuclear EFT to many-body systems.

\begin{figure} 
\centering
\includegraphics[width=1.0\columnwidth]{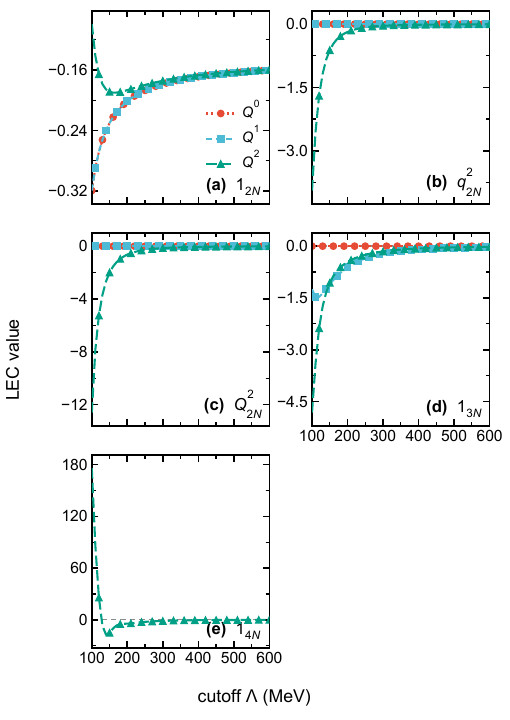}
\caption{LECs for the one-dimensional boson EFT fitted to few-body pseudodata generated by a delta potential with enhanced strength. Horizontal dashed lines at zero are shown to guide the eye. All results are given in lattice units (l.u.).}
\label{fig:LECs_enhanced_delta}
\end{figure}

Successful applications of nuclear EFT to many-body systems motivate
considering an alternative momentum estimate based on the binding
energy per particle.
We examine this possibility numerically within our one-dimensional
EFT framework.
We adopt the underlying $\delta$-function potential of Eq.~\eqref{eq:delta_potential}
and increase the magnitude of its attractive coupling, replacing $V_0$ by $1.5V_0$.
The stronger attraction increases the binding-energy magnitudes
and associated momentum scales, allowing us to examine convergence
as these scales approach the regulator cutoff.
The resulting pseudodata are $E_{2N}=-4.99$~MeV, $E_{3N}=-19.97$~MeV, $E_{4N}=-49.91$~MeV and 
$E_{5N,0}=-99.82$~MeV.
The two-body scattering parameters $a_0=2.88$~fm and $r_0=-1.01\times10^{-4}$~fm are also taken as input to constrain the LECs.
We perform calculations through order $\mathcal{O}(Q^2)$, retaining the GIR
term $O_{Q^2_{2N}}$ explicitly and following the same calibration
procedure as in the preceding sections.
The resulting LECs as functions of $\Lambda$ are shown in Fig.~\ref{fig:LECs_enhanced_delta}.

\begin{figure} 
\centering
\includegraphics[width=1.0\columnwidth]{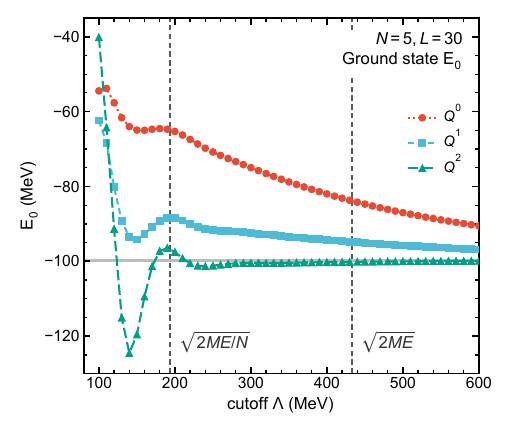}
\caption{Ground-state binding energy of five bosons in a periodic box, calculated with EFT interactions calibrated to data generated by a \(\delta\)-potential with an enhanced coupling constant. The horizontal gray line at \(E_{5N,0}=-99.82~\mathrm{MeV}\) gives the pseudodata binding energy from the underlying \(\delta\)-potential. The vertical dashed lines represent breakdown scales estimated using the total binding energy \(|E|\) and the binding energy per particle \(|E|/N\), respectively.}
\label{fig:energies_enhanced_delta}
\end{figure} 

The five-body ground-state energy is then predicted without
additional fitting.
Fig.~\ref{fig:energies_enhanced_delta} shows the results at successive EFT orders as functions
of the cutoff $\Lambda$.
The horizontal gray line indicates the pseudodata obtained
from the underlying $\delta$-function potential.
At order $\mathcal{O}(Q^2)$, the predictions agree well with the pseudodata
for $\Lambda\gtrsim200$~MeV.
The two vertical lines indicate the momentum estimates
\begin{equation}
    \kappa_{\mathrm{pp}}=\sqrt{\frac{2M|E|}{N}},
    \qquad
    \kappa_{\mathrm{tot}}=\sqrt{2M|E|},
\end{equation}
based on the binding energy per particle and the total binding
energy, respectively.
For $N=5$ and $E=E_{5N}\simeq-100$~MeV, these estimates are
approximately $194$ and $433$~MeV, differing by a factor
of $\sqrt{5}$.
As the cutoff is lowered, order-by-order convergence persists
below $\kappa_{\mathrm{tot}}$ and remains visible down to cutoffs
near $\kappa_{\mathrm{pp}}$.
When $\Lambda$ is reduced below $\kappa_{\mathrm{pp}}$,
this convergence pattern is lost.
For this benchmark, these results support a momentum estimate
based on the binding energy per particle and show that an estimate
based on the total binding energy can impose an unnecessarily
restrictive cutoff condition.

\section{Summary and perspective}
\label{sec:summary}

We present a constructive proof of the renormalization group (RG)
invariance of a Hamiltonian-based effective field theory (EFT) by deriving
step by step the unitary transformation that connects theories constructed
and calibrated at different scales. The essential prerequisite, which differs
from most previous EFT constructions, is the introduction of an
\emph{absolute} single-particle momentum
cutoff~\cite{Lu2022_PRL128-242501,Liu2025_EPJA61-85,Wu2025_PRC112-014009,Shi2026_PLB874-140303}.
This cutoff enables a clean decomposition of the Fock space and avoids the
high-momentum intermediate states that commonly occur for conventional
\emph{relative}-momentum cutoffs in many-body calculations. The resulting
unitary transformation applies simultaneously in both few- and many-body
subspaces, thereby ensuring that few-body-calibrated interactions remain
unitarily equivalent in many-body applications. Although this scheme
introduces a controlled breaking of Galilean invariance, such artifacts are
systematically restored via counterterms or extrapolation. These results have
several implications for nuclear \textit{ab initio} calculations. We also
clarify several conceptual issues concerning cutoffs in EFT and summarize
conclusions already discussed in previous sections.

\paragraph{Role of cutoff} Three questions can be asked regarding the cutoff in nuclear EFT. The first is whether well-defined predictions can be obtained in the infinite-cutoff limit, in the conventional field-theoretic sense of renormalization. Because EFTs contain nonrenormalizable interactions, renormalization generally requires an infinite tower of counterterms when all orders are considered. A suitable power counting nevertheless organizes perturbative calculations such that only a finite set of diagrams and counterterms contributes at any specified order. Divergences can then be absorbed order by order, allowing the cutoff to be formally taken to infinity~\cite{Weinberg1979PhysicaA96-327,Gomis1996_NPB469-473}. This renormalization framework is well established in perturbative EFTs, e.g., Standard Model EFT~\cite{Buchmuller1986_NPB268-621,Grzadkowski2010_JHEP10-085,Brivio2019PhysRept793-1} or chiral perturbation theory~\cite{Gasser1984_AnnPhys158-142,Gasser1985_NPB250-465}.
Its application to nuclear systems is more challenging. The nonperturbative nature of low-energy nuclear interactions requires an all-order resummation of the leading-order terms, while many-body observables generally must be evaluated numerically~\cite{Kaplan1998_PLB424-390,Kaplan1998_NPB534-329,Bedaque1999_PRL82-463}.

The second question is whether experimental data can be reproduced using a particular regularization scheme and cutoff. This practical viewpoint has guided a number of nuclear \textit{ab initio} calculations, which have made substantial progress in extending the reach of these methods to heavy systems, including tin and lead isotopes~\cite{Morris2018_PRL120-152503,Miyagi2022_PRC105-014302,Hu2022_NatPhys18-1196,Tichai2024_PLB851-138571,Hildenbrand2026_PRL136-062501}. In such applications, cutoff dependence is not necessarily treated as a central consistency test, and the choice of regulator and cutoff may itself be used to improve agreement with experiment~\cite{Hebeler2011_PRC83-031301,Simonis2017_PRC96-014303,Jiang2020_PRC102-054301}.
More recently, EFT Hamiltonians have also motivated phenomenological mean-field models employing the same interaction structures~\cite{Mishra2025_arXiv2504-20843}. In these models, EFT primarily guides the construction of forces, while its underlying field-theoretic structure and renormalization properties are not explicitly examined.

The third question, closely related to the first two, concerns how nonperturbative predictions respond to an infinitesimal change in the cutoff. Low-energy observables are expected to remain invariant under such changes when all symmetry-allowed local operators are retained and their couplings evolve consistently with the cutoff. Despite its widespread acceptance, this expectation has received limited systematic numerical scrutiny in nuclear many-body calculations.
Calculations for $A > 4$ at more than one cutoff can be used to examine RG evolution~\cite{Epelbaum2019_PRC99-024313,Maris2021_PRC103-054001,Maris2022_PRC106-064002,Tews2025_PRResearch7-033024}. However, usually only two or three cutoff values are considered, and many-body data are often used to calibrate the interactions~\cite{Hoppe2019_PRC100-024318,Huther2020_PLB808-135651,Jiang2020_PRC102-054301,Drischler2019_PRL122-042501,Drischler2020_PRC102-054315}. Firm conclusions about RG running are therefore difficult to draw.
Systematic calculations involving more than three cutoffs include NNLO\(_{\rm sim}\) calculations of $^{16}$O~\cite{Carlsson2016_PRX6-011019}, calculations of nuclear and neutron matter~\cite{Drischler2019_PRL122-042501}, and neutron-matter calculations with different coordinate-space regulators~\cite{Tews2025_PRResearch7-033024}.

In the present and related works~\cite{Lu2026_in_preparation}, we examine the conditions under which theories constructed and calibrated at different cutoffs yield equivalent low-energy predictions. We test these conditions using an exactly solvable toy model and calculations with realistic chiral interactions for nuclei up to $^{40}$Ca and nuclear matter. Our results suggest that applying a common regulator to all single-particle modes naturally yields RG-invariant many-body predictions from few-body input alone. The effects of Galilean-invariance breaking can be mostly absorbed into the three-body couplings, with residual impacts substantially smaller than anticipated from power counting.
The numerical results agree with the expectation from the standard Wilsonian RG analysis, either within the path-integral or unitary transformation formalism.

\paragraph{Connection to lattice regularization} The absolute momentum regularization used here and in related works is inspired in part by established principles of lattice field theory~\cite{Wilson1974_PRD10-2445,Kogut1983_RMP55-775,Thacker1991_PRD43-196,Lepage1992_PRD46-4052,Lee2004_PRC70-014007,Borasoy2007_EPJA31-105,Lee2009_PPNP63-117}. It can be viewed as a continuum implementation of lattice regularization. From the perspective of renormalization, the essential feature is that the regulator acts directly on field momentum modes, suppressing high single-particle momenta consistently across all particle-number sectors.
It is instructive to compare with relativistic field theories such as QCD, where particle number is not conserved and subspaces with different particle numbers are naturally coupled by the Hamiltonian and regularized on an equal footing. On a spatial lattice, the independent momentum modes are restricted to the first Brillouin zone, with each spatial momentum component satisfying $-\pi/a<p_i\le\pi/a$.
In the nonrelativistic nuclear EFT considered here, nucleon-number conservation decouples sectors with different nucleon numbers. Nevertheless, maintaining the predictive connection between few- and many-body systems requires these prescriptions to remain consistent with a common underlying EFT. A natural choice is to regulate all field modes uniformly, following the same field-level prescription used in relativistic lattice theories. This is achieved by the common single-particle regulator employed here, which provides a natural starting point for Wilsonian RG evolution. Conserved nucleon number plays a role analogous to that of baryon number in QCD.

\paragraph{Comparison with NRQCD} A closer comparison is provided by nonrelativistic QCD (NRQCD), an EFT of the strong interaction among heavy quarks~\cite{Caswell1986_PLB167-437,Bodwin1995_PRD51-1125,Manohar1997_PRD56-230,Luke2000_PRD61-074025,Brambilla2005_RMP77-1423}. NRQCD describes heavy quarks through a nonrelativistic expansion and performs calculations in sectors with fixed numbers of heavy quarks and antiquarks. NRQCD retains explicit gluon fields, while nuclear EFT includes pion fields. Both also contain short-distance contact interactions. The two theories differ in their power-counting hierarchies. Four-quark contact operators are subleading in standard quarkonium power counting, whereas nuclear $S$-wave contacts can be strongly enhanced to leading order. Moreover, most NRQCD applications involve only a few heavy constituents. Despite these differences, the treatment of NRQCD provides an important reference for understanding the renormalization of nuclear EFTs.

NRQCD provides a well-developed example of relating different regularization and renormalization prescriptions. In continuum formulations, renormalization can be performed perturbatively, order by order, using dimensional regularization or a suitable momentum cutoff. Bound-state dynamics may require appropriate resummations. The limit of physical dimension or infinite cutoff can be safely taken after the divergences have been absorbed by the corresponding counterterms~\cite{Manohar1997_PRD56-230,Luke1998_PRD57-413,Pineda1998_PRD58-114011,Luke2000_PRD61-074025}. Alternatively, in lattice NRQCD, observables are calculated nonperturbatively using a lattice action whose coefficients are fixed by matching. The lattice spacing must resolve the characteristic momentum $Q$ while remaining below the breakdown scale characterized by the heavy-quark mass $m_Q$, $Q\ll a^{-1}\lesssim m_Q$. Predictions should remain stable within this window when the operators, including discretization corrections, are consistently included~\cite{Thacker1991_PRD43-196,Lepage1992_PRD46-4052,Morningstar1993_PRD48-2265,Davies1994_PRD50-6963}. When implemented consistently, the low-momentum predictions from both schemes should agree within the expected EFT uncertainties.

In nuclear EFT, the KSW scheme resums the leading-order contact interactions to account for the large NN $S$-wave scattering lengths, while treating subleading contact interactions, pion exchange, and other corrections perturbatively according to a specified power counting~\cite{Kaplan1998_PLB424-390,Kaplan1998_NPB534-329}. Ultraviolet divergences are canceled order by order before removing the cutoff or taking the physical-dimension limit in dimensional regularization. This cancellation requires subleading interactions to be included consistently at the prescribed perturbative order. The procedure is analogous to perturbative renormalization in NRQCD.
When interactions beyond leading order are instead iterated nonperturbatively with bare couplings, a cutoff-independent $\Lambda\to\infty$ limit is generally not available within a fixed truncation. Causality constraints such as the Wigner bound can obstruct this limit, leading to singular or complex bare couplings~\cite{Wigner1955_PR98-145,Phillips1997_PLB390-7,Scaldeferri1997_PRC56-679,Beane1998_NPA632-445,Hammer2010_AnnPhys325-2212}. Most nuclear \textit{ab initio} calculations adopt a finite-cutoff, nonperturbative approach, which is analogous in this respect to lattice NRQCD. The cutoff is typically chosen at or below the chiral breakdown scale, $\Lambda_b\approx500$~MeV. These implementations must therefore be distinguished. A renormalization prescription that permits cutoff removal under a consistent perturbative treatment does not, by itself, justify taking $\Lambda$ far above $\Lambda_b$ when the same interactions are iterated nonperturbatively.

Our focus is the stability of nonperturbative predictions under finite or
infinitesimal cutoff variations, without adopting a particular perturbative
renormalization scheme or its associated power counting. We restrict our analysis to contact interactions and do not address
the renormalization of pion-exchange potentials. Even for contact
interactions alone, establishing consistent renormalization in many-body
systems with many-body forces remains an important open problem. Recent
examples include studies of pionless
EFT~\cite{Contessi2017_PLB772-839,Bansal2018_PRC98-054301,Drissi2020_EPJA56-119,Contessi2025_arXiv2505-09299}
and multi-boson
systems~\cite{Bazak2016_PRA94-052502,Carlson2017_PRL119-223002,Bazak2019_PRL122-143001,DeLeon2022_arXiv2211-00165,Contessi2024_PRA109-022814}.

\paragraph{Galilean and Lorentz symmetry} NRQCD illustrates how Lorentz symmetry can be realized order by order in a nonrelativistic theory through operators suppressed by powers of the particle mass, with relations among their coefficients enforced by reparametrization invariance~\cite{Luke1992_PLB286-348,Manohar1997_PRD56-230,Brambilla2003_PLB576-314,Heinonen2012_PRD86-094020,Berwein2019_PRD99-094008,Lepage1992_PRD46-4052}. In nuclear EFT, Galilean invariance represents the leading nonrelativistic limit of the underlying boost symmetry.
Relativistic corrections include the $p^4$ and $p^6$ contributions to the kinetic energy, spin-dependent terms associated with Wigner rotations under boosts, and interactions that depend explicitly on the center-of-mass momentum. Although some of these corrections share operator structures with the GIR terms introduced here, their physical origins are distinct. Relativistic corrections reflect the underlying Lorentz symmetry, whereas GIR counterterms compensate for spurious boost dependence induced by the regulator.
Unlike translation and rotation invariance, Galilean invariance is only an approximate symmetry of the underlying relativistic theory. Its exact preservation by the regulator is therefore not essential, provided that regulator artifacts suppressed by inverse powers of $\Lambda$ are compensated by appropriate counterterms and that relativistic corrections suppressed by inverse powers of the nucleon mass $M$ are included consistently.

\paragraph{Comparison of regularization schemes} The key distinction between our framework and conventional calculations lies in how few-body interactions are embedded into many-body sectors. In the conventional constructions, the interaction vertices and regulator functions are combined into Galilean-invariant few-body potentials. These regulated potentials are then embedded into larger systems through their dependence on relative or internal momenta. Galilean invariance is thus imposed on the regulator as well as on the interaction vertices. In our framework, the regulator instead acts universally on the underlying single-particle field modes across all particle-number sectors. Our numerical results demonstrate that these prescriptions can yield substantially different many-body predictions, as exemplified by the nuclear matter calculations in Ref.~\cite{Lu2026_in_preparation}, where the two regularization schemes yield exactly the same LECs and phase shifts in the two-body center-of-mass frame, but predict qualitatively different equations of state for nuclear matter.

Our focus is the equivalence of Hamiltonians constructed independently at different regulator cutoffs and calibrated to the same low-energy observables. The central question is how to construct and embed these interactions consistently across particle-number sectors so that their many-body predictions remain equivalent. In a unitary formulation, this equivalence can be demonstrated by constructing an underlying hidden decoupling transformation that relates the resulting Hamiltonians consistently in both few- and many-body sectors.

Our focus should be distinguished from the stability of predictions from
SRG-evolved interactions within the model space after including SRG-induced
many-body forces. The essential difference is that such studies typically
apply the unitary transformation to the Hamiltonian after embedding the
free-space interactions into the model space, which is spanned either by
plane waves for uniform systems~\cite{Hebeler2012_PRC85-021002,Wendt2013_PRC87-061001}
or by a harmonic-oscillator basis for finite
nuclei~\cite{Jurgenson2009_PRL103-082501,Jurgenson2011_PRC83-034301,Roth2011_PRL107-072501,Roth2014_PRC90-024325}.
This embedding is often carried out by treating both the vertices and the
regulator functions together as a Galilean-invariant potential. In contrast to
our scheme, the resulting matrix elements between basis wave functions differ
fundamentally. For example, for a plane-wave basis as in
Ref.~\cite{Hebeler2012_PRC85-021002}, the Galilean-invariant regulator allows
nonzero matrix elements between states with large center-of-mass momentum but
small relative momentum. This induces single-particle transitions to states
above $\Lambda$, as illustrated in Sec.~\ref{subsec:momentum-leakage} and Fig.~\ref{fig:partition}, whereas the
absolute-momentum regulator strictly forbids such transitions as well as the
corresponding matrix elements among these states. Therefore, starting from
exactly the same 2NF in the two-body center-of-mass frame, the different
embedding schemes lead to different Hamiltonian matrix elements in the model
space. The subsequent SRG evolution within the model space therefore acts on
the embedded Hamiltonian, so that the stability of observables reflects the
output of that same Hamiltonian. Consequently, such calculations do not
demonstrate the equivalence of Hamiltonians constructed and calibrated at
different scales, which is the focus of this and related works. In other
words, we attempt to build the hidden unitary transformation between
preexisting Hamiltonians, whereas the SRG studies explicitly apply the
unitary transformation to generate evolved Hamiltonians that are unknown
before the transformation. This difference is also related to the concept of
implicit and explicit renormalization in Ref.~\cite{RuizArriola2014_PLB728-596}.

\paragraph{Regulator effects in EFT predictions} 
It has long been known that the form of regularization has a significant impact on nuclear EFT predictions~\cite{Calci2016_PRC94-014322,Elhatisari2016_PRL117-132501,Soma2020_PRC101-014318,Hagen2014_PRC89-014319,Dyhdalo2016_PRC94-034001}.
In most calculations, however, the regulator forms are specifically chosen to facilitate either partial-wave analysis or many-body calculations. 
Direct comparisons of different regulators and cutoff values remain relatively rare and lack a systematic treatment.
We note that cutoff independence relies on a delicate cancellation between variations in contributions from different operators. 
Using operator-specific regulators and cutoffs thus risks disrupting these cancellations, which may lead to predictions that depend arbitrarily on the regularization choice.
We note that while the regulator dependence is usually attributed to missing higher-order terms, the consistency of the regularization scheme itself can also be an important source of regulator dependence.
We therefore suggest a systematic study comparing the various regularization schemes, particularly their consequences for many-body predictions.

\paragraph{Validity of EFT} An important issue in EFT regularization is the range of systems for which the EFT is applicable. For scattering, it is straightforward that the EFT applies whenever the incoming and outgoing momenta are all well below the cutoff. For many-body systems, the characteristic momentum can be estimated by either the average momentum or the Fermi momentum for fermions. Since the many-body wave function is much more complicated than in the two-body system, a natural question arises as to whether interacting particles within the medium can be close to or exceed the cutoff $\Lambda$, thereby violating the validity condition of the EFT. This question can be answered naturally within the universal regularization scheme introduced here.

Suppose that an underlying high-energy theory among nucleons, either phenomenological or constructed from QCD, is available at a relatively high momentum $\Lambda_b$. The exact wave function from such a theory may contain significant high-momentum components, including configurations in which a relatively large portion of the energy is concentrated on one or a few nucleons with momentum $p\sim\Lambda_b$. Such configurations occur with relatively low probability but seem to violate the EFT validity condition if we assume a cutoff $\Lambda \ll \Lambda_b$. They therefore cannot be described by such an EFT. However, by introducing a cutoff at $\Lambda_b$ and successively lowering it to $\Lambda$ through decoupling of the Hilbert spaces, we obtain a low-momentum EFT that reproduces the same low-momentum observables as the original high-momentum theory. The configurations containing high-momentum particles are thus decoupled and disappear from the final wave functions. The EFT wave function with a low cutoff $\Lambda$ contains only configurations with particle momenta no higher than $\Lambda$. We note that it is only valid to ask questions about low-momentum observables for such an EFT. The question of whether the wave function contains large-momentum particles with $p > \Lambda$ involves momentum scales beyond the capability of this EFT and cannot be meaningfully addressed in this context. In contrast, the same question can be asked for the original theory defined at $\Lambda_b$. We stress that the promise of EFT is that only low-momentum observables are preserved. To measure high-momentum components, one should consult the corresponding theories defined at the appropriate scale. We also note that the conventional Galilean-invariant regulator does lead to wave functions containing particles with momentum larger than $\Lambda$ in many-body calculations, as illustrated in Fig.~\ref{fig:partition}.

The role of the cutoff in the wave functions can be seen more clearly by inspecting the nucleon separation energies and wave-function tails. For the underlying theory at $\Lambda_b$, the wave function can oscillate rapidly inside the nucleus, while its tail varies more slowly and is governed by the single-particle separation momentum $\kappa \sim\sqrt{2mE}$, with $E$ the separation energy. The unitary transformation suppresses the oscillations within the nucleus without altering the wave-function tail. From the Schr\"odinger equation outside the nucleus, it immediately follows that the separation energy is unchanged. After successively applying the transformation, the oscillations within the nucleus are gradually removed, with the tails intact, until the cutoff reaches the separation momentum, $\kappa \sim \Lambda$. Therefore, an efficient method for exploring the validity range of the EFT is to lower the cutoff until the predictions show significant deviations, as performed for the one-dimensional toy model here and for realistic chiral forces in Ref.~\cite{Lu2026_in_preparation}.

\paragraph{Long-range interactions}
In this work and related studies, we focus on contact terms, which are particularly relevant to nuclear EFTs, where they appear at leading order and determine the nonperturbative dynamics of nuclear systems. This framework can also be applied to other nonrelativistic systems described by EFTs.
The same framework can be extended to long-range interactions, including pion-exchange potentials. Imposing momentum cutoffs on both pion and nucleon modes yields an OPEP regulated through a combination of local and nonlocal factors. Its cutoff dependence can be investigated using the same strategy as in the present work and related studies. A naive implementation can, however, break chiral symmetry and compromise consistent renormalization~\cite{Espriu1994_NPB418-494,Djukanovic2005_PRD72-045002,Long2016_PRC93-044003,Krebs2024_PRC110-044004}. This issue is analogous to the need to preserve gauge symmetry when discretizing QCD on a lattice. Implementing chiral symmetry consistently within this regularization framework therefore poses an important theoretical challenge.

The renormalization of the OPEP, particularly its short-distance singularity structure, has been extensively studied~\cite{Nogga2005_PRC72-054006,Birse2006_PRC74-014003}.
In one class of renormalization schemes, contact interactions are introduced systematically to absorb the sensitivity to the singular short-range behavior at large cutoffs, while subleading contact interactions are treated perturbatively~\cite{Long2011_PRC84-057001,Long2012_PRC85-034002,Long2012_PRC86-024001}.
Despite apparent convergence over broad cutoff intervals, such schemes can exhibit large deviations at exceptional cutoff values, associated with strongly oscillatory short-distance wave functions when the cutoff reaches the GeV scale~\cite{Gasparyan2023_PRC107-034001}.
These studies have largely focused on NN and few-body systems~\cite{Song2017_PRC96-024002}, and their implications for heavier nuclei remain incompletely understood~\cite{Yang2021_PRC103-054304}.

In practical nuclear \textit{ab initio} calculations beyond the lightest nuclei, the OPEP is typically regularized using either nonlocal regulators similar to those applied to contact interactions~\cite{Entem2003_PRC68-041001,Entem2017_PRC96-024004} or local regulators that preserve the pion-pole structure~\cite{Epelbaum2015_EPJA51-53,Reinert2018_EPJA54-86}.
The cutoffs are generally moderate and comparable to those used for the contact interactions, so the unregularized short-distance singularity is not directly probed.
Variations of the pion-exchange cutoffs must nevertheless be accompanied by a consistent refit of the relevant LECs to pion-nucleon or nucleon-nucleon data~\cite{Epelbaum2005_NPA747-362,Reinert2018_EPJA54-86}.
In the Supplemental Material of Ref.~\cite{Lu2026_in_preparation}, we demonstrate that varying the OPEP cutoff redistributes contributions between pion exchange and contact interactions while leaving the total binding energies of $^4$He and $^{16}$O nearly unchanged.
Demonstrating this compensation is important for establishing the regulator stability and predictive reliability of the theory.

\paragraph{Practical route towards predictive calculations} In practice, our prescription can be implemented in NLEFT and shell-model calculations by replacing relative-momentum regulators with a common regulator acting on absolute single-particle momenta across all particle-number sectors. The central point is that phase equivalence or agreement in few-body observables does not by itself establish many-body equivalence. Even identical few-body interactions can produce different many-body Hamiltonians when embedded using different regulator prescriptions. Equivalence must therefore be examined after embedding, through a unitary relation consistent across sectors or controlled numerical comparisons.

Single-particle regularization provides a continuum realization of the field-level regulation underlying lattice EFTs. Its role extends beyond numerical discretization by implementing a common separation of momentum modes across particle-number sectors. Regulator-induced Galilean-invariance breaking can be treated systematically with counterterms, following the logic of Symanzik improvement. Within our unitary construction, consistently matched low-energy Hamiltonians at different cutoffs yield equivalent predictions up to controlled truncation effects, independently of the \textit{ab initio} solver. Once these effects are accounted for, cutoff stability provides a stringent check of numerical accuracy.

\begin{acknowledgments}
We are grateful for discussions with S.-G. Zhou and members of the
Nuclear Lattice Effective Field Theory Collaboration. We thank Dean
Lee and Y.-Z. Ma for valuable remarks on the manuscript. This work was supported by the Science Challenge Project (No. TZ2025012), NSAF (No. U2330401), and the National Natural Science Foundation of China under Grant Nos. 12275259 and 12547105.
\end{acknowledgments}


\begin{thebibliography}{999}

\bibitem{Weinberg1979PhysicaA96-327}
S.~Weinberg, Phenomenological Lagrangians, Physica A \textbf{96}, 327 (1979).

\bibitem{Georgi1993ARNPS43-209}
H.~Georgi, Effective Field Theory, Annu. Rev. Nucl. Part. Sci. \textbf{43}, 209 (1993).

\bibitem{Brivio2019PhysRept793-1}
I.~Brivio and M.~Trott, The standard model as an effective field theory, Phys. Rep. \textbf{793}, 1 (2019).

\bibitem{Epelbaum2009RMP81-1773}
E.~Epelbaum, H.-W.~Hammer, and U.-G.~Mei\ss{}ner, Modern theory of nuclear forces, Rev. Mod. Phys. \textbf{81}, 1773 (2009).

\bibitem{Machleidt2011PhysRept503-1}
R.~Machleidt and D.~R.~Entem, Chiral effective field theory and nuclear forces, Phys. Rep. \textbf{503}, 1 (2011).

\bibitem{Hammer2020RMP92-025004}
H.-W.~Hammer, S.~K\"{o}nig, and U.~van Kolck, Nuclear effective field theory, Status and perspectives, Rev. Mod. Phys. \textbf{92}, 025004 (2020).

\bibitem{Braaten2006PhysRept428-259}
E.~Braaten and H.-W.~Hammer, Universality in few-body systems with large scattering length, Phys. Rep. \textbf{428}, 259 (2006).

\bibitem{Wilson1971PRB4-3174}
K.~G.~Wilson, Renormalization Group and Critical Phenomena. I. Renormalization Group and the Kadanoff Scaling Picture, Phys. Rev. B \textbf{4}, 3174 (1971).

\bibitem{Wilson1971PRB4-3184}
K.~G.~Wilson, Renormalization Group and Critical Phenomena. II. Phase-Space Cell Analysis of Critical Behavior, Phys. Rev. B \textbf{4}, 3184 (1971).

\bibitem{Wilson1974PhysRept12-75}
K.~G.~Wilson and J.~Kogut, The renormalization group and the $\epsilon$ expansion, Phys. Rep. \textbf{12}, 75 (1974).

\bibitem{Polchinski1984NPB231-269}
J.~Polchinski, Renormalization and effective lagrangians, Nucl. Phys. B \textbf{231}, 269 (1984).

\bibitem{Birse1999_PLB464-169}
M.~C.~Birse, J.~A.~McGovern, and K.~G.~Richardson, A renormalisation-group treatment of two-body scattering, Phys. Lett. B \textbf{464}, 169 (1999).

\bibitem{Valderrama2016_IJMPE25-1641007}
M.~Pav{\'o}n~Valderrama, Power counting and Wilsonian renormalization in nuclear effective field theory, Int. J. Mod. Phys. E \textbf{25}, 1641007 (2016).

\bibitem{Epelbaum2017_NPB925-161}
E. Epelbaum, J. Gegelia, and U.-G. Mei{\ss}ner, Wilsonian renormalization group versus subtractive renormalization in effective field theories for nucleon-nucleon scattering, Nucl. Phys. B \textbf{925}, 161 (2017).

\bibitem{Epelbaum2018_EPJA54-186}
E. Epelbaum, A. M. Gasparyan, J. Gegelia, and U.-G. Mei{\ss}ner, How (not) to renormalize integral equations with singular potentials in effective field theory, Eur. Phys. J. A \textbf{54}, 186 (2018).

\bibitem{Epelbaum2020_EPJA56-152}
E.~Epelbaum, A.~M.~Gasparyan, J.~Gegelia, U.-G.~Mei\ss{}ner, and X.-L.~Ren, How to renormalize integral equations with singular potentials in effective field theory, Eur. Phys. J. A \textbf{56}, 152 (2020).

\bibitem{Contessi2024_PLB856-138903}
L.~Contessi, M.~Pavon Valderrama, and U.~van Kolck, Limits on an improved action for contact effective field theory in two-body systems, Phys. Lett. B \textbf{856}, 138903 (2024).

\bibitem{Bedaque2000_NPA676-357}
P.~F.~Bedaque, H.-W.~Hammer, and U.~van Kolck, Effective theory of the triton, Nucl. Phys. A \textbf{676}, 357 (2000).

\bibitem{Harada2013_PRD87-085006}
K.~Harada, H.~Kubo, and I.~Yoshimoto, Wilsonian renormalization group analysis of nonrelativistic three-body systems without introducing dimerons, Phys. Rev. D \textbf{87}, 085006 (2013).

\bibitem{Epelbaum2017_EPJA53-98}
E.~Epelbaum, J.~Gegelia, U.-G.~Mei\ss{}ner, and D.-L.~Yao, Renormalization of the three-boson system with short-range interactions revisited, Eur. Phys. J. A \textbf{53}, 98 (2017).

\bibitem{Nogga2005_PRC72-054006}
A.~Nogga, R.~G.~E.~Timmermans, and U.~v.~Kolck, Renormalization of one-pion exchange and power counting, Phys. Rev. C \textbf{72}, 054006 (2005).

\bibitem{Long2011_PRC84-057001}
B.~Long and C.-J.~Yang, Renormalizing chiral nuclear forces, A case study of $^{3}P_{0}$, Phys. Rev. C \textbf{84}, 057001 (2011).

\bibitem{Long2012_PRC86-024001}
B.~Long and C.-J.~Yang, Short-range nuclear forces in singlet channels, Phys. Rev. C \textbf{86}, 024001 (2012).

\bibitem{Long2012_PRC85-034002}
B.~Long and C.-J.~Yang, Renormalizing chiral nuclear forces, Triplet channels, Phys. Rev. C \textbf{85}, 034002 (2012).

\bibitem{Kirscher2013_PLB721-335}
J.~Kirscher, Zero-energy neutron-triton and proton-Helium-3 scattering with EFT($\not\!\pi$), Phys. Lett. B \textbf{721}, 335 (2013).

\bibitem{Yang2025_PRC112-014004}
C.-J.~Yang, Further theoretical study on the renormalization group aspect of perturbative corrections, Phys. Rev. C \textbf{112}, 014004 (2025).

\bibitem{Epelbaum2009_EPJA41-341}
E.~Epelbaum and J.~Gegelia, Regularization, renormalization and ``peratization'' in effective field theory for two nucleons, Eur. Phys. J. A \textbf{41}, 341 (2009).

\bibitem{Epelbaum2015_EPJA51-53}
E.~Epelbaum, H.~Krebs, and U.-G.~Mei\ss{}ner, Improved chiral nucleon-nucleon potential up to next-to-next-to-next-to-leading order, Eur. Phys. J. A \textbf{51}, 53 (2015).

\bibitem{Reinert2018_EPJA54-86}
P.~Reinert, H.~Krebs, and E.~Epelbaum, Semilocal momentum-space regularized chiral two-nucleon potentials up to fifth order, Eur. Phys. J. A \textbf{54}, 86 (2018).

\bibitem{Kaplan1998_PLB424-390}
D.~B.~Kaplan, M.~J.~Savage, and M.~B.~Wise, A new expansion for nucleon-nucleon interactions, Phys. Lett. B \textbf{424}, 390 (1998).

\bibitem{Song2017_PRC96-024002}
Y.-H.~Song, R.~Lazauskas, and U.~van Kolck, Triton binding energy and neutron-deuteron scattering up to next-to-leading order in chiral effective field theory, Phys. Rev. C \textbf{96}, 024002 (2017).

\bibitem{Valderrama2017_PRC95-054001}
M.~Pav{\'o}n~Valderrama, M.~S{\'a}nchez~S{\'a}nchez, C.-J.~Yang, B.~Long, J.~Carbonell, and U.~van Kolck, Power counting in peripheral partial waves, The singlet channels, Phys. Rev. C \textbf{95}, 054001 (2017).

\bibitem{Kaplan1996_NPB478-629}
D.~B.~Kaplan, M.~J.~Savage, and M.~B.~Wise, Nucleon-nucleon scattering from effective field theory, Nucl. Phys. B \textbf{478}, 629 (1996).

\bibitem{Kirscher2010_EPJA44-239}
J.~Kirscher, H.~W.~Grie{\ss}hammer, D.~Shukla, and H.~M.~Hofmann, Universal correlations in pion-less EFT with the resonating group model, Three and four nucleons, Eur. Phys. J. A \textbf{44}, 239 (2010).

\bibitem{Entem2003_PRC68-041001}
D.~R.~Entem and R.~Machleidt, Accurate charge-dependent nucleon-nucleon potential at fourth order of chiral perturbation theory, Phys. Rev. C \textbf{68}, 041001(R) (2003).

\bibitem{Epelbaum2005_NPA747-362}
E. Epelbaum, W. Gl{\"o}ckle, and U.-G. Mei{\ss}ner, The two-nucleon system at next-to-next-to-next-to-leading order, Nucl. Phys. A \textbf{747}, 362 (2005).

\bibitem{Entem2017_PRC96-024004}
D.~R.~Entem, R.~Machleidt, and Y.~Nosyk, High-quality two-nucleon potentials up to fifth order of the chiral expansion, Phys. Rev. C \textbf{96}, 024004 (2017).

\bibitem{Bedaque1999_NPA646-444}
P.~F.~Bedaque, H.-W.~Hammer, and U.~van Kolck, The three-boson system with short-range interactions, Nucl. Phys. A \textbf{646}, 444 (1999).

\bibitem{Bedaque1999_PRL82-463}
P.~F.~Bedaque, H.-W.~Hammer, and U.~van Kolck, Renormalization of the Three-Body System with Short-Range Interactions, Phys. Rev. Lett. \textbf{82}, 463 (1999).

\bibitem{Platter2004_PRA70-052101}
L.~Platter, H.-W.~Hammer, and U.-G.~Mei\ss{}ner, Four-boson system with short-range interactions, Phys. Rev. A \textbf{70}, 052101 (2004).

\bibitem{Platter2005_PLB607-254}
L.~Platter, H.-W.~Hammer, and U.-G.~Mei\ss{}ner, On the correlation between the binding energies of the triton and the $\alpha$-particle, Phys. Lett. B \textbf{607}, 254 (2005).

\bibitem{Bazak2019_PRL122-143001}
B.~Bazak, J.~Kirscher, S.~K\"{o}nig, M.~Pav{\'o}n~Valderrama, N.~Barnea, and U.~van Kolck, Four-Body Scale in Universal Few-Boson Systems, Phys. Rev. Lett. \textbf{122}, 143001 (2019).

\bibitem{Avraham2026_PLB872-140100}
L.~H.~Avraham and B.~Bazak, Few-nucleon systems within finite-cutoff pionless EFT, Phys. Lett. B \textbf{872}, 140100 (2026).

\bibitem{Schiavilla2021_PRC103-054003}
R.~Schiavilla, L.~Girlanda, A.~Gnech, A.~Kievsky, A.~Lovato, L.~E.~Marcucci, M.~Piarulli, and M.~Viviani, Two- and three-nucleon contact interactions and ground-state energies of light- and medium-mass nuclei, Phys. Rev. C \textbf{103}, 054003 (2021).

\bibitem{Yang2021_PRC103-054304}
C.-J.~Yang, A.~Ekstr\"{o}m, C.~Forss\'{e}n, and G.~Hagen, Power counting in chiral effective field theory and nuclear binding, Phys. Rev. C \textbf{103}, 054304 (2021).

\bibitem{Sanchez2023_PRC107-064307}
M.~S{\'a}nchez~S{\'a}nchez, D.~D.~Dao, and L.~Bonneau, Cutoff effects in Hartree-Fock calculations at leading order of chiral effective field theory, Phys. Rev. C \textbf{107}, 064307 (2023).

\bibitem{Yang2023_EPJA59-233}
C.-J. Yang, A. Ekstr{\"o}m, C. Forss{\'e}n, G. Hagen, G. Rupak, and U. van Kolck, The importance of few-nucleon forces in chiral effective field theory, Eur. Phys. J. A \textbf{59}, 233 (2023).

\bibitem{Contessi2025_arXiv2505-09299}
L.~Contessi, M.~Sch{\"a}fer, A.~Gnech, A.~Lovato, and U.~van Kolck, A renormalizable theory for not-so-light nuclei, arXiv~2505.09299 [nucl-th] (2025).

\bibitem{Roth2014_PRC90-024325}
R.~Roth, A.~Calci, J.~Langhammer, and S.~Binder, Evolved chiral $NN+3N$ Hamiltonians for \textit{ab initio} nuclear structure calculations, Phys. Rev. C \textbf{90}, 024325 (2014).

\bibitem{Stroberg2021_PRL126-022501}
S.~R.~Stroberg, J.~D.~Holt, A.~Schwenk, and J.~Simonis, \textit{Ab~Initio} Limits of Atomic Nuclei, Phys. Rev. Lett. \textbf{126}, 022501 (2021).

\bibitem{Soma2020_PRC101-014318}
V.~Som{\`a}, P.~Navr{\'a}til, F.~Raimondi, C.~Barbieri, and T.~Duguet, Novel chiral Hamiltonian and observables in light and medium-mass nuclei, Phys. Rev. C \textbf{101}, 014318 (2020).

\bibitem{Maris2022_PRC106-064002}
P.~Maris, R.~Roth, E.~Epelbaum \textit{et al.}, Nuclear properties with semilocal momentum-space regularized chiral interactions beyond N$^2$LO, Phys. Rev. C \textbf{106}, 064002 (2022).

\bibitem{Dyhdalo2016_PRC94-034001}
A.~Dyhdalo, R.~J.~Furnstahl, K.~Hebeler, and I.~Tews, Regulator artifacts in uniform matter for chiral interactions, Phys. Rev. C \textbf{94}, 034001 (2016).

\bibitem{Epelbaum2019_PRC99-024313}
E.~Epelbaum, J.~Golak, K.~Hebeler \textit{et al.}, Few- and many-nucleon systems with semilocal coordinate-space regularized chiral two- and three-body forces, Phys. Rev. C \textbf{99}, 024313 (2019).

\bibitem{Maris2021_PRC103-054001}
P.~Maris, E.~Epelbaum, R.~J.~Furnstahl \textit{et al.}, Light nuclei with semilocal momentum-space regularized chiral interactions up to third order, Phys. Rev. C \textbf{103}, 054001 (2021).

\bibitem{Tsukiyama2011_PRL106-222502}
K.~Tsukiyama, S.~K.~Bogner, and A.~Schwenk, In-Medium Similarity Renormalization Group For Nuclei, Phys. Rev. Lett. \textbf{106}, 222502 (2011).

\bibitem{Hergert2016_PhysRept621-165}
H.~Hergert, S.~K.~Bogner, T.~D.~Morris, A.~Schwenk, and K.~Tsukiyama, The In-Medium Similarity Renormalization Group, A Novel \textit{Ab Initio} Method for Nuclei, Phys. Rep. \textbf{621}, 165 (2016).

\bibitem{Elhatisari2024_Nature630-59}
S.~Elhatisari \textit{et al.}, Wavefunction matching for solving quantum many-body problems, Nature \textbf{630}, 59 (2024).

\bibitem{Ekstrom2015_PRC91-051301}
A.~Ekstr\"{o}m, G.~R.~Jansen, K.~A.~Wendt, G.~Hagen, T.~Papenbrock, B.~D.~Carlsson, C.~Forss\'{e}n, M.~Hjorth-Jensen, P.~Navr\'{a}til, and W.~Nazarewicz, Accurate nuclear radii and binding energies from a chiral interaction, Phys. Rev. C \textbf{91}, 051301(R) (2015).

\bibitem{Drischler2019_PRL122-042501}
C.~Drischler, K.~Hebeler, and A.~Schwenk, Chiral Interactions up to Next-to-Next-to-Next-to-Leading Order and Nuclear Saturation, Phys. Rev. Lett. \textbf{122}, 042501 (2019).

\bibitem{Huther2020_PLB808-135651}
T.~H{\"u}ther, K.~Vobig, K.~Hebeler, R.~Machleidt, and R.~Roth, Family of chiral two- plus three-nucleon interactions for accurate nuclear structure studies, Phys. Lett. B \textbf{808}, 135651 (2020).

\bibitem{Jiang2024_PRC109-L061302}
W.~G.~Jiang, C.~Forss\'{e}n, T.~Dj\"{a}rv, and G.~Hagen, Nuclear-matter saturation and symmetry energy within $\Delta$-full chiral effective field theory, Phys. Rev. C \textbf{109}, L061302 (2024).

\bibitem{Hildenbrand2026_PRL136-062501}
F.~Hildenbrand, S.~Elhatisari, U.-G.~Mei{\ss}ner, H.~Meyer, Z.~Ren, A.~Herten, and M.~Bode, Lattice Calculation of the Sn Isotopes near the Proton Dripline, Phys. Rev. Lett. \textbf{136}, 062501 (2026).

\bibitem{Hoppe2019_PRC100-024318}
J.~Hoppe, C.~Drischler, K.~Hebeler, A.~Schwenk, and J.~Simonis, Probing chiral interactions up to next-to-next-to-next-to-leading order in medium-mass nuclei, Phys. Rev. C \textbf{100}, 024318 (2019).

\bibitem{Weinberg1990_PLB251-288}
S.~Weinberg, Nuclear forces from chiral lagrangians, Phys. Lett. B \textbf{251}, 288 (1990).

\bibitem{Weinberg1991_NPB363-3}
S.~Weinberg, Effective chiral lagrangians for nucleon-pion interactions and nuclear forces, Nucl. Phys. B \textbf{363}, 3 (1991).

\bibitem{Saha2023_PRC107-034002}
S.~K.~Saha, D.~R.~Entem, R.~Machleidt, and Y.~Nosyk, Local position-space two-nucleon potentials from leading to fourth order of chiral effective field theory, Phys. Rev. C \textbf{107}, 034002 (2023).

\bibitem{Somasundaram2024_PRC109-034005}
R.~Somasundaram, J.~E.~Lynn, L.~Huth, A.~Schwenk, and I.~Tews, Maximally local two-nucleon interactions at N$^3$LO in $\Delta$-less chiral effective field theory, Phys. Rev. C \textbf{109}, 034005 (2024).

\bibitem{Wu2025_PRC112-014009}
C.~Wu, T.~Wang, B.-N.~Lu, and N.~Li, Charge-dependent nucleon-nucleon interaction at $N {}^{3}LO$ in nuclear lattice effective field theory, Phys. Rev. C \textbf{112}, 014009 (2025).

\bibitem{Hu2026_PRL137-142502}
B.~S.~Hu, A.~Ekstr\"{o}m, C.~Forss\'{e}n, G.~Hagen, W.~G.~Jiang, T.~Miyagi, and T.~Papenbrock, Neutron Drip Line in Calcium Isotopes from a Chiral Interaction, Phys. Rev. Lett. \textbf{137}, 142502 (2026).

\bibitem{tHooft1972_NPB44-189}
G.~'t Hooft and M.~Veltman, Regularization and renormalization of gauge fields, Nucl. Phys. B \textbf{44}, 189 (1972).

\bibitem{Gasser1984_AnnPhys158-142}
J.~Gasser and H.~Leutwyler, Chiral perturbation theory to one loop, Ann. Phys. \textbf{158}, 142 (1984).

\bibitem{Kaplan1998_NPB534-329}
D.~B.~Kaplan, M.~J.~Savage, and M.~B.~Wise, Two-nucleon systems from effective field theory, Nucl. Phys. B \textbf{534}, 329 (1998).

\bibitem{Fleming2000_NPA677-313}
S.~Fleming, T.~Mehen, and I.~W.~Stewart, NNLO corrections to nucleon-nucleon scattering and perturbative pions, Nucl. Phys. A \textbf{677}, 313 (2000).

\bibitem{Valderrama2011_PRC83-024003}
M.~Pav\'{o}n Valderrama, Perturbative renormalizability of chiral two-pion exchange in nucleon-nucleon scattering, Phys. Rev. C \textbf{83}, 024003 (2011).

\bibitem{Vanasse2013_PRC88-044001}
J.~Vanasse, Fully perturbative calculation of $nd$ scattering to next-to-next-to-leading order, Phys. Rev. C \textbf{88}, 044001 (2013).

\bibitem{Wilson1974_PRD10-2445}
K.~G.~Wilson, Confinement of quarks, Phys. Rev. D \textbf{10}, 2445 (1974).

\bibitem{Kogut1975_PRD11-395}
J.~Kogut and L.~Susskind, Hamiltonian formulation of Wilson's lattice gauge theories, Phys. Rev. D \textbf{11}, 395 (1975).

\bibitem{Kogut1983_RMP55-775}
J.~B.~Kogut, The lattice gauge theory approach to quantum chromodynamics, Rev. Mod. Phys. \textbf{55}, 775 (1983).

\bibitem{Thacker1991_PRD43-196}
B.~A.~Thacker and G.~P.~Lepage, Heavy-quark bound states in lattice QCD, Phys. Rev. D \textbf{43}, 196 (1991).

\bibitem{Lepage1992_PRD46-4052}
G.~P.~Lepage, L.~Magnea, C.~Nakhleh, U.~Magnea, and K.~Hornbostel, Improved nonrelativistic QCD for heavy-quark physics, Phys. Rev. D \textbf{46}, 4052 (1992).

\bibitem{Davies1994_PRD50-6963}
C.~T.~H.~Davies, K.~Hornbostel, A.~Langnau, G.~P.~Lepage, A.~Lidsey, J.~Shigemitsu, and J.~Sloan, Precision $\Upsilon$ spectroscopy from nonrelativistic lattice QCD, Phys. Rev. D \textbf{50}, 6963 (1994).

\bibitem{Lee2004_PRC70-014007}
D.~Lee, B.~Borasoy, and T.~Schaefer, Nuclear lattice simulations with chiral effective field theory, Phys. Rev. C \textbf{70}, 014007 (2004).

\bibitem{Borasoy2007_EPJA31-105}
B.~Borasoy, E.~Epelbaum, H.~Krebs, D.~Lee, and U.-G.~Mei\ss{}ner, Lattice simulations for light nuclei, Chiral effective field theory at leading order, Eur. Phys. J. A \textbf{31}, 105 (2007).

\bibitem{Lee2009_PPNP63-117}
D.~Lee, Lattice simulations for few- and many-body systems, Prog. Part. Nucl. Phys. \textbf{63}, 117 (2009).

\bibitem{Epelbaum2011_PRL106-192501}
E.~Epelbaum, H.~Krebs, D.~Lee, and U.-G.~Mei\ss{}ner, \textit{Ab~Initio} Calculation of the Hoyle State, Phys. Rev. Lett. \textbf{106}, 192501 (2011).

\bibitem{Epelbaum2014_PRL112-102501}
E.~Epelbaum, H.~Krebs, T.~A.~L\"{a}hde, D.~Lee, U.-G.~Mei\ss{}ner, and G.~Rupak, \textit{Ab initio} calculation of the spectrum and structure of $^{16}$O, Phys. Rev. Lett. \textbf{112}, 102501 (2014).

\bibitem{Lahde2014_PLB732-110}
T.~A.~L\"{a}hde, E.~Epelbaum, H.~Krebs, D.~Lee, U.-G.~Mei\ss{}ner, and G.~Rupak, Lattice effective field theory for medium-mass nuclei, Phys. Lett. B \textbf{732}, 110 (2014).

\bibitem{Elhatisari2015_Nature528-111}
S.~Elhatisari, D.~Lee, G.~Rupak, E.~Epelbaum, H.~Krebs, T.~A.~L\"{a}hde, T.~Luu, and U.-G.~Mei\ss{}ner, Ab initio alpha-alpha scattering, Nature \textbf{528}, 111 (2015).

\bibitem{Elhatisari2016_PRL117-132501}
S.~Elhatisari, N.~Li, A.~Rokash \textit{et al.}, Nuclear Binding Near a Quantum Phase Transition, Phys. Rev. Lett. \textbf{117}, 132501 (2016).

\bibitem{Elhatisari2017_PRL119-222505}
S.~Elhatisari, E.~Epelbaum, H.~Krebs, T.~A.~L\"{a}hde, D.~Lee, N.~Li, B.-N.~Lu, U.-G.~Mei\ss{}ner, and G.~Rupak, \textit{Ab~initio} Calculations of the Isotopic Dependence of Nuclear Clustering, Phys. Rev. Lett. \textbf{119}, 222505 (2017).

\bibitem{Lu2019_PLB797-134863}
B.-N.~Lu, N.~Li, S.~Elhatisari, D.~Lee, E.~Epelbaum, and U.-G.~Mei\ss{}ner, Essential elements for nuclear binding, Phys. Lett. B \textbf{797}, 134863 (2019).

\bibitem{Lu2020_PRL125-192502}
B.-N.~Lu, N.~Li, S.~Elhatisari, D.~Lee, J.~E.~Drut, T.~A.~L\"{a}hde, E.~Epelbaum, and U.-G.~Mei\ss{}ner, \textit{Ab~Initio} Nuclear Thermodynamics, Phys. Rev. Lett. \textbf{125}, 192502 (2020).

\bibitem{Lu2022_PRL128-242501}
B.-N.~Lu, N.~Li, S.~Elhatisari, Y.-Z.~Ma, D.~Lee, and U.-G.~Mei{\ss}ner, Perturbative quantum Monte Carlo method for nuclear physics, Phys. Rev. Lett. \textbf{128}, 242501 (2022).

\bibitem{Niu2025_PRL135-222504}
Z.-W.~Niu and B.-N.~Lu, Sign-Problem-Free Nuclear Quantum Monte Carlo Simulation, Phys. Rev. Lett. \textbf{135}, 222504 (2025).

\bibitem{Klein2015_PLB747-511}
N.~Klein, D.~Lee, W.-T.~Liu, and U.-G.~Mei{\ss}ner, Regularization methods for nuclear lattice effective field theory, Phys. Lett. B \textbf{747}, 511 (2015).

\bibitem{Alarcon2017_EPJA53-83}
J.~M.~Alarc{\'o}n, D.~Du, N.~Klein, T.~A.~L{\"a}hde, D.~Lee, N.~Li, B.-N.~Lu, T.~Luu, and U.-G.~Mei{\ss}ner, Neutron-proton scattering at next-to-next-to-leading order in nuclear lattice effective field theory, Eur. Phys. J. A \textbf{53}, 83 (2017).

\bibitem{Klein2018_EPJA54-121}
N.~Klein, S.~Elhatisari, T.~A.~L{\"a}hde, D.~Lee, and U.-G.~Mei{\ss}ner, The Tjon band in nuclear lattice effective field theory, Eur. Phys. J. A \textbf{54}, 121 (2018).

\bibitem{Li2018_PRC98-044002}
N.~Li, S.~Elhatisari, E.~Epelbaum, D.~Lee, B.-N.~Lu, and U.-G.~Mei\ss{}ner, Neutron-proton scattering with lattice chiral effective field theory at next-to-next-to-next-to-leading order, Phys. Rev. C \textbf{98}, 044002 (2018).

\bibitem{Harada2016_PRC94-024004}
K.~Harada, S.~Sasabe, and M.~Yahiro, Numerical study of renormalization group flows of nuclear effective field theory without pions on a lattice, Phys. Rev. C \textbf{94}, 024004 (2016).

\bibitem{Shi2026_PLB874-140303}
J.-A.~Shi, C.-C.~Wang, and B.-N.~Lu, Observation of renormalization group invariance in symmetry-restored nuclear lattice effective field theory, Phys. Lett. B \textbf{874}, 140303 (2026).

\bibitem{Lu2026_in_preparation}
C.-C.~Wang, J.-A.~Shi, and B.-N.~Lu, \textit{et al.},  Predictive Nuclear Effective Field Theory
from Wilsonian Renormalization Group Principle I, Medium-mass nuclei and nuclear matter, arXiv submitted (2026).

\bibitem{Shankar1994_RMP66-129}
R.~Shankar, Renormalization-group approach to interacting fermions, Rev. Mod. Phys. \textbf{66}, 129 (1994).

\bibitem{Harada2006_PLB636-305}
K.~Harada, K.~Inoue, and H.~Kubo, Wilsonian RG and redundant operators in nonrelativistic effective field theory, Phys. Lett. B \textbf{636}, 305 (2006).

\bibitem{Kamefuchi1961_NP28-529}
S.~Kamefuchi, L.~O'Raifeartaigh, and A.~Salam, Change of variables and equivalence theorems in quantum field theories, Nucl. Phys. \textbf{28}, 529 (1961).

\bibitem{Arzt1995_PLB342-189}
C.~Arzt, Reduced effective lagrangians, Phys. Lett. B \textbf{342}, 189 (1995).

\bibitem{Furnstahl2001_NPA689-846}
R.~J.~Furnstahl, H.-W.~Hammer, and N.~Tirfessa, Field redefinitions at finite density, Nucl. Phys. A \textbf{689}, 846 (2001).

\bibitem{Okubo1954_PTP12-603}
S.~Okubo, Diagonalization of Hamiltonian and Tamm-Dancoff equation, Prog. Theor. Phys. \textbf{12}, 603 (1954).

\bibitem{Lee1980_PLB91-173}
S.~Y.~Lee and K.~Suzuki, The effective interaction of two nucleons in the s-d shell, Phys. Lett. B \textbf{91}, 173 (1980).

\bibitem{Suzuki1980_PTP64-2091}
K.~Suzuki and S.~Y.~Lee, Convergent theory for effective interaction in nuclei, Prog. Theor. Phys. \textbf{64}, 2091 (1980).

\bibitem{Suzuki1982_PTP68-246}
K.~Suzuki, Construction of Hermitian Effective Interaction in Nuclei, General Relation between Hermitian and Non-Hermitian Forms, Prog. Theor. Phys. \textbf{68}, 246 (1982).

\bibitem{Glazek1993_PRD48-5863}
S.~D.~G\l{}azek and K.~G.~Wilson, Renormalization of Hamiltonians, Phys. Rev. D \textbf{48}, 5863 (1993).

\bibitem{Wegner1994_AnnPhys506-77}
F.~Wegner, Flow-equations for Hamiltonians, Ann. Phys. (Leipzig) \textbf{3}, 77 (1994).

\bibitem{Bogner2007_PRC75-061001}
S.~K.~Bogner, R.~J.~Furnstahl, and R.~J.~Perry, Similarity renormalization group for nucleon-nucleon interactions, Phys. Rev. C \textbf{75}, 061001(R) (2007).

\bibitem{Jurgenson2009_PRL103-082501}
E.~D.~Jurgenson, P.~Navr\'{a}til, and R.~J.~Furnstahl, Evolution of Nuclear Many-Body Forces with the Similarity Renormalization Group, Phys. Rev. Lett. \textbf{103}, 082501 (2009).

\bibitem{Wegner1973_PRA8-401}
F.~J.~Wegner and A.~Houghton, Renormalization Group Equation for Critical Phenomena, Phys. Rev. A \textbf{8}, 401 (1973).

\bibitem{Anderson2008_PRC77-037001}
E.~Anderson, S.~K.~Bogner, R.~J.~Furnstahl, E.~D.~Jurgenson, R.~J.~Perry, and A.~Schwenk, Block diagonalization using similarity renormalization group flow equations, Phys. Rev. C \textbf{77}, 037001 (2008).

\bibitem{RuizArriola2014_PLB728-596}
E.~Ruiz Arriola, S.~Szpigel, and V.~S.~Tim\'{o}teo, Implicit vs explicit renormalization and effective interactions, Phys. Lett. B \textbf{728}, 596 (2014).

\bibitem{Szpigel1999_arXivnucl-th9906031}
S.~Szpigel and R.~J.~Perry, Simple applications of effective field theory and similarity renormalization group methods, arXiv~nucl-th/9906031 (1999).

\bibitem{Felline2003_PRC68-034003}
C.~Felline, N.~P.~Mehta, J.~Piekarewicz, and J.~R.~Shepard, Low-energy operators in effective theories, Phys. Rev. C \textbf{68}, 034003 (2003).

\bibitem{Valiente2015_FBS56-845}
M.~Valiente and N.~T.~Zinner, Effective Field Theory of Interactions on the Lattice, Few-Body Syst. \textbf{56}, 845 (2015).

\bibitem{Yong2018_PRA97-043630}
S.~Y.~Yong and D.~T.~Son, Effective-field theory for one-dimensional nonrelativistic particles with contact interaction, Phys. Rev. A \textbf{97}, 043630 (2018).

\bibitem{Klein2018_EPJA54-233}
N.~Klein, D.~Lee, and U.-G.~Mei{\ss}ner, Lattice improvement in lattice effective field theory, Eur. Phys. J. A \textbf{54}, 233 (2018).

\bibitem{Jurgenson2009_NPA818-152}
E.~D.~Jurgenson and R.~J.~Furnstahl, Similarity renormalization group evolution of many-body forces in a one-dimensional model, Nucl. Phys. A \textbf{818}, 152 (2009).

\bibitem{Lahde:2019npb}
T.~A.~L{\"a}hde and U.-G.~Mei{\ss}ner, \textit{Nuclear Lattice Effective Field Theory, An Introduction}, Lecture Notes in Physics, Vol.~957 (Springer, Cham, 2019).

\bibitem{Dean:2025}
D.~Lee, {Lattice Effective Field Theory Simulations of Nuclei}, Annu. Rev. Nucl. Part. Sci. \textbf{75}, 109 (2025).

\bibitem{Bogner2010_PPNP65-94}
S.~K.~Bogner, R.~J.~Furnstahl, and A.~Schwenk, From low-momentum interactions to nuclear structure, Prog. Part. Nucl. Phys. \textbf{65}, 94 (2010).

\bibitem{Nielsen1981NPB185-20}
H.~B.~Nielsen and M.~Ninomiya, Absence of neutrinos on a lattice. I. Proof by homotopy theory, Nucl. Phys. B \textbf{185}, 20 (1981).

\bibitem{Lu2014PRD90-034507}
B.-N.~Lu, T.~A.~L\"{a}hde, D.~Lee, and U.-G.~Mei\ss{}ner, Breaking and restoration of rotational symmetry on the lattice for bound state multiplets, Phys. Rev. D \textbf{90}, 034507 (2014).

\bibitem{Lu2015PRD92-014506}
B.-N.~Lu, T.~A.~L\"{a}hde, D.~Lee, and U.-G.~Mei\ss{}ner, Breaking and restoration of rotational symmetry for irreducible tensor operators on the lattice, Phys. Rev. D \textbf{92}, 014506 (2015).

\bibitem{symanzik1983NPB226-187}
K.~Symanzik, Continuum limit and improved action in lattice theories. I. Principles and $\phi^4$ theory, Nucl. Phys. B \textbf{226}, 187 (1983).

\bibitem{Li2019_PRC99-064001}
N.~Li, S.~Elhatisari, E.~Epelbaum, D.~Lee, B.-N.~Lu, and U.-G.~Mei\ss{}ner, Galilean invariance restoration on the lattice, Phys. Rev. C \textbf{99}, 064001 (2019).


\bibitem{schwartz2014book}
M.~D.~Schwartz, \textit{Quantum Field Theory and the Standard Model} (Cambridge University Press, Cambridge, 2014).

\bibitem{Ren2018cpc}
X.-L.~Ren, K.-W.~Li, L.-S.~Geng, B.~Long, P.~Ring, and J.~Meng, Leading order relativistic chiral nucleon-nucleon interaction, Chin. Phys. C \textbf{42}, 014103 (2018).

\bibitem{li2016prd}
K.-W.~Li, X.-L.~Ren, L.-S.~Geng, and B.~Long, Strangeness $S=-1$ hyperon-nucleon scattering in covariant chiral effective field theory, Phys. Rev. D \textbf{94}, 014029 (2016).

\bibitem{SHEN2019PPNP}
S.~Shen, H.~Liang, W.~H.~Long, J.~Meng, and P.~Ring, Towards an ab initio covariant density functional theory for nuclear structure, Prog. Part. Nucl. Phys. \textbf{109}, 103713 (2019).

\bibitem{Bodwin1995_PRD51-1125}
G.~T.~Bodwin, E.~Braaten, and G.~P.~Lepage, Rigorous QCD analysis of inclusive annihilation and production of heavy quarkonium, Phys. Rev. D \textbf{51}, 1125 (1995).

\bibitem{Luke1992_PLB286-348}
M.~Luke and A.~V.~Manohar, Reparameterisation invariance constraints on heavy particle effective field theories, Phys. Lett. B \textbf{286}, 348 (1992).

\bibitem{Manohar1997_PRD56-230}
A.~V.~Manohar, Heavy quark effective theory and nonrelativistic QCD Lagrangian to order $1/m^3$, Phys. Rev. D \textbf{56}, 230 (1997).

\bibitem{nogga2000prl}
A.~Nogga, H.~Kamada, and W.~Gl\"{o}ckle, Modern nuclear force predictions for the $\alpha$ particle, Phys. Rev. Lett. \textbf{85}, 944 (2000).

\bibitem{epelbaum2001prl}
E.~Epelbaum, H.~Kamada, A.~Nogga, H.~Wita\l{}a, W.~Gl\"{o}ckle, and U.-G.~Mei\ss{}ner, Three- and Four-Nucleon Systems from Chiral Effective Field Theory, Phys. Rev. Lett. \textbf{86}, 4787 (2001).

\bibitem{Kamada2001PRC}
H.~Kamada, A.~Nogga, W.~Gl\"{o}ckle \textit{et al.}, Benchmark test calculation of a four-nucleon bound state, Phys. Rev. C \textbf{64}, 044001 (2001).

\bibitem{Binder2016_PRC93-044002}
S.~Binder, A.~Calci, E.~Epelbaum \textit{et al.}, Few-nucleon systems with state-of-the-art chiral nucleon-nucleon forces, Phys. Rev. C \textbf{93}, 044002 (2016).

\bibitem{wang2024prc}
P.~Y.~Wang, J.~G.~Li, S.~Zhang, Q.~Yuan, M.~R.~Xie, and W.~Zuo, \textit{Ab initio} calculations with a new local chiral N$^3$LO nucleon-nucleon force, Phys. Rev. C \textbf{109}, 064316 (2024).

\bibitem{Shirokov2016_PLB761-87}
A.~M.~Shirokov, I.~J.~Shin, Y.~Kim, M.~Sosonkina, P.~Maris, and J.~P.~Vary, $\mathrm{N}^{3}\mathrm{LO}$ $NN$ interaction adjusted to light nuclei in ab exitu approach, Phys. Lett. B \textbf{761}, 87 (2016).

\bibitem{Logoteta2016_PLB758-449}
D.~Logoteta, I.~Bombaci, and A.~Kievsky, Nuclear matter saturation with chiral three-nucleon interactions fitted to light nuclei properties, Phys. Lett. B \textbf{758}, 449 (2016).

\bibitem{Ekstrom2013_PRL110-192502}
A.~Ekstr\"{o}m, G.~Baardsen, C.~Forss\'{e}n \textit{et al.}, Optimized Chiral Nucleon-Nucleon Interaction at Next-to-Next-to-Leading Order, Phys. Rev. Lett. \textbf{110}, 192502 (2013).

\bibitem{Binder2014_PLB736-119}
S.~Binder, J.~Langhammer, A.~Calci, and R.~Roth, Ab initio path to heavy nuclei, Phys. Lett. B \textbf{736}, 119 (2014).

\bibitem{Machleidt2023FewBodySyst64-77}
R.~Machleidt, What is ab initio?, Few-Body Syst. \textbf{64}, 77 (2023).

\bibitem{Hebeler2011_PRC83-031301}
K.~Hebeler, S.~K.~Bogner, R.~J.~Furnstahl, A.~Nogga, and A.~Schwenk, Improved nuclear matter calculations from chiral low-momentum interactions, Phys. Rev. C \textbf{83}, 031301(R) (2011).

\bibitem{Simonis2017_PRC96-014303}
J.~Simonis, S.~R.~Stroberg, K.~Hebeler, J.~D.~Holt, and A.~Schwenk, Saturation with chiral interactions and consequences for finite nuclei, Phys. Rev. C \textbf{96}, 014303 (2017).

\bibitem{Liu2025_EPJA61-85}
J.~Liu, T.~Wang, and B.-N.~Lu, Perturbative quantum Monte Carlo calculation with high-fidelity nuclear forces, Eur. Phys. J. A \textbf{61}, 85 (2025).

\bibitem{Gomis1996_NPB469-473}
J.~Gomis and S.~Weinberg, Are nonrenormalizable gauge theories renormalizable?, Nucl. Phys. B \textbf{469}, 473 (1996).

\bibitem{Buchmuller1986_NPB268-621}
W.~Buchm{\"u}ller and D.~Wyler, Effective Lagrangian analysis of new interactions and flavour conservation, Nucl. Phys. B \textbf{268}, 621 (1986).

\bibitem{Grzadkowski2010_JHEP10-085}
B.~Grzadkowski, M.~Iskrzy{\'n}ski, M.~Misiak, and J.~Rosiek, Dimension-six terms in the Standard Model Lagrangian, J. High Energy Phys. \textbf{10} (2010) 085.

\bibitem{Gasser1985_NPB250-465}
J.~Gasser and H.~Leutwyler, Chiral perturbation theory, Expansions in the mass of the strange quark, Nucl. Phys. B \textbf{250}, 465 (1985).

\bibitem{Morris2018_PRL120-152503}
T.~D.~Morris, J.~Simonis, S.~R.~Stroberg, C.~Stumpf, G.~Hagen, J.~D.~Holt, G.~R.~Jansen, T.~Papenbrock, R.~Roth, and A.~Schwenk, Structure of the Lightest Tin Isotopes, Phys. Rev. Lett. \textbf{120}, 152503 (2018).

\bibitem{Miyagi2022_PRC105-014302}
T.~Miyagi, S.~R.~Stroberg, P.~Navr{\'a}til, K.~Hebeler, and J.~D.~Holt, Converged \textit{ab initio} calculations of heavy nuclei, Phys. Rev. C \textbf{105}, 014302 (2022).

\bibitem{Hu2022_NatPhys18-1196}
B.~Hu \textit{et al.}, Ab initio predictions link the neutron skin of $^{208}$Pb to nuclear forces, Nat. Phys. \textbf{18}, 1196 (2022).

\bibitem{Tichai2024_PLB851-138571}
A.~Tichai, P.~Demol, and T.~Duguet, Towards heavy-mass ab initio nuclear structure, Open-shell Ca, Ni and Sn isotopes from Bogoliubov coupled-cluster theory, Phys. Lett. B \textbf{851}, 138571 (2024).

\bibitem{Jiang2020_PRC102-054301}
W.~G.~Jiang, A.~Ekstr\"{o}m, C.~Forss\'{e}n, G.~Hagen, G.~R.~Jansen, and T.~Papenbrock, Accurate bulk properties of nuclei from $A=2$ to $\infty$ from potentials with $\Delta$ isobars, Phys. Rev. C \textbf{102}, 054301 (2020).

\bibitem{Mishra2025_arXiv2504-20843}
C.~Mishra, A.~Ekstr{\"o}m, G.~Hagen, M.~Menickelly, T.~Papenbrock, and S.~M.~Wild, A nuclear mass model rooted in chiral effective field theory, arXiv~2504.20843 [nucl-th] (2025).

\bibitem{Tews2025_PRResearch7-033024}
I.~Tews, R.~Somasundaram, D.~Lonardoni, H.~G\"{o}ttling, R.~Seutin, J.~Carlson, S.~Gandolfi, K.~Hebeler, and A.~Schwenk, Neutron matter from local chiral effective field theory interactions at large cutoffs, Phys. Rev. Research \textbf{7}, 033024 (2025).

\bibitem{Drischler2020_PRC102-054315}
C.~Drischler, J.~A.~Melendez, R.~J.~Furnstahl, and D.~R.~Phillips, Quantifying uncertainties and correlations in the nuclear-matter equation of state, Phys. Rev. C \textbf{102}, 054315 (2020).

\bibitem{Carlsson2016_PRX6-011019}
B.~D.~Carlsson, A.~Ekstr\"{o}m, C.~Forss\'{e}n, D.~F.~Str\"{o}mberg, G.~R.~Jansen, O.~Lilja, M.~Lindby, B.~A.~Mattsson, and K.~A.~Wendt, Uncertainty Analysis and Order-by-Order Optimization of Chiral Nuclear Interactions, Phys. Rev. X \textbf{6}, 011019 (2016).

\bibitem{Caswell1986_PLB167-437}
W.~E.~Caswell and G.~P.~Lepage, Effective lagrangians for bound state problems in QED, QCD, and other field theories, Phys. Lett. B \textbf{167}, 437 (1986).

\bibitem{Luke2000_PRD61-074025}
M.~E.~Luke, A.~V.~Manohar, and I.~Z.~Rothstein, Renormalization group scaling in nonrelativistic QCD, Phys. Rev. D \textbf{61}, 074025 (2000).

\bibitem{Brambilla2005_RMP77-1423}
N.~Brambilla, A.~Pineda, J.~Soto, and A.~Vairo, Effective-field theories for heavy quarkonium, Rev. Mod. Phys. \textbf{77}, 1423 (2005).

\bibitem{Luke1998_PRD57-413}
M.~Luke and M.~J.~Savage, Power counting in dimensionally regularized nonrelativistic QCD, Phys. Rev. D \textbf{57}, 413 (1998).

\bibitem{Pineda1998_PRD58-114011}
A.~Pineda and J.~Soto, Matching at one loop for the four-quark operators in NRQCD, Phys. Rev. D \textbf{58}, 114011 (1998).

\bibitem{Morningstar1993_PRD48-2265}
C.~J.~Morningstar, Heavy-quark self-energy in nonrelativistic lattice QCD, Phys. Rev. D \textbf{48}, 2265 (1993).

\bibitem{Wigner1955_PR98-145}
E.~P.~Wigner, Lower limit for the energy derivative of the scattering phase shift, Phys. Rev. \textbf{98}, 145 (1955).

\bibitem{Phillips1997_PLB390-7}
D.~R.~Phillips and T.~D.~Cohen, How short is too short? Constraining zero-range interactions in nucleon-nucleon scattering, Phys. Lett. B \textbf{390}, 7 (1997).

\bibitem{Scaldeferri1997_PRC56-679}
K.~A.~Scaldeferri, D.~R.~Phillips, C.-W.~Kao, and T.~D.~Cohen, Short-range interactions in an effective field theory approach for nucleon-nucleon scattering, Phys. Rev. C \textbf{56}, 679 (1997).

\bibitem{Beane1998_NPA632-445}
S.~R.~Beane, T.~D.~Cohen, and D.~R.~Phillips, The potential of effective field theory in NN scattering, Nucl. Phys. A \textbf{632}, 445 (1998).

\bibitem{Hammer2010_AnnPhys325-2212}
H.-W.~Hammer and D.~Lee, Causality and the effective range expansion, Ann. Phys. \textbf{325}, 2212 (2010).

\bibitem{Contessi2017_PLB772-839}
L.~Contessi, A.~Lovato, F.~Pederiva, A.~Roggero, J.~Kirscher, and U.~van Kolck, Ground-state properties of $^4$He and $^{16}$O extrapolated from lattice QCD with pionless EFT, Phys. Lett. B \textbf{772}, 839 (2017).

\bibitem{Bansal2018_PRC98-054301}
A.~Bansal, S.~Binder, A.~Ekstr\"{o}m, G.~Hagen, G.~R.~Jansen, and T.~Papenbrock, Pion-less effective field theory for atomic nuclei and lattice nuclei, Phys. Rev. C \textbf{98}, 054301 (2018).

\bibitem{Drissi2020_EPJA56-119}
M.~Drissi, T.~Duguet, and V.~Som\`{a}, Renormalization of pionless effective field theory in the A-body sector, Eur. Phys. J. A \textbf{56}, 119 (2020).

\bibitem{Bazak2016_PRA94-052502}
B.~Bazak, M.~Eliyahu, and U.~van Kolck, Effective field theory for few-boson systems, Phys. Rev. A \textbf{94}, 052502 (2016).

\bibitem{Carlson2017_PRL119-223002}
J.~Carlson, S.~Gandolfi, U.~van Kolck, and S.~A.~Vitiello, Ground-State Properties of Unitary Bosons, From Clusters to Matter, Phys. Rev. Lett. \textbf{119}, 223002 (2017).

\bibitem{DeLeon2022_arXiv2211-00165}
H.~De-Leon and F.~Pederiva, Equation of state of a strongly interacting many-boson system from an effective interaction, arXiv~2211.00165 [cond-mat.quant-gas] (2022).

\bibitem{Contessi2024_PRA109-022814}
L.~Contessi, M.~Sch\"{a}fer, and U.~van Kolck, Improved action for contact effective field theory, Phys. Rev. A \textbf{109}, 022814 (2024).

\bibitem{Brambilla2003_PLB576-314}
N.~Brambilla, D.~Gromes, and A.~Vairo, Poincar\'{e} invariance constraints on NRQCD and potential NRQCD, Phys. Lett. B \textbf{576}, 314 (2003).

\bibitem{Heinonen2012_PRD86-094020}
J.~Heinonen, R.~J.~Hill, and M.~P.~Solon, Lorentz invariance in heavy particle effective theories, Phys. Rev. D \textbf{86}, 094020 (2012).

\bibitem{Berwein2019_PRD99-094008}
M.~Berwein, N.~Brambilla, S.~Hwang, and A.~Vairo, Poincar\'{e} invariance in NRQCD and potential NRQCD revisited, Phys. Rev. D \textbf{99}, 094008 (2019).

\bibitem{Hebeler2012_PRC85-021002}
K.~Hebeler, Momentum-space evolution of chiral three-nucleon forces, Phys. Rev. C \textbf{85}, 021002(R) (2012).

\bibitem{Wendt2013_PRC87-061001}
K.~A.~Wendt, Similarity renormalization group evolution of three-nucleon forces in a hyperspherical momentum representation, Phys. Rev. C \textbf{87}, 061001(R) (2013).

\bibitem{Jurgenson2011_PRC83-034301}
E.~D.~Jurgenson, P.~Navr\'{a}til, and R.~J.~Furnstahl, Evolving nuclear many-body forces with the similarity renormalization group, Phys. Rev. C \textbf{83}, 034301 (2011).

\bibitem{Roth2011_PRL107-072501}
R.~Roth, J.~Langhammer, A.~Calci, S.~Binder, and P.~Navr\'{a}til, Similarity-transformed chiral $NN+3N$ interactions for the \textit{ab initio} description of $^{12}$C and $^{16}$O, Phys. Rev. Lett. \textbf{107}, 072501 (2011).

\bibitem{Calci2016_PRC94-014322}
A.~Calci and R.~Roth, Sensitivities and correlations of nuclear structure observables emerging from chiral interactions, Phys. Rev. C \textbf{94}, 014322 (2016).

\bibitem{Hagen2014_PRC89-014319}
G.~Hagen, T.~Papenbrock, A.~Ekstr\"{o}m, K.~A.~Wendt, G.~Baardsen, S.~Gandolfi, M.~Hjorth-Jensen, and C.~J.~Horowitz, Coupled-cluster calculations of nucleonic matter, Phys. Rev. C \textbf{89}, 014319 (2014).

\bibitem{Espriu1994_NPB418-494}
D.~Espriu and J.~Matias, Regularization methods in chiral perturbation theory, Nucl. Phys. B \textbf{418}, 494 (1994).

\bibitem{Djukanovic2005_PRD72-045002}
D.~Djukanovic, M.~R.~Schindler, J.~Gegelia, and S.~Scherer, Improving the ultraviolet behavior in baryon chiral perturbation theory, Phys. Rev. D \textbf{72}, 045002 (2005).

\bibitem{Long2016_PRC93-044003}
B.~Long and Y.~Mei, Cutoff regulators in chiral nuclear effective field theory, Phys. Rev. C \textbf{93}, 044003 (2016).

\bibitem{Krebs2024_PRC110-044004}
H.~Krebs and E.~Epelbaum, Toward consistent nuclear interactions from chiral Lagrangians. II. Symmetry preserving regularization, Phys. Rev. C \textbf{110}, 044004 (2024).

\bibitem{Birse2006_PRC74-014003}
M.~C.~Birse, Power counting with one-pion exchange, Phys. Rev. C \textbf{74}, 014003 (2006).

\bibitem{Gasparyan2023_PRC107-034001}
A.~M.~Gasparyan and E.~Epelbaum, ``Renormalization-group-invariant effective field theory'' for few-nucleon systems is cutoff dependent, Phys. Rev. C \textbf{107}, 034001 (2023).

\end{thebibliography}

\end{document}